\RequirePackage[bookmarksnumbered,unicode]{hyperref}
\documentclass[manuscript,screen]{acmart}
\AtBeginDocument{%
  \providecommand\BibTeX{{%
    \normalfont B\kern-0.5em{\scshape i\kern-0.25em b}\kern-0.8em\TeX}}}

\setcopyright{acmcopyright}
\copyrightyear{2026}
\acmYear{2026}
\acmDOI{XXXXXXX.XXXXXXX}

\acmJournal{TOSEM}
\acmVolume{1}
\acmNumber{1}
\acmArticle{1}
\acmMonth{1}

\usepackage[utf8]{inputenc}
\usepackage{tcolorbox}
\usepackage{multirow}

\usepackage{tikz}
\usetikzlibrary{positioning,arrows.meta}

\usepackage{tabularx}
\usepackage{booktabs}
\usepackage{longtable}
\usepackage{needspace}
\usepackage{float}
\usepackage{placeins}

\newlength{\ResultBarMax}
\newcommand{\ResultBar}[2]{\textcolor{black!60}{\rule{\dimexpr\ResultBarMax*#1/#2\relax}{1.0ex}}}
\newcommand{\ResultStudies}[2]{\makebox[\ResultBarMax][l]{\ResultBar{#1}{#2}}\hspace{0.3em}\makebox[1.7em][r]{#1}}

\newcommand{\head}[1]{\noindent\textbf{#1.}}

\usepackage{ifthen}
\usepackage{xcolor}

\newboolean{showcomments}
\setboolean{showcomments}{true}

\ifthenelse{\boolean{showcomments}}
{
    \definecolor{myyellow}{RGB}{255,228,26}
    \definecolor{myblue}{RGB}{50,50,220}

    \newcommand{\nb}[2]{%
        {\sf
        \fcolorbox{myyellow}{yellow}{\scriptsize\textbf{#1}}%
        $\blacktriangleright$%
        {\color{myblue}\fontsize{7pt}{8pt}\selectfont\textbf{#2}}%
        }%
    }
}
{
    \newcommand{\nb}[2]{}
}

\begin{document}

\title{The Reality Gap in Automated Driving Systems Testing: Landscape, Challenges, and Research Directions}

\author{Stefano Carlo Lambertenghi}
\email{lambertenghi@fortiss.org}
\orcid{1234-5678-9012}
\affiliation{%
  \institution{Technical University of Munich; fortiss GmbH}
  \streetaddress{Boltzmannstraße 3}
  \city{Garching}
  \state{Bavaria}
  \country{Germany}
  \postcode{85748}
}
\author{Oliver Wei{\ss}l}
\orcid{0009-0008-7575-0187}
\affiliation{%
  \institution{Technical University of Munich; NorCom GmbH \& Co. KGaA}
  \streetaddress{Boltzmannstraße 3}
  \city{Garching}
  \state{Bavaria}
  \country{Germany}
  \postcode{85748}
}
\author{Johannes Betz}
\orcid{0000-0001-9197-2849}
\affiliation{%
  \institution{Technical University of Munich}
  \streetaddress{Boltzmannstraße 15}
  \city{Garching}
  \state{Bavaria}
  \country{Germany}
  \postcode{85748}
}
\author{Andrea Stocco}
\orcid{0000-0001-8956-3894}
\affiliation{%
  \institution{Technical University of Munich; fortiss GmbH}
  \streetaddress{Boltzmannstraße 3}
  \city{Garching}
  \state{Bavaria}
  \country{Germany}
  \postcode{85748}
}

\newcommand\myworries[1]{\textcolor{red}{#1}}

\begin{abstract}
Simulation is essential for testing automated driving systems (ADS), but discrepancies between simulation and reality can undermine testing conclusions. It remains unclear which discrepancies the literature addresses and what evidence supports correspondence claims.
We present a systematic mapping of 235 primary studies that measure or mitigate the reality gap in ADS testing, selected from 4,195 publications. We characterise objectives, targets, approaches, experimental setups, and evaluation evidence. Our classification distinguishes simulated and real-world execution across six ADS test components, separately for each approach and its evaluation reference. This distinguishes real-world information used to construct an approach from evidence used to evaluate its claims.
Of the included studies, 76\% address mitigation, primarily targeting scenario and actor realism and perception, while 94\% report validation or measurement procedures. Only 61 studies report metrics from multiple families covering sensor/perception correspondence, ADS behaviour, vehicle dynamics, or safety outcomes; this coverage does not establish causal links between discrepancies and test verdicts. The synthesis highlights that real-world content and physical execution must be interpreted against the claim evaluated. The findings motivate claim-specific evaluation and component-level reporting to establish which discrepancies affect testing conclusions and when simulation evidence supports conclusions about real-world ADS behaviour.
\end{abstract}


\maketitle

\section{Introduction}\label{sec:intro}

Advanced driver assistance systems (ADAS) and automated driving systems (ADS) operate in an open and largely unconstrained environment~\cite{yurtsever2020survey}. Their behaviour depends on learned perception components, planning and control software, and physical actuators. On-road testing alone cannot establish that such a system is acceptably safe. Demonstrating a fatality rate comparable to human drivers with statistical confidence would require hundreds of millions of kilometres of testing~\cite{RR-1478-RC}. Software changes can require renewed evidence, while on-road testing exposes road users to the failures that the campaign aims to find. Simulation makes rare and hazardous situations reproducible, less expensive to test, and executable in parallel. It is therefore widely used in ADS research and scenario-based safety assessment~\cite{zhong2021survey,sotif,ul4600,arxiv.2204.07874}.

Simulation introduces a different problem. A simulated test relies on models of the road, traffic, sensors, and vehicle dynamics, each approximating reality. The difference between a simulated execution and the corresponding real-world execution is the \emph{simulation-to-reality gap}, or reality gap. This gap can undermine a test verdict in two ways. A simulator that is more benign than reality can create \emph{false confidence} by failing to expose failures that occur in the real system. Conversely, it can create \emph{false alarms} by reporting failures that do not reproduce in reality, directing engineering effort towards simulator artefacts rather than defects of the system under test. These concerns are supported by empirical observations. Offline and online evaluations can disagree on which inputs cause failures~\cite{2020-Haq-ICST,briand-offline-emse,Codevilla}. Failures found in simulation may transfer only partially to the corresponding physical platform~\cite{stocco2023MindGap,lambertenghi2025MultimodalityEvaluation}. Different simulators can produce different results for the same ADAS and differ from reality~\cite{digitalTwinAreNotMonozygotic,agendaMultiSim,biagiola2023better}. Simulator non-determinism can introduce further differences~\cite{AminiFlaky2024}.

Existing approaches include physics-based sensor models, calibrated vehicle dynamics, scenario reconstruction from driving logs and crash reports, learned and generative content synthesis, hardware-, vehicle-, and mixed-reality-in-the-loop testing, domain randomisation, and policy transfer~\cite{sim2real_gans_cit_2,wayve-sim2real,nerf,fathy2006review,chen2019NovelIntegrated,9561240,domain_rando,hu2023sim2real}. Across software engineering, robotics, computer vision, and vehicle engineering, researchers use partially overlapping or inconsistent terms such as \emph{reality gap}, \emph{fidelity}, \emph{realism}, \emph{sim2real}, and \emph{validity}. These terms can concern the faithfulness of a sensor signal, the plausibility of a traffic scenario, the accuracy of a dynamics model, or the transferability of a control policy.

Existing secondary studies cover different parts of this problem. Surveys address ADS testing techniques~\cite{ADS_testing_survey_2022,ZHAO2024122836}, scenario specification and generation~\cite{zhong2021survey}, testing of learned components~\cite{RiccioEMSE20,zhangTSE22}, and simulation platforms and capabilities~\cite{DBLP:journals/corr/abs-2101-05337,s19030648}. The closest work to our scope surveys sim2real and digital twins in autonomous driving~\cite{hu2023sim2real}, primarily organising the field around technologies for transfer and virtual representation. These studies do not jointly characterise the reality gap as a testing problem, the real and simulated execution used by individual studies, and the evaluation references supporting their claims.

This systematic mapping examines which parts of an ADS or its environment are targeted by reality-gap research, how studies measure or mitigate discrepancies, and how real and simulated elements are combined in experimental setups. Following established guidelines for systematic reviews~\cite{ISTguidelines,Petersen-IST-2015}, we analyse 235 primary studies selected from 4{,}195 publications. We characterise their research \emph{context}, proposed \emph{approaches}, \emph{evaluation}, and \emph{experimental components}.
We distinguish \emph{sim2real} approaches, which make simulation-derived artefacts applicable to reality, from \emph{real2sim} approaches, which use real-world information to construct or calibrate simulated content. We also classify execution separately for the proposed approach and its evaluation reference across six components: scenario, actors, perception, localization, actuation, and ECU/control. This separation makes it possible to distinguish how real-world information contributes to an approach from how its claims are evaluated.

Our results show that most studies address mitigation, with research concentrating on scenario and actor realism and perception. Sim2real and real2sim approaches occur at similar overall rates, and experimental setups combine simulation, recorded observations, and physical equipment in diverse ways. These combinations must be interpreted against the research target and evaluation claim. Reducing a discrepancy or including real-world content does not by itself establish correspondence of ADS behaviour or agreement between simulated and physical test verdicts.

This paper makes the following contributions:
\begin{itemize}
    \item a systematic mapping of 235 primary studies on the simulation-to-reality gap in ADAS and ADS testing;
    \item a characterisation of existing approaches, distinguishing sim2real and
          real2sim strategies;
    \item a component-level taxonomy of real and simulated execution across scenario,
          actors, perception, localization, actuation, and ECU/control, applied separately
          to the proposed approach and its evaluation reference;
    \item a characterisation of reported evaluation techniques, references, and metrics
          used to measure the gap or validate mitigation; 
    \item a synthesis relating research targets, mitigation strategies, execution
          configurations, and evaluation evidence, motivating research on how
          discrepancies affect ADS behaviour and testing conclusions. We provide a replication package containing the search results, extracted data, and analysis scripts~\cite{replication-package}.
\end{itemize}

\section{Background}
\label{sec:background}

\subsection{Automated Driving Systems}
\label{sec:bg:ads}

SAE J3016 distinguishes six levels of driving automation, from no automation to full automation~\cite{SAE2021J3016}. Although ADAS and ADS have different responsibilities under this classification, our mapping concerns their testing configurations. We therefore use ADS broadly to include the driving automation systems evaluated in the literature, including systems formally classified as ADAS.

ADS commonly comprise sensing, perception, localization, planning, and control functions~\cite{yurtsever2020survey}. In modular architectures, perception identifies objects and road structures from sensor observations, localization estimates vehicle position and state, planning determines trajectories or manoeuvres, and control produces steering, braking, and acceleration commands. End-to-end architectures replace parts or all of this pipeline with learned mappings from observations to driving decisions or commands~\cite{bojarski2016end,chen2023endtoend}. Other approaches incorporate vision-language models~\cite{2026-Gao-OJ-ITS} or learn policies through reinforcement learning~\cite{kiran2021deep}. Regardless of architecture, closed-loop behaviour depends on interactions between the ADS, vehicle, and environment.

\subsection{Simulation and Reality in ADS Development and Testing}
\label{sec:bg:simulation}

Simulation and synthetic data support component development, training, and evaluation as well as system-level testing. The distinction between simulated and real execution therefore extends beyond the use of a driving simulator.

\head{Synthetic and real-world data}
Synthetic sensor observations, vehicle states, and annotations can complement physical measurements, including recorded datasets such as KITTI~\cite{Geiger2013IJRR} and the Waymo Open Dataset~\cite{Sun2020CVPR}. These sources can also be combined: recordings can be replayed, driving logs can support scenario reconstruction, and physical measurements can calibrate models. Processing recorded real-world observations differs from conducting a new physical execution. The observations originate from reality, but the ADS may process them offline or within an otherwise simulated setup.

\head{Sensor simulation and learned generation}
Conventional sensor simulation models how a sensor observes a virtual environment. Discrepancies can arise from both the environment representation and the sensing model. Learned approaches can instead generate or transform observations, for example through image translation~\cite{sim2real_gans_cit_2,wayve-sim2real}, diffusion models, or neural rendering of reconstructed scenes~\cite{nerf}. An observation remains synthetic even when its appearance or underlying representation is learned from real-world measurements.

\head{Scenario and closed-loop simulation}
A scenario describes the road and environmental conditions, participating actors, and their behaviour over time. Scenarios can be specified manually, generated from parameterised spaces, or reconstructed from logs and reports~\cite{zhong2021survey}. Driving simulators such as CARLA~\cite{Dosovitskiy17}, SVL/LGSVL~\cite{rong2020lgsvl}, and BeamNG~\cite{gambi-beamng} provide environments for executing them. In closed-loop testing, ADS actions affect vehicle motion and subsequent observations. This feedback distinguishes closed-loop execution from processing a fixed sequence of recorded inputs.

\head{Physical and mixed execution}
In-the-loop configurations combine simulation with different forms of system execution. In \emph{model-in-the-loop} (MiL) testing, the driving logic and environment are computational models. In \emph{software-in-the-loop} (SiL) testing, ADS software interacts with a simulated environment. In \emph{hardware-in-the-loop} (HiL) testing, physical hardware, such as an ECU, interacts with simulated inputs or environments~\cite{fathy2006review,chen2019NovelIntegrated}. In \emph{vehicle-in-the-loop} (ViL) testing, a physical vehicle interacts with a partly or fully virtual environment. Mixed- and augmented-reality approaches similarly combine physical execution with virtual actors or stimuli~\cite{9561240}. These labels do not uniquely specify which components are simulated or real.

\begin{figure}[t]
\centering
\includegraphics[width=0.80\linewidth, trim=0 2cm 0 4cm, clip]{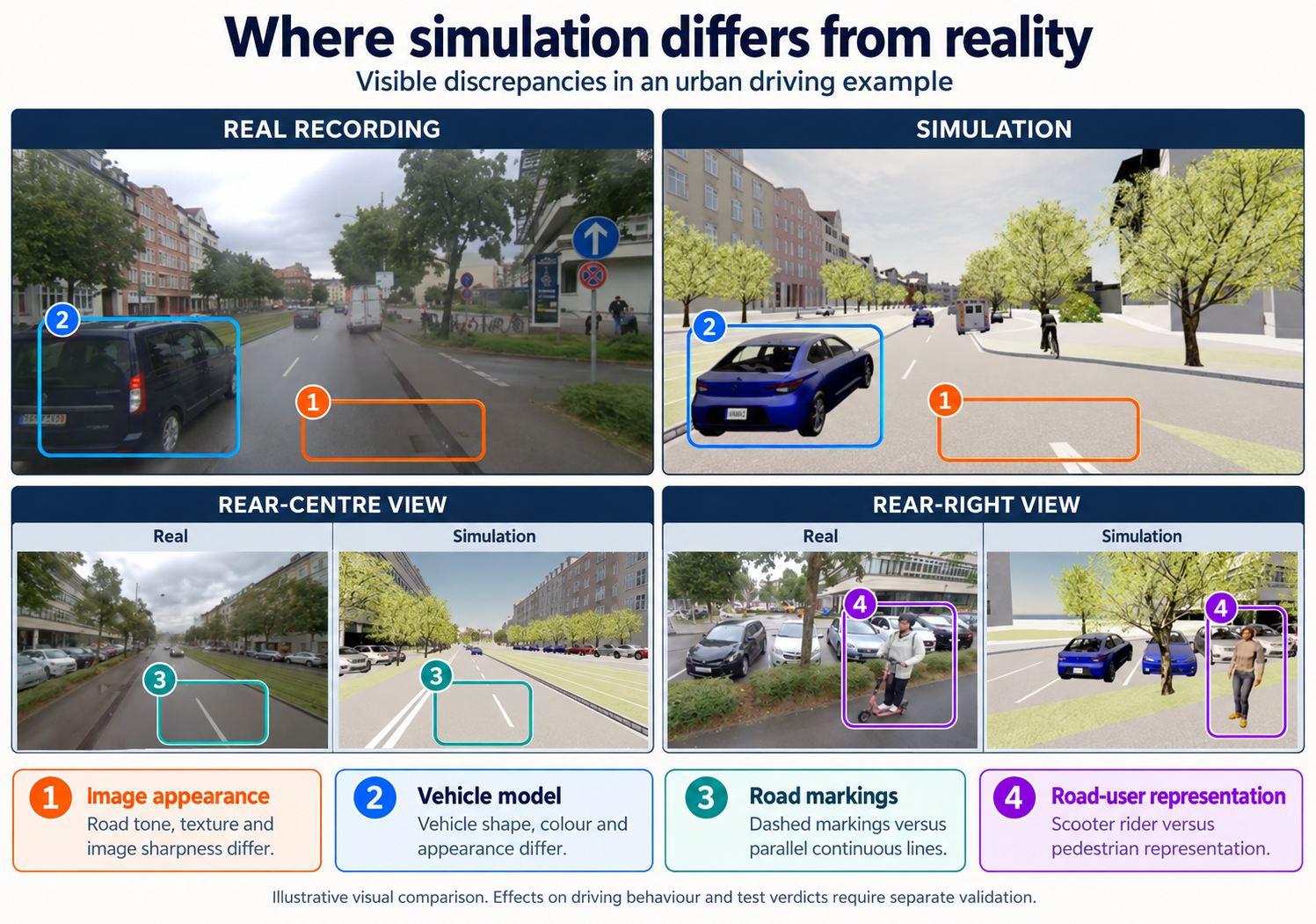}
\caption{Four visual discrepancies between a real camera recording and a comparable simulation: (1)~image appearance, (2)~vehicle model, (3)~road markings, and (4)~road-user representation. Boxes highlight the differences; their impact on ADS behaviour and test verdicts requires separate investigation.}
\label{fig:placeholder}
\end{figure}

Scaled platforms, such as DonkeyCar~\cite{donkeycar2023} and F1TENTH~\cite{pmlr-v123-o-kelly20a}, enable physical closed-loop experiments using real sensors, computing hardware, and actuators at reduced size and cost. Their physical execution does not by itself establish transfer to full-size vehicles. Digital twins provide another connection between domains by associating a physical system or environment with a corresponding virtual representation~\cite{hu2023sim2real}; their implementations differ in the information exchanged and its direction.

\subsection{The Simulation-to-Reality Gap}
\label{sec:bg:gap}

The \emph{simulation-to-reality gap}, or \emph{reality gap}, is the difference between an artifact, behavior, or observation obtained using a simulated or synthetic representation and its corresponding real-world counterpart. It can concern sensor observations, environmental and actor models, or
system properties such as dynamics, timing, and actuation (\autoref{fig:placeholder} illustrates discrepancies at the sensor and scene level). For testing, these discrepancies can affect individual verdicts and the assessment of test adequacy.

Simulation and reality are therefore not necessarily properties of an entire experiment. A physical ECU can process simulated observations, and a real vehicle can interact with virtual traffic. To describe these combinations, we distinguish six components:

\begin{itemize}
    \item \emph{Scenario}: static road characteristics and environmental conditions;
    \item \emph{Actors}: other vehicles, pedestrians, cyclists, and their behaviour;
    \item \emph{Perception}: sensing and the observations provided to the ADS;
    \item \emph{Localization}: estimation of the ego vehicle's position and state;
    \item \emph{Actuation}: steering, braking, throttle, and vehicle dynamics; and
    \item \emph{ECU/control}: the execution platform and driving software.
\end{itemize}

This decomposition describes the experimental setup, rather than the internal architecture of the ADS. It applies to both modular and end-to-end systems. Planning and decision-making remain part of the system under test rather than separate execution components. The operational classification is defined in \autoref{methodology}.

\head{Sim2real and real2sim}
We distinguish two directions in addressing the gap. \emph{Sim2real} approaches make simulation-derived models, data, behaviours, or systems applicable to reality. Examples include improving simulator fidelity and adapting policies trained on synthetic data. \emph{Real2sim} approaches use real-world information to construct, calibrate, or reconstruct simulated content. The directions can coexist: a study may calibrate a simulator using physical measurements and then evaluate whether its outcomes transfer to reality.

\head{Related notions}
\emph{Fidelity} describes how accurately a model represents selected properties of its reference system. \emph{Realism} more broadly concerns resemblance to real observations, scenarios, or behaviours. \emph{Domain shift} concerns differences between data distributions, particularly those used for training and deployment. \emph{Robustness} concerns stability under changes or perturbations and does not itself establish correspondence with reality. Simulator \emph{non-determinism}, or flakiness, causes repeated simulated executions to produce different outcomes~\cite{AminiFlaky2024}; it can complicate gap measurement but does not itself constitute the gap.

We use simulation-to-reality gap as an umbrella term for discrepancies between simulated or synthetic representations and their real-world counterparts. The related notions retain the distinct meanings defined above.

\section{Research Questions} \label{questions}

This survey investigates how research on ADS testing and validation addresses the gap between simulation and real-world execution. We analyse the selected studies from five perspectives: context, proposed approaches, evaluation, and experimental components. The research questions for each perspective are presented below and summarised in~\autoref{tab:table_RQs}.

\begin{table*}[tbp]
\caption{Research questions and what each one captures from a study.}
\label{tab:table_RQs}
\centering

\footnotesize
\renewcommand{\arraystretch}{1.08}
\setlength{\tabcolsep}{4pt}

\begin{tabularx}{\textwidth}{
    >{\raggedright\arraybackslash}p{1.5cm}
    >{\raggedright\arraybackslash}p{7.2cm}
    >{\raggedright\arraybackslash}X
}
\toprule
\textbf{Category} &
\textbf{Research Question} &
\textbf{What it captures} \\
\midrule

Context
& \textbf{RQ\textsubscript{1.1}} What is the primary objective of the study?
& Whether the study measures the gap, mitigates it, or does both. \\
& \textbf{RQ\textsubscript{1.2}} Which aspect(s) of the simulation--reality gap are addressed?
& The aspect of the simulation--reality gap addressed by the study, including combinations of aspects. \\
& \textbf{RQ\textsubscript{1.3}} What is the direction of the simulation--reality relationship?
& Whether the study addresses the gap from simulation to reality, from reality to simulation, or in both directions. \\

\midrule

Approach
& \textbf{RQ\textsubscript{2.1}} Which research target does the study's simulation--reality gap concern?
& The specific capability, component, or phenomenon considered in relation to the simulation--reality gap. \\
& \textbf{RQ\textsubscript{2.2}} How is the proposed simulation--reality mitigation implemented?
& How the gap is mitigated, from the specific technique to its broader family. \\
& \textbf{RQ\textsubscript{2.3}} What are the execution levels of the proposed approach?
& The execution level of each component and functional group across simulation and reality. \\

\midrule

Evaluation
& \textbf{RQ\textsubscript{3.1}} How is the simulation--reality relationship evaluated?
& How the study measures the gap or validates its mitigation, including cases with no explicit evaluation. \\
& \textbf{RQ\textsubscript{3.2}} What are the execution levels of the evaluation reference?
& The same execution-level classification as RQ\textsubscript{2.3}, applied to the evaluation reference. \\
& \textbf{RQ\textsubscript{3.3}} Which metrics are used to quantify the simulation--reality gap or the effectiveness of the proposed approach?
& The quantities the study measures, and whether any quantitative measurement is reported at all. \\

\midrule

Experimental 
& \textbf{RQ\textsubscript{4.1}} Which simulation platforms and reference datasets are employed?
& The simulators and reference datasets used by the study. \\
components 
& \textbf{RQ\textsubscript{4.2}} Which hardware platforms are used?
& The physical equipment involved, by its role in the experimental setup. \\
& \textbf{RQ\textsubscript{4.3}} Which system under test is evaluated?
& The driving system or component being exercised, and the specific technology implementing it. \\
& \textbf{RQ\textsubscript{4.4}} How are evaluation scenarios or test cases generated?
& The origin of the tested scenarios, ranging from manually specified scenarios to scenarios reconstructed from real driving. \\
& \textbf{RQ\textsubscript{4.5}} Which testing artefacts are publicly available?
& Whether code, data, both, or nothing is released. \\



\bottomrule
\end{tabularx}
\end{table*}

\subsection{Context}

We distinguish the study objective, the aspect of the gap addressed, and the direction between simulation and reality.

RQ\textsubscript{1.1} identifies whether a study measures the simulation--reality gap, mitigates it, or does both, because measurement and mitigation studies make different claims.
RQ\textsubscript{1.2} identifies which aspect of the simulation--reality gap each study addresses. The gap may originate from different parts of an ADS testing setup, such as the representation of the driving scenario, the behaviour of traffic participants, the sensing process, localization, or vehicle actuation. A study may also address several of these aspects together or consider the system as a whole.
RQ\textsubscript{1.3} records direction separately because sim2real and real2sim, as defined in \autoref{sec:background}, are independent of whether a study measures or mitigates the gap.

\subsection{Approach}

We characterise each approach by its research target, mitigation technique, and execution across simulation and reality.

RQ\textsubscript{2.1} identifies the specific research target considered in relation to the simulation--reality gap. While RQ\textsubscript{1.2} identifies the broader aspect of the gap, RQ\textsubscript{2.1} captures the particular capability, component, or phenomenon investigated by the authors. For example, studies addressing a perception gap may focus on camera imagery, LiDAR sensing, object detection, or combinations of sensing modalities.
RQ\textsubscript{2.2} characterises how studies that aim to mitigate the gap implement their approach. Different studies may address similar research targets using substantially different techniques. We describe how the approach reduces the gap, from the specific technique to its broader family.
RQ\textsubscript{2.3} characterises how the proposed approach is executed across simulation and reality. We decompose the experimental setup into the six execution components introduced in \autoref{sec:background}: scenario, actors, perception, localization, actuation, and ECU/control. This captures experiments in which different components operate at different levels of simulation and reality. We record the execution level separately from the execution source, such as a live vehicle or a recorded dataset, because the same level can be reached in different ways.
We also group the six components into three functional groups to simplify presentation: scene (scenario and actors), sensing (perception and localization), and control (actuation and ECU/control). Each component and functional group is characterised according to how it is executed across simulation and reality. The execution-level classification and its definitions are provided in \autoref{methodology}.

\subsection{Evaluation}

We characterise how each study evaluates the simulation--reality relationship, including the reference used for comparison and the metrics used to quantify differences. What the evidence establishes depends on the study objective. For ``measure only'' studies, we examine how the simulation--reality gap is measured. For ``mitigate only'' studies, we examine how the proposed mitigation is validated. Measure and mitigate studies contribute to both analyses.

RQ\textsubscript{3.1} records how each study evaluates the simulation--reality relationship, including cases with no explicit evaluation.
An evaluation reference may be real for some components and simulated for others, just like the proposed approach. RQ\textsubscript{3.2} classifies this reference using the same components and execution categories. We compare scene with scene, sensing with sensing, and control with control, and inspect the individual components within each group. We do not assume that a reference with more physical components provides stronger evidence.
RQ\textsubscript{3.3} identifies the metrics used to quantify the simulation--reality gap or evaluate the effectiveness of the proposed mitigation. These metrics quantify the simulation-to-reality gap or the effectiveness of the proposed mitigation. They include measures of perception performance, trajectory behaviour, and sensor similarity.

\subsection{Experimental Components}

These approaches are implemented and evaluated using different simulators, hardware, systems under test, and test cases. Together, these define the experimental setting from which the evidence is obtained. We record the simulation platforms and reference datasets (\textbf{RQ\textsubscript{4.1}}), hardware (\textbf{RQ\textsubscript{4.2}}), system under test (\textbf{RQ\textsubscript{4.3}}), origin of evaluation scenarios or test cases (\textbf{RQ\textsubscript{4.4}}), and publicly available testing artefacts (\textbf{RQ\textsubscript{4.5}}).



\section{Methodology}
\label{methodology}

We follow the systematic mapping guidelines of Kitchenham and Charters~\cite{ISTguidelines}. The methodology covers the identification and selection of primary studies, data extraction, and data analysis. We first establish the review procedure using a pilot set, refine it through a preliminary selection phase, and then apply the resulting procedure to the full corpus.

\subsection{Identification of Research}

\subsubsection{Pilot Study}

Before the systematic search, we selected an initial set of 10 studies, listed in~\autoref{tab:pilot}, to establish and validate the review procedure. The purpose of the pilot is to identify missing or ambiguous categories rather than to provide a representative sample of the literature. We therefore selected studies already known to the authors that covered different objectives, directions, gap aspects, approaches, and execution levels.
The set covers all three objectives and directions. It spans perception, ego behaviour~\cite{stocco2023MindGap}, scenario and traffic realism~\cite{cao2024ReinforcementLearning}, actuation~\cite{oh2024AdvilsImplementation}, and combinations of aspects~\cite{amini2022Vista20}. Its approaches include sensor physics modelling~\cite{vargasrivero2022EffectSpray,manivasagam2020LidarsimRealistic}, generative image techniques~\cite{lambertenghi2024AssessingQuality}, data-driven simulator construction~\cite{amini2022Vista20}, learned traffic behaviour~\cite{cao2024ReinforcementLearning}, and real-equipment integration~\cite{shen2023SimonwheelsPhysical,oh2024AdvilsImplementation}.

The spread of experimental setups was the primary selection criterion because the execution-level classification defined in~\autoref{methodology} must capture different combinations of simulation and reality. VISTA 2.0~\cite{amini2022Vista20} combines sensor observations synthesised from real driving data with simulated vehicle behaviour and deployment on a full-size vehicle. The spray study~\cite{vargasrivero2022EffectSpray} calibrates simulated LiDAR and spray models using real measurements. The feature-embedding study~\cite{gadipudi2022SyntheticReal} compares real and synthetic image datasets without a vehicle or control loop.
Sim-on-Wheels~\cite{shen2023SimonwheelsPhysical} combines a real vehicle, sensors, and control with virtual traffic participants. TrafficRLHF~\cite{cao2024ReinforcementLearning} remains within simulation but uses human preferences and real driving data to improve traffic behaviour. The LiDAR domain-shift study~\cite{huch2023QuantifyingLidar} compares real and simulated sensing while controlling scenario and actor motion, whereas Stocco et al.~\cite{stocco2023MindGap} compare a small-scale physical vehicle with its digital twin. LiDARsim~\cite{manivasagam2020LidarsimRealistic} reconstructs scenes and actors from real captures to generate synthetic LiDAR data. The neural reality-gap study~\cite{lambertenghi2024AssessingQuality} uses image translation and compares ADS behaviour across real, simulated, and translated inputs. Finally, AD-VILS~\cite{oh2024AdvilsImplementation} combines physical actuation, localization, and ECU/control with simulated traffic, actors, and perception inputs.

Together, these studies cover dataset-only comparisons, fully simulated experiments, simulation calibrated or reconstructed from real data, mixed real and simulated execution, and comparisons with physical vehicles. This diversity exposed the need to classify each execution component separately rather than treating an entire study as simulated or real.
Three authors of this paper independently analysed the pilot studies using a preliminary version of the classification scheme. The resulting classifications were compared to identify ambiguities, inconsistencies, and missing categories. The research questions, extraction fields, and coding rules were then refined until consensus was reached. This produced the initial review protocol, which was further refined during the preliminary phase described in~\autoref{selection}.

\begin{table}[tbp]
\centering
\caption{The 10 pilot studies used to calibrate the review protocol and the coded values that motivated their selection.}
\label{tab:pilot}

\scriptsize
\renewcommand{\arraystretch}{1.08}
\begin{tabularx}{\linewidth}{>{\raggedright\arraybackslash}X
>{\raggedright\arraybackslash}p{2.1cm}
>{\raggedright\arraybackslash}p{1cm}
>{\raggedright\arraybackslash}p{1.5cm}}
\toprule
\textbf{Study} & \textbf{Objective} & \textbf{Direction} & \textbf{Gap aspect} \\
\midrule
VISTA 2.0: An Open, Data-driven Simulator for Multimodal Sensing and Policy Learning for Autonomous Vehicles~\cite{amini2022Vista20} & Mitigate & Real2sim & Perception, Behaviour \\
The Effect of Spray Water on an Automotive LIDAR Sensor: A Real-Time Simulation Study~\cite{vargasrivero2022EffectSpray} & Mitigate & Real2sim & Perception \\
Synthetic to Real Gap Estimation of Autonomous Driving Datasets using Feature Embedding~\cite{gadipudi2022SyntheticReal} & Measure & Sim2real & Perception \\
Sim-on-Wheels: Physical World in the Loop Simulation for Self-Driving~\cite{shen2023SimonwheelsPhysical} & Mitigate & Both & Perception \\
Reinforcement Learning with Human Feedback for Realistic Traffic Simulation~\cite{cao2024ReinforcementLearning} & Mitigate & Sim2real & Scenario \\
Quantifying the LiDAR Sim-to-Real Domain Shift: A Detailed Investigation Using Object Detectors and Analysing Point Clouds at Target-Level~\cite{huch2023QuantifyingLidar} & Measure & Sim2real & Perception \\
Mind the Gap! A Study on the Transferability of Virtual Versus Physical-World Testing of Autonomous Driving Systems~\cite{stocco2023MindGap} & Measure & Sim2real & Behaviour \\
LiDARsim: Realistic LiDAR simulation by leveraging the real world~\cite{manivasagam2020LidarsimRealistic} & Mitigate & Real2sim & Perception \\
Assessing Quality Metrics for Neural Reality Gap Input Mitigation in Autonomous Driving Testing~\cite{lambertenghi2024AssessingQuality} & Measure and Mitigate & Sim2real & Perception \\
AD-VILS: Implementation and Reliability Validation of Vehicle-in-the-Loop Simulation Platform for Evaluating Autonomous Driving Systems~\cite{oh2024AdvilsImplementation} & Measure and Mitigate & Both & Actuation \\
\bottomrule
\end{tabularx}
\end{table}

\subsubsection{Database Selection and Search Query}
\label{databases}

The literature search was conducted using Scopus~\cite{Scopus2026}, a bibliographic database of peer-reviewed scientific literature. Scopus was selected for its coverage of engineering and computer science and its support for Boolean searches over titles, abstracts, and keywords.
The search query was developed iteratively using terminology identified during the pilot study. It was refined until all 10 pilot studies were retrieved while limiting the number of irrelevant results.
The query combines three groups of search terms:

\head{Automated driving systems} Terms describing ADAS and ADS, such as \emph{autonomous vehicle}, \emph{self-driving}, and \emph{ADAS}.

\head{Testing and validation} Terms related to software testing and validation, including \emph{test*}, \emph{quality}, and \emph{validat*}.

\head{Simulation-reality relationship} Terms describing the reality gap and related concepts, including \emph{sim2real}, \emph{real2sim}, \emph{reality gap}, \emph{realism}, and in-the-loop testing approaches.

Terms within each group are connected using the \verb|OR| operator, while the three groups are combined using the \verb|AND| operator. The search is restricted to titles, abstracts, and keywords.

\subsubsection{Additional Selection Constraints}

We apply the following constraints:

\head{SC1 (Language)} Only publications written in English.

\head{SC2 (Publication year)} Publications from 2018 onward, to focus on recent research.

\head{SC3 (Peer review)} Only publications from a publisher or venue with a documented, verifiable peer-review process.

The language and publication year constraints were applied through the Scopus search interface~\cite{Scopus2026} as part of the query shown below. The peer-review constraint was applied during post-processing of the exported Scopus records.

SC3 applies a reproducible classification rule to Scopus publisher and source-title metadata. Retained venues include IEEE and ACM publications; peer-reviewed journals from MDPI, Elsevier, Wiley, SAGE, Taylor \& Francis, Springer/Nature, and SAE; IFAC-PapersOnLine; ASME IMECE; SPIE proceedings; ASCE transportation conferences; Springer Proceedings in Advanced Robotics; Proceedings of Machine Learning Research; Journal of Machine Learning Research; Advances in Neural Information Processing Systems; and Driving Simulation Association proceedings.

We exclude venues whose peer-review process cannot be verified, including non-reviewed technical-paper series, edited-book chapters, and generic proceedings series with unconfirmed conference-level review. The replication package provides the executable metadata-based classification rule and supporting evidence for each publisher and venue decision. 
The resulting Scopus query is shown below.

\begin{tcolorbox}
TITLE-ABS-KEY( “test*” OR “assessing” OR “quality” OR “validat*” ) AND TITLE-ABS-KEY(( “simulation” AND ( “real-world” OR “real” ) ) OR “sim2real” OR “real2sim” OR “sim-to-real” OR “real-to-sim” OR “sim to real” OR “real to sim” OR “simulation gap” OR “simulation-gap” OR “reality gap” OR “reality-gap” OR “synthetic gap” OR “realism” OR “vil” OR “hil” OR “scil” OR “sil” OR “in the loop” OR “-in-the-loop”) AND TITLE-ABS-KEY( “self-driving” OR “autonomous vehicle” OR “adaptive cruise control” OR “acc” OR “adas” OR “advanced driver assistance systems” OR “automated driving” OR “ads” OR “automated-driving” OR “self driving” ) AND PUBYEAR > 2017 AND ( LIMIT-TO ( LANGUAGE,“English” ) )
\end{tcolorbox}

When multiple versions of the same publication were available, the most complete version was retained. This affects two included studies: for each, the search retrieved a conference version, but we extracted and cited the extended journal version. Their demographic metadata in RQ\textsubscript{5} therefore also refers to the journal versions.

\subsubsection{Search Results}

The search was performed on 4 May 2026, applying the language and publication year constraints described above. The Scopus query returned 4,195 publications. Applying the peer-review constraint (SC3) excluded 1,009 publications, leaving 3,186 candidate publications for study selection.

\subsection{Study Selection}
\label{selection}

\begin{figure*}[tbp]
\centering
\small

\definecolor{exclusionfill}{RGB}{250,232,232}
\definecolor{surveyfill}{RGB}{232,240,250}
\definecolor{includedfill}{RGB}{235,245,229}
\definecolor{processfill}{RGB}{248,248,248}

\resizebox{\textwidth}{!}{%
\begin{tikzpicture}[
    box/.style={draw, rounded corners=2pt, align=center, inner sep=4pt},
    phase/.style={box, fill=processfill, text width=3.1cm,
                  minimum height=1.1cm},
    outcome/.style={box, text width=2.35cm, minimum height=1.8cm},
    exclusion/.style={outcome, fill=exclusionfill},
    survey/.style={outcome, fill=surveyfill},
    included/.style={outcome, fill=includedfill},
    summary/.style={box, text width=4.1cm, minimum height=1.2cm},
    arrow/.style={thick, -{Latex[length=1.5mm]}},
    combine/.style={thick, dashed},
    crossing/.style={preaction={draw=white, line width=4pt}},
    junction/.style={fill=white, inner sep=2pt, font=\large\bfseries}
]

\node[phase] (search) at (-5.5,0) {
    \textbf{Scopus search}\\
    SC1: English, SC2: Year $\geq$ 2018\\
    4,195 publications
};
\node[phase] (sc3) at (0,0) {\textbf{SC3}\\Peer review};
\node[phase] (candidates) at (5.5,0) {
    \textbf{Candidate corpus}\\3,186 publications
};
\draw[arrow] (search) -- (sc3);
\draw[arrow] (sc3) -- (candidates);

\node[phase] (preliminary) at (-3.6,-2.3) {
    \textbf{Preliminary stage}\\1,112 publications
};
\node[phase] (remaining) at (3.6,-2.3) {
    \textbf{Remaining}\\2,074 publications
};
\draw[thick] (candidates.south) -- (5.5,-1.2) -- (-3.6,-1.2);
\draw[arrow] (-3.6,-1.2) -- (preliminary.north);
\draw[arrow] (3.6,-1.2) -- (remaining.north);

\node[exclusion] (preexcluded) at (-7.25,-5.65) {
    \textit{Preliminary}\\
    \textbf{Relevance exclusions}\\
    999 publications\\second reviewer check
};
\node[exclusion] (fullexcluded) at (-4.35,-5.65) {
    \textit{Full selection}\\
    \textbf{Relevance exclusions}\\1,894 publications
};
\node[survey] (presurveys) at (-1.45,-5.65) {
    \textit{Preliminary}\\
    \textbf{Surveys/reviews}\\12 publications
};
\node[survey] (fullsurveys) at (1.45,-5.65) {
    \textit{Full selection}\\
    \textbf{Surveys/reviews}\\40 publications
};
\node[included] (initial) at (4.35,-5.65) {
    \textit{Preliminary}\\
    \textbf{Included}\\10 pilot + 91 selected\\$-$6 removed: 95 studies
};
\node[included] (additional) at (7.25,-5.65) {
    \textit{Full selection}\\
    \textbf{Included}\\140 studies
};

\draw[thick] (preliminary.south) -- (-3.6,-3.5);
\draw[thick] (-7.25,-3.5) -- (4.35,-3.5);
\foreach \x/\target in {-7.25/preexcluded,-1.45/presurveys,4.35/initial}
    \draw[arrow] (\x,-3.5) -- (\target.north);

\draw[thick,crossing] (remaining.south) -- (3.6,-4.15);
\draw[thick,crossing] (-4.35,-4.15) -- (7.25,-4.15);
\foreach \x/\target in {-4.35/fullexcluded,1.45/fullsurveys,7.25/additional}
    \draw[arrow] (\x,-4.15) -- (\target.north);

\coordinate (exjoin) at (-5.8,-7.3);
\coordinate (survjoin) at (0,-7.3);
\coordinate (incjoin) at (5.8,-7.3);

\node[summary,fill=exclusionfill] (excludedtotal) at (-5.8,-8.65) {
    \textbf{Total relevance exclusions}\\
    999 + 6 + 1,894\\\textbf{2,899 publications}
};
\node[summary,fill=surveyfill] (surveystotal) at (0,-8.65) {
    \textbf{Total surveys/reviews}\\
    12 + 40\\\textbf{52 publications}
};
\node[summary,fill=includedfill] (final) at (5.8,-8.65) {
    \textbf{Final corpus}\\
    95 + 140\\\textbf{235 studies}
};

\foreach \left/\right/\join/\total in {
    preexcluded/fullexcluded/exjoin/excludedtotal,
    presurveys/fullsurveys/survjoin/surveystotal,
    initial/additional/incjoin/final}
{
    \draw[combine] (\left.south) |- (\join);
    \draw[combine] (\right.south) |- (\join);
    \draw[combine,arrow] (\join) -- (\total.north);
    \node[junction] at (\join) {+};
}

\node[font=\bfseries] at (0,-9.7) {Overall outcome};
\end{tikzpicture}%
}

\caption{Study identification, preliminary selection, and full selection process.}
\label{fig:study-selection}
\end{figure*}
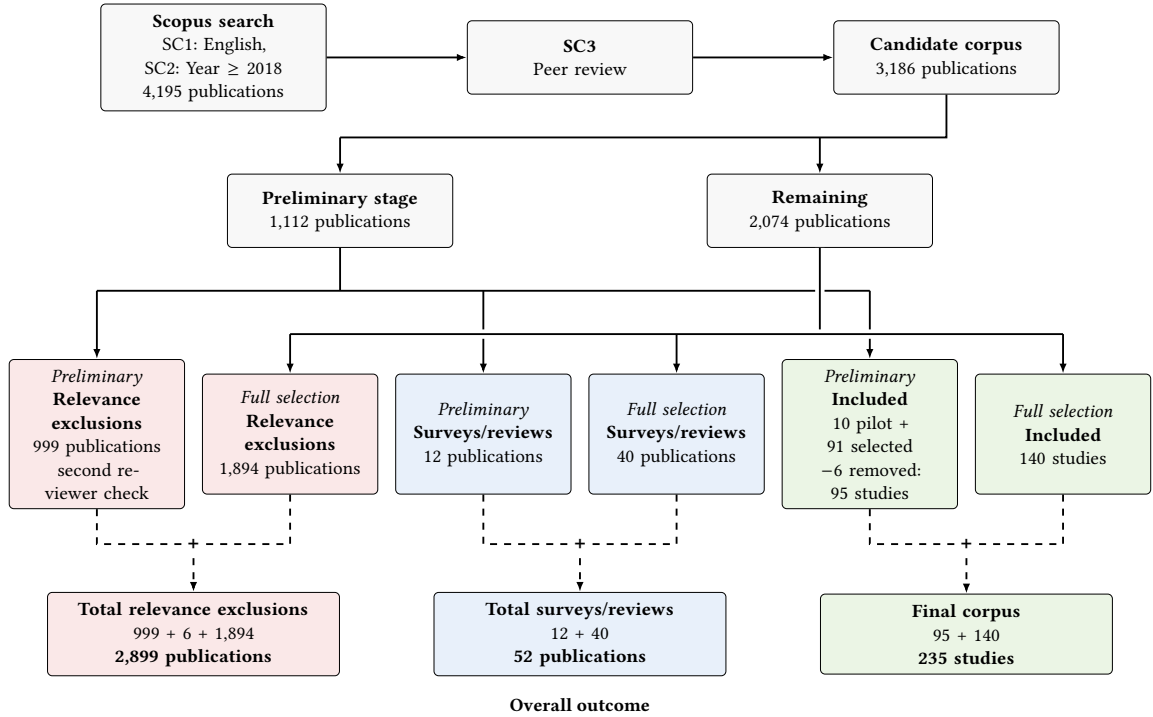

Study selection was performed in two phases. The preliminary phase was used to obtain a larger set of included studies and refine the review procedure. The full selection phase then applied the refined procedure to the remaining candidate publications. Throughout both phases, publications failing the relevance criteria were distinguished from surveys and reviews excluded because they were not primary studies. The complete selection process is summarised in~\autoref{fig:study-selection}.

\subsubsection{Preliminary Phase}

The candidate publications were divided equally among the three reviewers. Each reviewer independently screened their assigned publications until identifying roughly 30 primary studies that satisfied the selection criteria. These 91 newly included studies, together with the 10 pilot studies, formed a first set of 101 studies.

The preliminary phase accounted for 1,112 publications: the 10 pilot studies, 91 newly included studies, 999 publications excluded based on the relevance criteria, and 12 surveys or reviews excluded as secondary studies. This left 2,074 candidate publications for the full selection phase.
We excluded a publication based on the relevance criteria if its primary focus was not on ADS, if it did not measure or mitigate a simulation-to-reality difference relevant to evaluating or testing an ADS, or if it focused only on training or improving ADS without addressing this difference. Surveys, reviews, tutorials, and position papers without an original testing or validation approach were excluded as secondary studies.

The three reviewers then jointly examined the 91 newly included studies. They discussed the interpretation and classification of each study and refined the extraction and coding rules where necessary. This examination showed that six of the 91 did not belong in the corpus: five were not relevant on closer examination, and one was the conference version of a paper whose journal version was also selected. The initial corpus therefore comprised 95 studies. The 999 publications excluded based on the relevance criteria during this phase were independently checked by a second reviewer to reduce the risk of false-negative exclusions.

\subsubsection{Full Selection}

The remaining 2,074 candidate publications were divided equally among the three reviewers and screened independently using the refined selection criteria. Uncertain cases were discussed by all three reviewers until consensus was reached. This phase resulted in 140 additional included studies, 1,894 relevance exclusions, and 40 surveys or reviews excluded as secondary studies.
A small number of candidates shared the Driving Simulation Association's publisher and venue metadata with its peer-reviewed Driving Simulation Conference proceedings but were themselves vendor product demonstrations or conference front-matter listings that were not peer-reviewed publications. These were identified manually from their content during screening and excluded without changing the SC3 rule.

Across the 3,186 candidate publications, 2,899 were excluded based on the relevance criteria (999 during the preliminary phase, 6 at the joint examination of the selected studies, and 1,894 during full selection), while 52 surveys or reviews were excluded as secondary studies (12 and 40, respectively). The initial corpus of 95 studies and the 140 additional studies identified during full selection therefore form the final corpus of 235 primary studies.

\subsection{Data Extraction}
\label{extraction}

Each included study is analysed to answer the research questions defined in~\autoref{questions}. The extracted information is recorded in a structured spreadsheet, with one row per study and one column per extraction field. The spreadsheet was defined before extraction, calibrated using the pilot studies, further refined during the preliminary phase, and then applied to the full corpus.
The extraction fields use either predefined categorical values or free-text descriptions. Predefined fields include the study objective, gap aspect, simulation-to-reality direction, and execution classifications. For fields whose possible values could not be defined in advance, the reviewers instead recorded a short description of what the study reported. These free-text fields include the research target, proposed technique, evaluation procedure, and evaluation metrics. Their categories were derived later during data analysis, as described in~\autoref{analysis}.

During the pilot study, all three reviewers independently coded the 10 pilot studies to calibrate the extraction process. Inter-rater agreement was measured using Fleiss' $\kappa$~\cite{Fleiss:1971} before disagreements were resolved. The agreement calculation covered the primary objective, direction, and execution level of each proposed-approach and evaluation-reference component. The pooled agreement across the 10 pilot studies was $\kappa = 0.79$, indicating substantial agreement. The reviewers then discussed disagreements and refined the extraction guidelines until consensus was reached.
The extraction procedure was further refined during the preliminary phase by jointly examining the 91 newly included studies. The three reviewers read and discussed these studies together, resolving ambiguities in the extraction fields, classifications, and coding rules. The resulting procedure was then applied to the remaining studies included during full selection. Ambiguities and borderline cases encountered during full extraction were discussed by all three reviewers until consensus was reached.

\subsection{Data Analysis}
\label{analysis}
\label{analysis-conventions}

Predefined categorical fields are analysed using the categories defined during data extraction. 
For each free-text field, we first record a short description of what each study reports. At this stage, the description is not assigned to a predefined category. For example, the research-target field describes what the study investigates in relation to the simulation--reality gap. The proposed-technique field describes how the study attempts to mitigate the gap. The evaluation-procedure field describes how the study measures the gap or validates a mitigation. The evaluation-metrics field records the quantities used in that evaluation.
After extraction, descriptions referring to the same concept are grouped together. Different terminology does not create a separate group when it describes the same concept. Conversely, descriptions that refer to different concepts remain separate even when they occur in the same study. A study can therefore contribute to more than one group when it reports several distinct targets, techniques, procedures, or metrics.

The three reviewers then jointly examine the complete set of groups for each field. They compare their meaning, merge groups that represent the same concept, separate groups that contain distinct concepts, and refine their names and boundaries until consensus is reached. The resulting groups form the lowest level of the classification used in the analysis.
Where needed, related groups are then organised into broader categories. For RQ\textsubscript{2.1}, related research targets are organised into macro categories and subcategories. For RQ\textsubscript{2.2}, mitigation techniques are organised into macro categories, subcategories, and, where useful, subsubcategories. RQ\textsubscript{3.1} and RQ\textsubscript{3.3} use separate classifications derived in the same way from their evaluation procedures and metrics. The hierarchy is therefore used to organise the categories after they have been derived from the extracted descriptions. 

After the category structure is finalised, the assignments are checked against the original extracted descriptions and the final category definitions are applied consistently across the corpus. The categories are then used to compute the counts and distributions reported in the results.
A study may have several values for a field when the corresponding research question allows multiple assignments. It is counted once in each applicable category and at most once within the same category. Counts across non-mutually-exclusive categories can therefore exceed the number of studies included in the corresponding analysis.
\autoref{tab:rq_data} summarises the data used for each research question. The remainder of this subsection defines the analysis performed for each field.

\begin{table}[tbp]
\centering
\caption{Type and content of the data used for each research question.}
\label{tab:rq_data}

\footnotesize
\renewcommand{\arraystretch}{1.08}
\begin{tabularx}{\linewidth}{p{1.0cm} p{2.9cm} >{\raggedright\arraybackslash}X}
\toprule
\textbf{RQ} & \textbf{Data type} & \textbf{Data content} \\
\midrule
RQ\textsubscript{1.1} & Predefined categorical & Measure only, Mitigate only, Measure and mitigate \\
RQ\textsubscript{1.2} & Predefined categorical, combinable & Scenario, Behaviour, Perception, Localization, Actuation, Entire; a study addressing several aspects equally is recorded as their combination \\
RQ\textsubscript{1.3} & Predefined categorical & Sim to real, Real to sim, Both \\
\midrule
RQ\textsubscript{2.1} & Free text, categorised & Research target, grouped into macro categories and subcategories \\
RQ\textsubscript{2.2} & Free text, categorised & Mitigation technique, grouped into macro categories, subcategories, and optional subsubcategories \\
RQ\textsubscript{2.3} & Predefined categorical, one value per component & Scenario, Actors, Perception, Localization, Actuation, ECU/control, each recorded as Not modelled, Simulated, Pseudo-sim, Pseudo-real, Mixed, Both, or Real; Partially real additionally arises at group level; each setup also carries one or more execution-source tags (real vehicle, real hardware, real dataset, simulator, simulated dataset), or None \\
\midrule
RQ\textsubscript{3.1} & Free text, categorised & Evaluation procedure, including its explicit absence \\
RQ\textsubscript{3.2} & Predefined categorical, one value per component & The same components and execution levels as RQ\textsubscript{2.3}; the evaluation reference has its own execution-source tags \\
RQ\textsubscript{3.3} & Free text, categorised & Evaluation metrics, grouped by the quantity being measured \\
\midrule
RQ\textsubscript{4.1} & Free text, named entities & Simulation platforms and reference datasets \\
RQ\textsubscript{4.2} & Free text, categorised & Hardware, grouped by its role in the experimental setup \\
RQ\textsubscript{4.3} & Free text, categorised & System under test and the specific technology implementing it \\
RQ\textsubscript{4.4} & Free text, categorised & Origin of evaluation scenarios or test cases \\
RQ\textsubscript{4.5} & Predefined categorical & Code, Dataset, Code and dataset, Other artefact, None; mutually exclusive and assigned by priority \\
\bottomrule
\end{tabularx}
\end{table}

\subsubsection{RQ\textsubscript{1}: Context}

For RQ\textsubscript{1.1}, the predefined primary objective categories are measure, mitigate, and both. For RQ\textsubscript{1.2}, the predefined gap aspect categories are scenario, behaviour, perception, localization, actuation, and entire. When a study addresses several aspects with equal importance, their combination is recorded directly rather than derived through the free-text categorisation used elsewhere. For RQ\textsubscript{1.3}, the predefined direction categories are sim to real, real to sim, and both.

\subsubsection{RQ\textsubscript{2}: Approaches}

RQ\textsubscript{2.1} analyses the research-target descriptions extracted from the studies. Descriptions referring to the same research target are grouped together using the procedure described above. Related targets are then organised into macro categories and subcategories. The classification describes what the study investigates in relation to the simulation--reality gap, rather than the technique used to address it.

RQ\textsubscript{2.2} analyses the techniques used to mitigate the simulation--reality gap and therefore applies only to studies whose objective includes mitigation. Descriptions referring to the same mitigation technique are first grouped together. Related techniques are then organised into six macro categories: Real-equipment integration, Generative content synthesis, Traffic/agent behaviour modelling, Real-world evidence reconstruction, Sensor physics modelling, and Ego policy/control transfer. These are further divided into subcategories and, where recurring techniques require an additional distinction, subsubcategories. 
A study may use more than one mitigation technique. It is assigned to every applicable macro category and may also appear in several subcategories within the same macro category.

RQ\textsubscript{2.3} analyses the execution of the proposed approach across six components: scenario, actors, perception, localization, actuation, and ECU/control. Unlike RQ\textsubscript{2.1} and RQ\textsubscript{2.2}, these categories are predefined. Each component is assigned one of the execution levels defined in~\autoref{tab:execution_levels}. This classification describes how a component is provided during execution, not how it was created. For example, calibrating a simulator or a generative model with real-world data does not by itself change how the resulting component is provided.

\begin{table}[tbp]
\footnotesize
\centering
\caption{Categorical execution levels used for RQ\textsubscript{2.3} and RQ\textsubscript{3.2}.}
\label{tab:execution_levels}
\begin{tabular}{p{0.20\linewidth} p{0.70\linewidth}}
\toprule
\textbf{Category} & \textbf{Definition} \\
\midrule
Not modelled & The component is not present. \\
Simulated & The component is provided by a general-purpose simulation platform or physics model. This holds even if the platform is calibrated, tuned, or partly trained using real-world data. \\
Pseudo-sim & The component is provided by a standalone model or generator, not a general-purpose simulation platform, whose representation is not constructed from real-world observations or measured real-world characteristics. \\
Pseudo-real & The component is provided by a standalone model, reconstruction, or generator whose representation is constructed from real-world observations, measurements, or measured real-world characteristics, but the real-world component is not present live during execution. \\
Mixed & Real and simulated elements are combined within the same execution. \\
Both & Separate simulated and real-world executions are performed in parallel. \\
Real & The component uses entirely real-world content. \\
Partially real & Group level only: one component is Real while the other is neither Real, Both, nor Mixed. \\
\bottomrule
\end{tabular}
\end{table}

A general-purpose simulation platform remains Simulated even when it is calibrated, tuned, or partly trained using real-world data. The classification depends on whether the platform provides the component, not on how its parameters were set.
Pseudo-sim and Pseudo-real both describe standalone models or generators used without such a platform. They differ in the source of their representation. For example, a physical model built from first-principles equations and only validated against real measurements is Pseudo-sim. The same type of model is Pseudo-real if its error or degradation behaviour is fitted to real measurements. This distinction applies across all six components.
A component is Mixed when real and simulated elements are combined within the same execution. Examples include a real vehicle or sensor operating with simulated content, or a general-purpose simulator combined with real-world-derived content within the same component.

The six components are also grouped into three functional groups: scene, comprising scenario and actors; sensing, comprising perception and localization; and control, comprising actuation and ECU/control. Execution is analysed at both component and group level.

We use \emph{Partially real} at group level when one component is Real while the other is neither Real, Both, nor Mixed. For example, real perception combined with simulated localization produces a Partially real sensing group. Mixed and Both describe the use of simulation and reality for the same component. Mixed combines them within one execution, while Both uses separate simulated and real executions. Partially real instead describes a difference between the two components of the same functional group. This category is derived from the component classifications and is never assigned to an individual component.

Component execution levels and execution sources are recorded separately. The execution level describes how a component is provided, while the execution source records the type of setup used to obtain it. \emph{Real} describes real-world component content and does not by itself establish physical execution. For example, a study can use a recorded real-world dataset without operating a physical vehicle. The corresponding component is then Real, while its execution source is a Real dataset.

For both the proposed approach and its evaluation reference, we assign any applicable study-level execution sources:

\head{Real vehicle} A physical vehicle operated in the study, including small-scale vehicles;

\head{Real hardware} A live sensor, controller, rig, or participant without an operated vehicle;

\head{Real dataset} Pre-existing recorded real-world data;

\head{Simulator} A model or generator executed by the study, including the generation of a new synthetic dataset;

\head{Simulated dataset} Pre-existing synthetic data used without executing its generator.

\head{None} It marks the absence of an execution source for the proposed approach or the absence of a comparison reference for evaluation. Recordings collected through vehicle or hardware experiments performed in the study retain the corresponding physical-source tag. Recordings from an earlier campaign are classified as a Real dataset. Approach tags include sources used for training or calibration, whereas reference tags describe sources used for comparison.

\subsubsection{RQ\textsubscript{3}: Evaluation}

RQ\textsubscript{3.1} analyses the free-text description of each study's evaluation procedure. Descriptions of equivalent evaluation procedures are grouped using the procedure defined at the beginning of~\autoref{analysis}. Studies with no recorded evaluation are checked again to distinguish an explicitly absent evaluation from missing extraction information.
The same classification of evaluation procedures is used for all study objectives, but its interpretation depends on the objective. For a Measure only study, the procedure describes how the study measures the simulation--reality gap. For a Mitigate only study, it describes how the study validates the proposed mitigation. For a Measure and Mitigate study, the measurement of the gap and the validation of the mitigation are considered separately.

RQ\textsubscript{3.2} analyses the evaluation reference using the same predefined components, functional groups, and execution levels as RQ\textsubscript{2.3}. We match the proposed approach and evaluation reference within each study and compare the same component on both sides. Component-wise comparisons include only studies in which that component is present in the proposed approach.
This restriction controls for differences in the components included in experimental setups. However, the presence of a component does not establish that it is the research target. We therefore interpret these comparisons alongside RQ\textsubscript{2.1} and do not treat an absent reference component as inadequate evidence without considering the study's claim.
The meaning of the evaluation reference depends on the study objective. For a mitigation study, it is the reference against which the proposed mitigation is validated. For a measurement study, it is the system or execution treated as the reference when measuring the simulation--reality gap.
\emph{Not modelled} ordinarily indicates that a component is absent from an otherwise present reference setup. When a study performs no evaluation, no reference setup exists. At component level, we represent this case using Not modelled for all components rather than introducing another execution-level category. The separate \emph{None} source tag distinguishes the absence of an evaluation reference from the absence of an individual component.

RQ\textsubscript{3.3} analyses the free-text descriptions of the reported evaluation metrics. Metrics are first grouped according to the quantity they measure rather than the mathematical statistic used. Metrics that measure the same type of quantity therefore belong to the same higher-level category even when they use different numerical measures. Where needed, the specific metric is retained as a lower-level type. This separates what is being evaluated from the particular measure used to quantify it.

\subsubsection{RQ\textsubscript{4}: Experimental Components}

RQ\textsubscript{4.1} records each named simulation platform and reference dataset used by a study. These are retained as named entities rather than grouped into derived categories.
RQ\textsubscript{4.2} analyses the free-text hardware descriptions. Hardware with the same role in the experimental setup is grouped into the same category.
RQ\textsubscript{4.3} analyses the free-text description of the system under test. Systems performing the same function are grouped together, while the specific model, architecture, or technique is retained where reported.
RQ\textsubscript{4.4} analyses how the evaluation scenarios or test cases are obtained. The extracted descriptions are grouped into six categories: predefined test cases, simulation-based generation, reconstruction from real-world data, replay of recorded data, live real-world execution, and hybrid or mixed generation.
RQ\textsubscript{4.5} assigns each study exactly one of five mutually exclusive artefact categories: Code and dataset when both are reported, Code or Dataset when only one is reported, Other artefact when the reported artefact is neither, and None when no artefact is reported. Categories are assigned in that priority order.

\subsection{Threats to Validity}
\label{sec:threats}

\subsubsection{Search and selection validity}

The corpus may omit relevant studies because the search relies on Scopus and requires terms related to ADS, testing or validation, and the simulation--reality relationship in titles, abstracts, or keywords. Studies using different terminology may therefore remain undiscovered. Retrieving all ten pilot studies provides an initial check of search coverage, but this golden set cannot establish completeness. Restricting the corpus to English-language publications from 2018 onward and venues with verifiable peer review also excludes earlier work, industrial reports, and potentially relevant publications whose review process cannot be established from venue metadata. Our findings therefore characterise the retrieved peer-reviewed literature within these boundaries.

\subsubsection{Reviewer bias and reliability}

Study selection and extraction involve judgement, particularly when distinguishing reality gap research from work concerned only with ADS training or improvement. We mitigate this threat through pilot calibration, joint examination of the preliminary corpus, independent checking of preliminary relevance exclusions, and consensus discussions. However, full-phase screening was divided among reviewers rather than independently duplicated. The reported Fleiss' $\kappa=0.79$ concerns selected categorical fields in ten pilot studies and does not establish agreement across the full corpus or the subsequently derived categories. Residual differences in interpretation may therefore affect inclusion decisions and category assignments.

\subsubsection{Construct validity}

Concepts such as realism, fidelity, and the reality gap are used inconsistently across disciplines. Mapping them onto common categories requires interpretation and may obscure distinctions between studies. Separately classifying the proposed approach and its evaluation reference across six components reduces ambiguity, but execution categories describe experimental configurations rather than evidence adequacy. Similarly, physical equipment, recorded observations, or multiple metric families do not by themselves establish correspondence in ADS behaviour or test outcomes. Incomplete reporting may also constrain classification: an unreported element does not necessarily indicate its absence from the underlying experiment.

\subsubsection{Synthesis and generalisability}

The reported frequencies describe publications rather than independent approaches or demonstrations of effectiveness. Related studies may reuse methods, datasets, or platforms, and a study may contribute to several categories. Category counts are therefore non-exclusive and should not be interpreted as comparative effectiveness or methodological quality. The synthesis also depends on published evidence and may underrepresent unsuccessful approaches or unpublished industrial practice. Consequently, the identified patterns apply primarily to the included literature. The mapping does not establish causal relationships between gap mitigation, behavioural correspondence, and agreement between simulated and physical test verdicts.
\section{Results}
\label{results}

\subsection{Context}
\label{sec:results-context}

\subsubsection{RQ\textsubscript{1.1}: Primary Objective}

Mitigation is the most common objective; it accounts for 103 studies (43.8\%), followed by Measure and mitigate with 76 studies (32.3\%) and Measure only with 56 studies (23.8\%). Overall, 76.2\% of the studies include some form of mitigation.

\subsubsection{RQ\textsubscript{1.2}: Addressed Gap Aspect}

A study is assigned either a single gap aspect or, when it addresses several aspects equally, a combination of them, such as scenario and behaviour. The single aspects are scenario (the representation of the driving scenario), behaviour (the behaviour of traffic participants), perception (the sensing process), localization, actuation (vehicle actuation), and the entire gap (the system as a whole). Each study appears once, so the counts across all single aspects and combinations sum to 235 studies.

A single aspect is addressed by 193 studies (82.1\%). Perception is the most common with 84 studies (35.7\%), followed by scenario with 51 (21.7\%). The entire gap, rather than a specific aspect, is addressed by 22 studies (9.4\%), while 21 (8.9\%) focus on actuation and 15 (6.4\%) on behaviour. No study addresses localization, alone or in combination.

The remaining 42 studies (17.9\%) address several aspects, in seven distinct combinations. Scenario and behaviour is the most common, with 10 studies (4.3\%), followed by scenario and actuation, and perception and behaviour (8 each, 3.4\%). Perception and scenario, and perception, scenario and actuation, occur in 5 studies each (2.1\%), while perception and actuation, and actuation and behaviour, occur in 3 each (1.3\%). Scenario is the aspect most often combined with others (28 of the 42 studies), ahead of perception and behaviour (21 each) and actuation (19).

\subsubsection{RQ\textsubscript{1.3}: Direction}

Real2sim accounts for 97 studies (41.3\%) and sim2real for 95 studies (40.4\%), while 43 studies (18.3\%) address both directions.


\subsection{Approach}
\label{sec:results-approach}


\subsubsection{RQ\textsubscript{2.1}: Research Target}

The research targets and the corresponding primary studies are shown in~\autoref{tab:rq2-1}. 


\head{Scenario and actor realism (109 studies)} Traffic/NPC realism is the most common target with 50 studies, followed by test scenario validity with 26. Crash scenario validity accounts for 13 studies, pedestrian/VRU realism for 12, and road/environment realism for 11. Langer et al.~\cite{langer2021CalibrationAssessment} calibrate an urban microscopic traffic simulation against real traffic data before using it as a testing environment. Fremont et al.~\cite{fremont2020FormalScenariobased} specify scenarios in simulation and re-execute matching scenarios on a real vehicle. Oladimeji et al.~\cite{oladimeji2024EnhancedApproaches} reconstruct real crash scenarios and compare them with the original accidents. Zofka et al.~\cite{zofka2018TrafficParticipants} place a real pedestrian inside a mixed-reality testbed. Bateman et al.~\cite{bateman2024ExploringReal} test an HD-map prediction model on real-world map changes not seen during training.

\begin{table}[t]
\caption{RQ\textsubscript{2.1} research targets and supporting primary studies. Bars show the number of studies assigned to each target, and the final column lists the corresponding primary studies. Studies may be assigned to several targets, so counts across categories are not mutually exclusive and subcategory counts may not sum to the parent count.}
\label{tab:rq2-1}
\centering
\footnotesize
\begin{tabular}{@{}p{0.35\textwidth} p{0.20\textwidth} p{0.4\textwidth}@{}}
\toprule
\textbf{Research target} & \textbf{Studies} & \textbf{Primary studies} \\
\midrule
\hangindent=1em\hangafter=1 \textbf{Scenario and actor realism} & \ResultStudies{109}{109} &  \\
\hangindent=2em\hangafter=1\hspace{1em}Traffic/\hspace{0pt}NPC realism & \ResultStudies{50}{109} & \cite{liu2021ApplicationDigital,langer2021CalibrationAssessment,kim2024DesigningNaturalistic,lichtle2024FromSim,shuguang2022VehicleintheloopIntelligent,amini2020LearningRobust,bergamini2021SimnetLearning,cao2024AresTextdriven,gu2023CopulabasedInteractive,tettamanti2018VehicleintheloopTest,yang2021OnlineAdaptation,wagner2019TowardsCrossverification,tenbrock2021ConscendDataset,shi2022IntegratedTraffic,zhang2022TrajgenGenerating,xu2023BitsBilevel,sun2024DrivescenegenGenerating,stadler2022CredibilityAssessment,queiroz2024DrivervehicleModel,barker2018LowcostHardwareintheloop,che2021TestMethod,liu2024IntelligentCosimulation,samak2024MetaverseSafer,suo2023MixsimHierarchical,guo2025ScenedmConsistent,huang2025CadreControllable,neis2025DatadrivenLane,ameli2025DesignPreparation,yan2025DistributionallyConsistent,gao2026DrivenceRealistic,zhang2025EndtoendLearning,naing2025FinegrainedTrajectory,wu2025FusingTemporal,li2025GenerativeReconstruction,buzdugan2025HumancenteredSimulations,wang2025RadeLearning,jing2025EfficientHighrisk,bi2025TowardFullstack,wang2026UtdtBaseline,huang2026VersatileBehavior,cao2024ReinforcementLearning,khaire2025AugmentedRealityenhanced,wang2022AutomaticTraffic,liang2025DttfsimDigital,drechsler2022DynamicVehicleintheloop,gao2025MsietransformerNovel,coppola2025VirtualTesting,daza2023SimtorealTransfer,leung2025RoadCode,sigl2021ExperiencingDatadriven} \\
\hangindent=2em\hangafter=1\hspace{1em}Test-scenario validity & \ResultStudies{26}{109} & \cite{liu2021ApplicationDigital,nalic2019DevelopmentCosimulation,peintner2021MixedReality,kim2022QueryingLabelled,bai2024ArbasedMeta,chen2020MixedTest,fremont2020FormalScenariobased,gao2020HardwareintheloopSimulation,gupta2023XintheloopXil,schyr2022VehicleintheloopTestinga,zofka2018SleepwalkerFramework,wagner2019TowardsCrossverification,stadler2022TowardsMultivariate,wu2024RealityBites,solmaz2020VehicleintheloopMethodology,lan2024AssessingReliability,reckenzaun2024TransnationalTesting,chen2025SnerfAutonomous,wu2025ToolBenchmarking,kolb2025AutomaticallyImproving,shen2023SimonwheelsPhysical,arango2026ClosingSim2real,toth2023DevelopmentFunctional,tian2026HgsimHighfidelity,elgharbawy2019OntologybasedAdaptive,guo2021SeamlessModel} \\
\hangindent=2em\hangafter=1\hspace{1em}Crash-scenario validity & \ResultStudies{13}{109} & \cite{oladimeji2024EnhancedApproaches,bashetty2020DeepcrashtestTurning,gambi2019GeneratingEffective,xinxin2020CsgCritical,zhang2023BuildingCritical,zhou2024EvaluatingAutonomous,zhang2025HighriskTrajectories,wei2025InteractiveCritical,zhao2025TachographbasedApproach,abbasi2025Acr2xodrFrom,hu2025SimulationbasedTesting,shoukat2024AutonomousDriving,wei2025GeneratingRisky} \\
\hangindent=2em\hangafter=1\hspace{1em}Pedestrian/\hspace{0pt}VRU realism & \ResultStudies{12}{109} & \cite{zofka2018TrafficParticipants,drechsler2022MireMixed,larter2022HierarchicalPedestrian,long2024GeneratingRealistic,fu2024ImmersiveTraffic,fu2025ControllablePedestrian,liu2026EnhancingAutonomous,li2025TowardIntelligent,martinserrano2025BehaviouralGap,wei2025GeneratingRisky,steimle2025HumanvehicleintheloopTest,yang2020RecoveringSimulating} \\
\hangindent=2em\hangafter=1\hspace{1em}Road/\hspace{0pt}environment realism & \ResultStudies{11}{109} & \cite{bateman2024ExploringReal,brogle2019HardwareintheloopAutonomous,szalai2020MixedReality,zhang2022SelfdrivingSimulation,rezaldi20213dOrthomosaic,li2025AutonomousDriving,sun2025TestingAutonomous,schulz2025VirtualKitti,wang2025DigitalTwinbased,jekl2025SpeedPlanner,park2020VehicleintheloopGlobal} \\ [0.5em]
\hangindent=1em\hangafter=1 \textbf{Perception} & \ResultStudies{105}{109} &  \\
\hangindent=2em\hangafter=1\hspace{1em}Camera perception & \ResultStudies{56}{109} & \cite{nalic2019DevelopmentCosimulation,pahk2023EffectsSim2real,pahk2024LaneSegmentation,reway2022SimulationbasedTest,reway2020TestMethod,amini2020LearningRobust,argui2023BuildingVisionbased,drechsler2022MireMixed,duminil2024AssessingFidelity,stocco2023ModelVs,hsiang2022DevelopmentSimulationbased,vonbernuth2018RenderingPhysically,thieling2019NeuralNetworks,tim2020SimulationReal,talwar2020EvaluatingValidity,xu2021ReliabilityGan,soni2021DeepLearning,ulhas2024GanbasedDomain,oh2024ReliabilityEvaluation,lan2024AssessingReliability,amini2024BridgingGap,ulreich2024NovelTest,fu2025ControllablePedestrian,zeng2025ControllableRainy,steinhauser2025DataQuality,li2025DatadrivenHeadlight,li2025DgnrDensityguided,zhou2026DiffusionModels,wang2025DrivegenCoevaluating,simakov2025EffectsSimulator,baresi2025EfficientDomain,pasios2025EnhancingPhotorealism,nosheen2025EnhancingSynthetic,wang2025GenerativeTesting,zhou2026HugsimRealtime,zuo2025RaladBridging,gao2025DatacentricImage,elmquist2025MethodologyQuantify,zhang2025PixellevelImage,eberhardt2025BeyondSmudge,harhammer2025SimtorealDomain,samak2026Sim2realDiffusion,yu2026TrajectorybasedVideo,ulreich2025ValidatingCamera,otto2025ValidationAibased,li2025XldCrosslane,lambertenghi2024AssessingQuality,shen2023SimonwheelsPhysical,gadipudi2022SyntheticReal,duminil2024ComprehensiveExploration,cardozo2025CarlaHardwareintheloop,argui2025ExploringAugmentation,tian2026HgsimHighfidelity,segonne2023QualificationPavin,song2026VisionlanguageModelbased,dauner2024Navsim} \\
\hangindent=2em\hangafter=1\hspace{1em}LiDAR perception & \ResultStudies{23}{109} & \cite{rosenberger2018AnalysisReal,rosenberger2019BenchmarkingFunctional,saleh2019DomainAdaptation,kettelgerdes2023InfieldMeasurement,manivasagam2023TowardsZero,chen2022AnalysisRealtime,zofka2018SleepwalkerFramework,wallace2022ValidatingSimulation,huang2024HierarchicalLidar,gaddam2025LidarSensor,peng2025MethodApplications,saad2025DigitalTwin,haghighi2026UnifiedGenerative,anand2026TowardClosing,schabus2025ObjectbasedLidar,manivasagam2020LidarsimRealistic,huch2023QuantifyingLidar,vargasrivero2022EffectSpray,balm2025BenchmarkingDomain,hofrichter2025IntroducingDevelopment,li2024MultilevelMetrics,wang2022Cadsim,yang2020RecoveringSimulating} \\
\hangindent=2em\hangafter=1\hspace{1em}Radar perception & \ResultStudies{10}{109} & \cite{ngo2021DeepEvaluation,jayapalgowdu2022EvaluationScenariobased,wachtel2021ValidationRadar,zhu2022MillimeterwaveRadar,haider2024MethodologyVerifying,ngo2021MultilayeredApproach,wachtel2024SimulationValidation,haider2025ModelingSimulation,magosi2023NovelApproach,magosi2022EvaluationMethodology} \\
\hangindent=2em\hangafter=1\hspace{1em}Camera + LiDAR perception & \ResultStudies{8}{109} & \cite{capdevila2024SmartsafeMobility,yang2023UnisimNeural,chen2025SnerfAutonomous,gao2026DrivenceRealistic,lambertenghi2025MultimodalityEvaluation,hess2025SplatadRealtime,deter2021SimulatingAutonomous,wang2023WhenSimulator} \\
\hangindent=2em\hangafter=1\hspace{1em}Camera + LiDAR + Radar perception & \ResultStudies{6}{109} & \cite{liu2021ApplicationDigital,hasirlioglu2018ModelbasedApproach,chen2019NovelIntegrated,hwang2024ImplementationSimulationbased,lee2024StudyFull,kuroda2025AnalysisSnow} \\
\hangindent=2em\hangafter=1\hspace{1em}Camera + LiDAR + Event camera perception & \ResultStudies{1}{109} & \cite{amini2022Vista20} \\
\hangindent=2em\hangafter=1\hspace{1em}Multi-sensor perception & \ResultStudies{1}{109} & \cite{arango2026ClosingSim2real} \\ [0.5em]
\hangindent=1em\hangafter=1 \textbf{Ego system behaviour and planning} & \ResultStudies{62}{109} &  \\
\hangindent=2em\hangafter=1\hspace{1em}Ego-vehicle dynamics/\hspace{0pt}actuation & \ResultStudies{37}{109} & \cite{lu2022DesignVehicle,peintner2021MixedReality,samak2024TowardsValidation,parra2020ValidationRealtime,krecht2023ValidationProcess,amini2020LearningRobust,bai2024ArbasedMeta,brogle2019HardwareintheloopAutonomous,bui2024ValidationAdaptive,dona2022VirtualTesting,drechsler2022HowSimulation,gao2020HardwareintheloopSimulation,schyr2022VehicleintheloopTestinga,tettamanti2018VehicleintheloopTest,yeo2019MaximMixedreality,durukal2022DigitalTwin,jiang2021HighaccuracyFramework,chen2019NovelIntegrated,oh2024ReliabilityEvaluation,fei2024CorrelationSoftwareintheloop,lee2024StudyFull,reckenzaun2024TransnationalTesting,drechsler2024VehicleDynamics,allamaa2025LearningbasedNmpc,darocha2025MicroautowareAutoware,lambertenghi2025MultimodalityEvaluation,drechsler2025AcceleratingApproval,kim2025AutonomousVehicle,waheed2025QuantifyingSim2real,shen2023SimonwheelsPhysical,wang2025DigitalTwinbased,arango2026ClosingSim2real,daza2023SimtorealTransfer,jekl2025SpeedPlanner,park2020VehicleintheloopGlobal,reisgys2021ArgumentationADAS,heikkinen2019HardwareintheloopPlatform} \\
\hangindent=2em\hangafter=1\hspace{1em}SUT behaviour realism & \ResultStudies{28}{109} & \cite{reway2020TestMethod,samak2024TowardsValidation,manivasagam2023TowardsZero,bai2024ArbasedMeta,brogle2019HardwareintheloopAutonomous,bui2024ValidationAdaptive,chen2020MixedTest,drechsler2022MireMixed,drechsler2022HowSimulation,briand-offline-emse,szalai2020MixedReality,szalay2019ProofConcept,oh2024ReliabilityEvaluation,che2021TestMethod,solmaz2020VehicleintheloopMethodology,fei2024CorrelationSoftwareintheloop,samak2024MetaverseSafer,capdevila2024SmartsafeMobility,lee2024StudyFull,reckenzaun2024TransnationalTesting,zhou2026HugsimRealtime,stangherlin2025XintheloopApproach,lee2025ModelcentricQuantitative,oh2024AdvilsImplementation,stocco2023MindGap,shen2023SimonwheelsPhysical,guo2021SeamlessModel,wang2023WhenSimulator} \\
\hangindent=2em\hangafter=1\hspace{1em}ADS-policy transfer & \ResultStudies{9}{109} & \cite{gutierrezmoreno2024DecisionMaking,yuan2024FromNaturalistic,chen2019NovelIntegrated,fei2024CorrelationSoftwareintheloop,ganesan2025BridgingSimtoreal,wang2025DrivegenCoevaluating,hong2025EffectiveLearning,amini2022Vista20,song2026VisionlanguageModelbased} \\ [0.5em]
\hangindent=1em\hangafter=1 \textbf{Ego localization/\hspace{0pt}pose estimation} & \ResultStudies{7}{109} & \cite{barker2018LowcostHardwareintheloop,chen2019NovelIntegrated,verma2025EgoVehicle,jung2025RealvirtualSynchronized,stangherlin2025XintheloopApproach,shoukat2024AutonomousDriving,wang2023WhenSimulator} \\ [0.5em]
\hangindent=1em\hangafter=1 \textbf{Communication/\hspace{0pt}network fidelity} & \ResultStudies{3}{109} & \cite{hoppe2022AgentbasedAutonomous,zofka2023PavingWays,stangherlin2025XintheloopApproach} \\ [0.5em]
\hangindent=1em\hangafter=1 \textbf{Small scale vehicle suitability} & \ResultStudies{1}{109} & \cite{petrovic2025SuitabilitySmallsized} \\
\bottomrule
\end{tabular}
\end{table}

\head{Perception (105 studies)} Camera perception is the most common target with 56 studies, followed by LiDAR perception with 23 and radar perception with 10. The remaining 16 studies combine multiple sensing modalities, including Camera + LiDAR + Radar, Camera + LiDAR, and Camera + LiDAR + Event camera. Reway et al.~\cite{reway2020TestMethod} compare the performance of a camera-based object detector on simulated and real driving footage. Rosenberger et al.~\cite{rosenberger2018AnalysisReal} analyse real-world LiDAR sensor behaviour to improve the physical fidelity of a simulated LiDAR sensor model. Wachtel et al.~\cite{wachtel2021ValidationRadar} validate a radar sensor model under non-ideal conditions.

\head{Ego system behaviour and planning (62 studies)} Ego vehicle dynamics/actuation is the most common target with 37 studies, followed by SUT behaviour realism with 28 and ADS policy transfer with 9. Lu et al.~\cite{lu2022DesignVehicle} build a vehicle-in-the-loop test system in which a real vehicle responds to an L3 autopilot operating against simulated surroundings. Tettamanti et al.~\cite{tettamanti2018VehicleintheloopTest} drive a real vehicle inside a microscopic traffic simulation and compare its behaviour with the same manoeuvres performed in simulation. Guti\'errez-Moreno et al.~\cite{gutierrezmoreno2024DecisionMaking} adapt a decision-making policy developed in simulation for transfer to a real autonomous driving stack.

\head{Ego localization/pose estimation (7 studies)} These studies target the estimation of the ego vehicle's position/pose.

\head{Communication/network fidelity (3 studies)} These studies target V2X or vehicle network communication behaviour.

\head{Small scale vehicle suitability (1 study)} This study investigates whether a small-scale or robot platform is a valid substitute for a full-size vehicle.

\subsubsection{RQ\textsubscript{2.2}: Mitigation Techniques}

Of the 179 studies with a mitigation objective, 164 use techniques from one mitigation family and 15 use techniques from several families. Studies using several families are counted in each, so the family counts in~\autoref{tab:rq2-2} exceed 179. The table also shows the techniques within each family; the corresponding primary studies are listed in our Appendix. A study can also use several techniques within the same family, so subcategory counts can exceed the family total.

\begin{table}[!htbp]
\caption{RQ\textsubscript{2.2} mitigation technique families and techniques, among the studies with a mitigation objective. Bars show the number of studies using each family or technique. A study can use several families and several techniques within a family, so counts are not mutually exclusive and technique counts may exceed the family total.}
\label{tab:rq2-2}
\centering
\footnotesize
\setlength{\ResultBarMax}{1.6cm}
\begin{tabular}{@{}>{\raggedright\arraybackslash}p{\dimexpr0.95\columnwidth-\dimexpr\ResultBarMax+2.3em\relax-2\tabcolsep\relax} p{\dimexpr\ResultBarMax+2.3em\relax}@{}}
\toprule
\textbf{Mitigation technique} & \textbf{Studies} \\
\midrule
\hangindent=1em\hangafter=1 \textbf{Real-equipment integration (ViL/\hspace{0pt}HiL/\hspace{0pt}MR/\hspace{0pt}XiL)} & \ResultStudies{56}{56} \\
\hangindent=2em\hangafter=1\hspace{1em}Vehicle-in-the-loop (on-road) & \ResultStudies{28}{56} \\
\hangindent=2em\hangafter=1\hspace{1em}Mixed reality & \ResultStudies{21}{56} \\
\hangindent=2em\hangafter=1\hspace{1em}Controller/\hspace{0pt}ECU hardware-in-the-loop & \ResultStudies{6}{56} \\
\hangindent=2em\hangafter=1\hspace{1em}Co-simulation & \ResultStudies{5}{56} \\
\hangindent=2em\hangafter=1\hspace{1em}Vehicle-in-the-loop (testbench) & \ResultStudies{4}{56} \\
\hangindent=2em\hangafter=1\hspace{1em}Camera sensor-in-the-loop & \ResultStudies{3}{56} \\
\hangindent=2em\hangafter=1\hspace{1em}Human-in-the-loop & \ResultStudies{3}{56} \\
\hangindent=2em\hangafter=1\hspace{1em}Physical scale-model /\hspace{0pt} surrogate-platform validation & \ResultStudies{2}{56} \\
\hangindent=2em\hangafter=1\hspace{1em}LiDAR sensor-in-the-loop & \ResultStudies{1}{56} \\
\hangindent=2em\hangafter=1\hspace{1em}Modular X-in-the-loop framework & \ResultStudies{1}{56} \\
\hangindent=2em\hangafter=1\hspace{1em}Radar sensor-in-the-loop & \ResultStudies{1}{56} \\
\hangindent=2em\hangafter=1\hspace{1em}Small-scale vehicle-in-the-loop & \ResultStudies{1}{56} \\
\hangindent=2em\hangafter=1\hspace{1em}Steering-actuator hardware-in-the-loop & \ResultStudies{1}{56} \\
\hangindent=2em\hangafter=1\hspace{1em}XiL/\hspace{0pt}HIL & \ResultStudies{1}{56} \\
\midrule
\hangindent=1em\hangafter=1 \textbf{Generative content synthesis} & \ResultStudies{41}{56} \\
\hangindent=2em\hangafter=1\hspace{1em}Diffusion-based generation & \ResultStudies{14}{56} \\
\hangindent=2em\hangafter=1\hspace{1em}GAN-based generation/\hspace{0pt}translation & \ResultStudies{12}{56} \\
\hangindent=2em\hangafter=1\hspace{1em}Neural rendering (NeRF/\hspace{0pt}3DGS) & \ResultStudies{6}{56} \\
\hangindent=2em\hangafter=1\hspace{1em}Synthetic data generation/\hspace{0pt}augmentation & \ResultStudies{4}{56} \\
\hangindent=2em\hangafter=1\hspace{1em}Synthetic perturbation/\hspace{0pt}augmentation & \ResultStudies{3}{56} \\
\hangindent=2em\hangafter=1\hspace{1em}Image compositing & \ResultStudies{1}{56} \\
\hangindent=2em\hangafter=1\hspace{1em}Physics-based multi-sensor object insertion & \ResultStudies{1}{56} \\
\hangindent=2em\hangafter=1\hspace{1em}Point-cloud/\hspace{0pt}image reprojection \& warping & \ResultStudies{1}{56} \\
\midrule
\hangindent=1em\hangafter=1 \textbf{Traffic/\hspace{0pt}agent behaviour modelling} & \ResultStudies{29}{56} \\
\hangindent=2em\hangafter=1\hspace{1em}Learning-based (imitation/\hspace{0pt}RL) traffic models & \ResultStudies{7}{56} \\
\hangindent=2em\hangafter=1\hspace{1em}Traffic/\hspace{0pt}behaviour-model calibration & \ResultStudies{5}{56} \\
\hangindent=2em\hangafter=1\hspace{1em}Route/\hspace{0pt}trajectory-conditioned behaviour modelling & \ResultStudies{4}{56} \\
\hangindent=2em\hangafter=1\hspace{1em}Learned/\hspace{0pt}generative trajectory models & \ResultStudies{3}{56} \\
\hangindent=2em\hangafter=1\hspace{1em}Rule-based/\hspace{0pt}behaviour-tree traffic models & \ResultStudies{3}{56} \\
\hangindent=2em\hangafter=1\hspace{1em}Optimization/\hspace{0pt}search-based generation & \ResultStudies{2}{56} \\
\hangindent=2em\hangafter=1\hspace{1em}Statistical/\hspace{0pt}distributional traffic calibration & \ResultStudies{2}{56} \\
\hangindent=2em\hangafter=1\hspace{1em}Field-test/\hspace{0pt}traffic-flow calibration & \ResultStudies{1}{56} \\
\hangindent=2em\hangafter=1\hspace{1em}Human-reactive traffic models & \ResultStudies{1}{56} \\
\hangindent=2em\hangafter=1\hspace{1em}Traffic/\hspace{0pt}NPC behaviour modelling & \ResultStudies{1}{56} \\
\midrule
\hangindent=1em\hangafter=1 \textbf{Real-world evidence reconstruction} & \ResultStudies{28}{56} \\
\hangindent=2em\hangafter=1\hspace{1em}Recorded driving/\hspace{0pt}sensor-log reconstruction & \ResultStudies{12}{56} \\
\hangindent=2em\hangafter=1\hspace{1em}Crash/\hspace{0pt}accident-evidence reconstruction & \ResultStudies{6}{56} \\
\hangindent=2em\hangafter=1\hspace{1em}Map/\hspace{0pt}aerial-derived reconstruction & \ResultStudies{5}{56} \\
\hangindent=2em\hangafter=1\hspace{1em}Scenario-generation framework/\hspace{0pt}toolchain & \ResultStudies{2}{56} \\
\hangindent=2em\hangafter=1\hspace{1em}Data-driven simulator construction & \ResultStudies{1}{56} \\
\hangindent=2em\hangafter=1\hspace{1em}Domain adaptation & \ResultStudies{1}{56} \\
\hangindent=2em\hangafter=1\hspace{1em}Reconstruction-strategy comparison (classical vs. neural assets) & \ResultStudies{1}{56} \\
\hangindent=2em\hangafter=1\hspace{1em}Retrieval-augmented domain adaptation & \ResultStudies{1}{56} \\
\midrule
\hangindent=1em\hangafter=1 \textbf{Sensor physics modelling} & \ResultStudies{28}{56} \\
\hangindent=2em\hangafter=1\hspace{1em}LiDAR sensor modelling & \ResultStudies{14}{56} \\
\hangindent=2em\hangafter=1\hspace{1em}Camera sensor modelling & \ResultStudies{6}{56} \\
\hangindent=2em\hangafter=1\hspace{1em}Radar sensor modelling & \ResultStudies{5}{56} \\
\hangindent=2em\hangafter=1\hspace{1em}Camera+LiDAR+radar sensor modelling & \ResultStudies{2}{56} \\
\hangindent=2em\hangafter=1\hspace{1em}Compute/\hspace{0pt}timing-fidelity emulation & \ResultStudies{2}{56} \\
\midrule
\hangindent=1em\hangafter=1 \textbf{Ego policy/\hspace{0pt}control transfer} & \ResultStudies{13}{56} \\
\hangindent=2em\hangafter=1\hspace{1em}Reinforcement-learning-based policy transfer & \ResultStudies{5}{56} \\
\hangindent=2em\hangafter=1\hspace{1em}Vehicle-dynamics/\hspace{0pt}controller calibration & \ResultStudies{5}{56} \\
\hangindent=2em\hangafter=1\hspace{1em}Imitation-learning-based policy transfer & \ResultStudies{2}{56} \\
\hangindent=2em\hangafter=1\hspace{1em}Generative-video-assisted policy transfer & \ResultStudies{1}{56} \\
\bottomrule
\end{tabular}
\end{table}

\head{Real-equipment integration (56 studies)} These approaches combine simulation with physical vehicles, sensors, controllers, or human participants during testing. On-road vehicle-in-the-loop and mixed reality are the two largest subcategories. They include virtual traffic or actors surrounding a real vehicle, studies of real-vehicle testing platforms, and virtual or synthetic content injected into a vehicle's sensing or planning stack. Smaller subcategories include controller/ECU hardware-in-the-loop, co-simulation, camera sensor-in-the-loop, non-road vehicle-in-the-loop testbenches, and human-in-the-loop testing.

\head{Generative content synthesis (41 studies)} These approaches create or modify sensor observations or scene content to make simulated inputs more representative of reality. Diffusion-based and GAN-based methods are the two largest subcategories. Diffusion approaches generate or refine sensor imagery, video, traffic scenarios, or multi-agent trajectories. GAN-based approaches mainly modify camera inputs for downstream perception or detection tasks, with one LiDAR case. Saleh et al.~\cite{saleh2019DomainAdaptation} use a GAN to adapt simulated bird's-eye-view LiDAR point clouds toward the real domain before training a vehicle detector. A further subset uses neural rendering, including neural radiance fields (NeRF) and three-dimensional Gaussian splatting (3DGS).

\head{Traffic/agent behaviour modelling (29 studies)} These approaches model how surrounding vehicles and other road users move and react. Techniques include learned behaviour, predefined rules, and calibration from observed traffic. Learned approaches include imitation learning from demonstrations, reinforcement learning through interaction or feedback, and trajectory-generation models. Trajectory-conditioned models generate behaviour subject to a specified path or motion pattern. Bergamini et al.~\cite{bergamini2021SimnetLearning} learn a reactive traffic-agent simulator from real-world driving logs.

\head{Real-world evidence reconstruction (28 studies)} These approaches use driving recordings, crash evidence, or other real-world information to construct simulation content or adapt models across domains. Recorded driving or sensor-log reconstruction is the largest subcategory, followed by crash or accident-evidence reconstruction and map- or aerial-derived reconstruction. Most studies reconstruct trajectories or agent behaviour from recorded logs. Others reconstruct complete scenes or road geometry, evaluate domain-adaptation techniques, or generate scenarios from real driving or accident databases. Kim et al.~\cite{kim2024DesigningNaturalistic} extract naturalistic driving situations from real recordings to build training scenarios for an automated driving system.

\head{Sensor physics modelling (28 studies)} These approaches model how sensors produce measurements and how objects or environmental conditions affect them. LiDAR sensor modelling is the largest subcategory. The models cover effects such as weather-related attenuation, intensity, reflectance, and point dropout and are calibrated or validated against real LiDAR measurements. Smaller subsets model camera, radar, or combined camera, LiDAR, and radar sensing. Two studies model compute-hardware timing fidelity: one calibrates it against real CPU measurements rather than a live-integrated ECU, and one injects model and edge computation latency. Rosenberger et al.~\cite{rosenberger2019BenchmarkingFunctional} decompose an automotive LiDAR sensor model into benchmarked functional components representing physical sensor behaviour.

\head{Ego policy/control transfer (13 studies)} These approaches adapt driving policies or control models so that behaviour developed in simulation transfers to a physical vehicle. Guti\'errez-Moreno et al.~\cite{gutierrezmoreno2024DecisionMaking} adapt a decision-making policy developed in simulation for transfer to a real autonomous driving stack.

\subsubsection{RQ\textsubscript{2.3}: Execution Level of the Proposed Approach}

\begin{figure*}[tbp]
\centering
\includegraphics[width=0.85\linewidth]{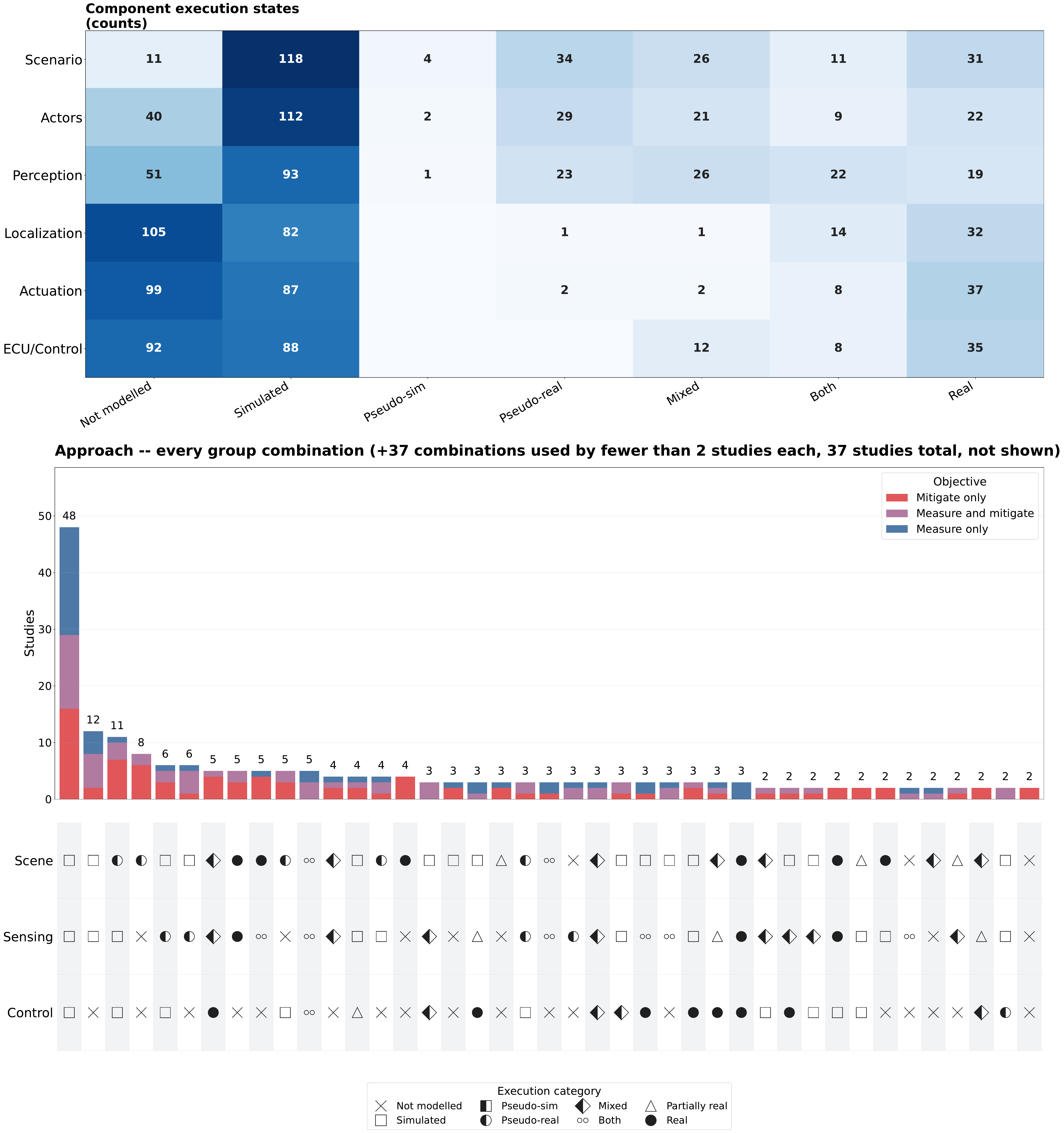}
\caption{RQ\textsubscript{2.3} execution level of the proposed approach: execution levels of the six individual components (top), and the observed combinations for Scene, Sensing, and Control ranked by frequency (bottom). In the lower panel, colours encode study objective and icons encode execution level.}
\label{fig:rq2-3}
\end{figure*}

\autoref{fig:rq2-3} shows how the proposed approaches are executed across the six components and the three functional groups. At the functional-group level, 48 of 235 studies (20.4\%) use simulated content across Scene, Sensing, and Control. This is the most common configuration. Only 3 studies (1.3\%) use real content across all three groups. The remaining studies use many different combinations of simulated, real, and intermediate execution levels.

Simulation is the most common execution level at the component level. Across the 1,410 component classifications, 580 (41.1\%) are Simulated. Another 398 (28.2\%) are Not modelled, while 176 (12.5\%) are Real. The remaining classifications are Pseudo-real (89, 6.3\%), Mixed (88, 6.2\%), Both (72, 5.1\%), and Pseudo-sim (7, 0.5\%).

The distribution also differs substantially between components. Scenario and Actors are most often Simulated, with 118 and 112 studies, respectively. Only 11 studies do not model Scenario, while 40 do not model Actors. Perception is also commonly Simulated (93 studies), but it has a more varied distribution. It is Pseudo-real in 23 studies, Mixed in 26, Both in 22, and Real in 19.

Later components are more often absent from the experimental setup. Localization is Not modelled in 105 studies, Actuation in 99, and ECU/control in 92. Nevertheless, simulation remains common when these components are included. Localization is Simulated in 82 studies, while Actuation is Simulated in 87 studies and ECU/control in 88. Real execution is also more common for these components than for Perception. The Real counts are 32 for Localization, 37 for Actuation, and 35 for ECU/control, compared with 19 for Perception.

The intermediate execution levels represent a substantial part of the corpus. Pseudo-real, Mixed, and Both together account for 249 of the 1,410 component classifications (17.7\%). Their use also differs across components. Pseudo-real occurs most often for Scenario (34), Actors (29), and Perception (23). Mixed follows a similar pattern, with 26 studies for Scenario, 21 for Actors, and 26 for Perception. Both is most common for Perception (22), followed by Localization (14). Pseudo-sim is rare, with only 7 classifications across all six components.

These categories capture different relationships between simulation and reality. Pseudo-real represents real-world-derived content without the corresponding real component being present live. Mixed combines real and simulated elements within the same execution. Both instead represents separate simulated and real executions of the same component. Their frequency shows that a binary simulated-versus-real classification would hide a substantial part of the experimental setups in the corpus.

The lower panel of \autoref{fig:rq2-3} shows the combinations of execution levels across Scene, Sensing, and Control. The fully simulated configuration is the largest single group with 48 studies, but the overall distribution is fragmented. The next configurations contain only 12, 11, and 8 studies. A further 37 combinations occur in only one study each. Thus, no execution profile other than the fully simulated configuration represents a large share of the corpus.

These results describe component content rather than the physical setup used to obtain it. In particular, Real does not necessarily mean live physical execution. A component can be classified as Real when it uses recorded real-world data.

\begin{table}[tbp]
\caption{RQ\textsubscript{3.1} evaluation techniques and supporting primary studies. Bars show the number of studies using each technique, and the final column lists the corresponding primary studies. Studies may use several techniques, so counts are not mutually exclusive.}
\label{tab:rq3-1}
\centering
\footnotesize
\begin{tabular}{@{}p{0.30\textwidth} p{0.20\textwidth} p{0.45\textwidth}@{}}
\toprule
\textbf{Evaluation technique} & \textbf{Studies} & \textbf{Primary studies} \\
\midrule
\hangindent=1em\hangafter=1 Direct or matched sim-to-real comparison & \ResultStudies{132}{132} & \cite{rosenberger2018AnalysisReal,rosenberger2019BenchmarkingFunctional,langer2021CalibrationAssessment,ngo2021DeepEvaluation,pahk2023EffectsSim2real,jayapalgowdu2022EvaluationScenariobased,lichtle2024FromSim,pahk2024LaneSegmentation,peintner2021MixedReality,kim2022QueryingLabelled,reway2020TestMethod,manivasagam2023TowardsZero,wachtel2021ValidationRadar,parra2020ValidationRealtime,krecht2023ValidationProcess,shuguang2022VehicleintheloopIntelligent,amini2020LearningRobust,argui2023BuildingVisionbased,bashetty2020DeepcrashtestTurning,bateman2024ExploringReal,bergamini2021SimnetLearning,brogle2019HardwareintheloopAutonomous,dona2022VirtualTesting,drechsler2022HowSimulation,fremont2020FormalScenariobased,gupta2023XintheloopXil,gutierrezmoreno2024DecisionMaking,briand-offline-emse,stocco2023ModelVs,hoppe2022AgentbasedAutonomous,hsiang2022DevelopmentSimulationbased,schyr2022VehicleintheloopTestinga,tettamanti2018VehicleintheloopTest,zofka2018SleepwalkerFramework,wagner2019TowardsCrossverification,xu2021ReliabilityGan,wallace2022ValidatingSimulation,shi2022IntegratedTraffic,zhu2022MillimeterwaveRadar,xu2023BitsBilevel,yuan2024FromNaturalistic,stadler2022CredibilityAssessment,durukal2022DigitalTwin,queiroz2024DrivervehicleModel,huang2024HierarchicalLidar,larter2022HierarchicalPedestrian,jiang2021HighaccuracyFramework,haider2024MethodologyVerifying,hasirlioglu2018ModelbasedApproach,ngo2021MultilayeredApproach,chen2019NovelIntegrated,oh2024ReliabilityEvaluation,che2021TestMethod,solmaz2020VehicleintheloopMethodology,lan2024AssessingReliability,amini2024BridgingGap,fu2024ImmersiveTraffic,hwang2024ImplementationSimulationbased,suo2023MixsimHierarchical,ulreich2024NovelTest,wachtel2024SimulationValidation,lee2024StudyFull,reckenzaun2024TransnationalTesting,yang2023UnisimNeural,drechsler2024VehicleDynamics,chen2025SnerfAutonomous,guo2025ScenedmConsistent,ganesan2025BridgingSimtoreal,li2025DatadrivenHeadlight,neis2025DatadrivenLane,ameli2025DesignPreparation,li2025DgnrDensityguided,zhou2026DiffusionModels,yan2025DistributionallyConsistent,gao2026DrivenceRealistic,verma2025EgoVehicle,liu2026EnhancingAutonomous,nosheen2025EnhancingSynthetic,naing2025FinegrainedTrajectory,wu2025FusingTemporal,wei2025InteractiveCritical,allamaa2025LearningbasedNmpc,gaddam2025LidarSensor,zuo2025RaladBridging,jung2025RealvirtualSynchronized,saad2025DigitalTwin,lambertenghi2025MultimodalityEvaluation,zhao2025TachographbasedApproach,haghighi2026UnifiedGenerative,harhammer2025SimtorealDomain,samak2026Sim2realDiffusion,bi2025TowardFullstack,anand2026TowardClosing,wang2026UtdtBaseline,lee2025ModelcentricQuantitative,waheed2025QuantifyingSim2real,oh2024AdvilsImplementation,stocco2023MindGap,manivasagam2020LidarsimRealistic,shen2023SimonwheelsPhysical,huch2023QuantifyingLidar,vargasrivero2022EffectSpray,magosi2023NovelApproach,wang2025DigitalTwinbased,khaire2025AugmentedRealityenhanced,wang2022AutomaticTraffic,balm2025BenchmarkingDomain,cardozo2025CarlaHardwareintheloop,arango2026ClosingSim2real,liang2025DttfsimDigital,toth2023DevelopmentFunctional,drechsler2022DynamicVehicleintheloop,magosi2022EvaluationMethodology,argui2025ExploringAugmentation,wei2025GeneratingRisky,tian2026HgsimHighfidelity,steimle2025HumanvehicleintheloopTest,hofrichter2025IntroducingDevelopment,gao2025MsietransformerNovel,li2024MultilevelMetrics,coppola2025VirtualTesting,elgharbawy2019OntologybasedAdaptive,segonne2023QualificationPavin,daza2023SimtorealTransfer,jekl2025SpeedPlanner,park2020VehicleintheloopGlobal,wang2022Cadsim,yang2020RecoveringSimulating,leung2025RoadCode,guo2021SeamlessModel,sigl2021ExperiencingDatadriven,deter2021SimulatingAutonomous} \\
\hangindent=1em\hangafter=1 Task/\hspace{0pt}SUT performance comparison & \ResultStudies{130}{132} & \cite{nalic2019DevelopmentCosimulation,saleh2019DomainAdaptation,pahk2023EffectsSim2real,oladimeji2024EnhancedApproaches,jayapalgowdu2022EvaluationScenariobased,pahk2024LaneSegmentation,peintner2021MixedReality,reway2022SimulationbasedTest,reway2020TestMethod,zofka2018TrafficParticipants,wachtel2021ValidationRadar,shuguang2022VehicleintheloopIntelligent,amini2020LearningRobust,argui2023BuildingVisionbased,bai2024ArbasedMeta,bashetty2020DeepcrashtestTurning,bateman2024ExploringReal,brogle2019HardwareintheloopAutonomous,bui2024ValidationAdaptive,chen2022AnalysisRealtime,chen2020MixedTest,drechsler2022MireMixed,drechsler2022HowSimulation,fremont2020FormalScenariobased,gambi2019GeneratingEffective,gao2020HardwareintheloopSimulation,gupta2023XintheloopXil,gutierrezmoreno2024DecisionMaking,briand-offline-emse,stocco2023ModelVs,hoppe2022AgentbasedAutonomous,hsiang2022DevelopmentSimulationbased,schyr2022VehicleintheloopTestinga,tettamanti2018VehicleintheloopTest,yang2021OnlineAdaptation,vonbernuth2018RenderingPhysically,tim2020SimulationReal,talwar2020EvaluatingValidity,tenbrock2021ConscendDataset,xu2021ReliabilityGan,zhang2022SelfdrivingSimulation,shi2022IntegratedTraffic,zhang2023BuildingCritical,stadler2022CredibilityAssessment,queiroz2024DrivervehicleModel,huang2024HierarchicalLidar,jiang2021HighaccuracyFramework,barker2018LowcostHardwareintheloop,haider2024MethodologyVerifying,hasirlioglu2018ModelbasedApproach,ngo2021MultilayeredApproach,chen2019NovelIntegrated,oh2024ReliabilityEvaluation,che2021TestMethod,solmaz2020VehicleintheloopMethodology,lan2024AssessingReliability,amini2024BridgingGap,fu2024ImmersiveTraffic,suo2023MixsimHierarchical,reckenzaun2024TransnationalTesting,yang2023UnisimNeural,drechsler2024VehicleDynamics,chen2025SnerfAutonomous,fu2025ControllablePedestrian,zeng2025ControllableRainy,steinhauser2025DataQuality,li2025DatadrivenHeadlight,zhou2026DiffusionModels,wang2025DrivegenCoevaluating,hong2025EffectiveLearning,simakov2025EffectsSimulator,baresi2025EfficientDomain,zhang2025EndtoendLearning,pasios2025EnhancingPhotorealism,nosheen2025EnhancingSynthetic,li2025GenerativeReconstruction,wang2025GenerativeTesting,zhang2025HighriskTrajectories,zhou2026HugsimRealtime,allamaa2025LearningbasedNmpc,gaddam2025LidarSensor,peng2025MethodApplications,darocha2025MicroautowareAutoware,zuo2025RaladBridging,gao2025DatacentricImage,saad2025DigitalTwin,elmquist2025MethodologyQuantify,lambertenghi2025MultimodalityEvaluation,zhang2025PixellevelImage,haghighi2026UnifiedGenerative,drechsler2025AcceleratingApproval,li2025AutonomousDriving,kim2025AutonomousVehicle,harhammer2025SimtorealDomain,sun2025TestingAutonomous,stangherlin2025XintheloopApproach,li2025TowardIntelligent,wang2026UtdtBaseline,ulreich2025ValidatingCamera,otto2025ValidationAibased,huang2026VersatileBehavior,schulz2025VirtualKitti,haider2025ModelingSimulation,schabus2025ObjectbasedLidar,petrovic2025SuitabilitySmallsized,stocco2023MindGap,manivasagam2020LidarsimRealistic,lambertenghi2024AssessingQuality,shen2023SimonwheelsPhysical,huch2023QuantifyingLidar,amini2022Vista20,wang2025DigitalTwinbased,shoukat2024AutonomousDriving,martinserrano2025BehaviouralGap,balm2025BenchmarkingDomain,arango2026ClosingSim2real,toth2023DevelopmentFunctional,argui2025ExploringAugmentation,tian2026HgsimHighfidelity,gao2025MsietransformerNovel,segonne2023QualificationPavin,daza2023SimtorealTransfer,jekl2025SpeedPlanner,park2020VehicleintheloopGlobal,song2026VisionlanguageModelbased,dauner2024Navsim,wang2022Cadsim,yang2020RecoveringSimulating,reisgys2021ArgumentationADAS,deter2021SimulatingAutonomous} \\
\hangindent=1em\hangafter=1 Quantitative or distributional comparison & \ResultStudies{65}{132} & \cite{langer2021CalibrationAssessment,kettelgerdes2023InfieldMeasurement,reway2020TestMethod,dona2022VirtualTesting,duminil2024AssessingFidelity,gambi2019GeneratingEffective,gu2023CopulabasedInteractive,hsiang2022DevelopmentSimulationbased,wagner2019TowardsCrossverification,tim2020SimulationReal,stadler2022TowardsMultivariate,shi2022IntegratedTraffic,zhang2022TrajgenGenerating,sun2024DrivescenegenGenerating,stadler2022CredibilityAssessment,durukal2022DigitalTwin,jiang2021HighaccuracyFramework,haider2024MethodologyVerifying,oh2024ReliabilityEvaluation,lan2024AssessingReliability,amini2024BridgingGap,fei2024CorrelationSoftwareintheloop,suo2023MixsimHierarchical,ulreich2024NovelTest,yang2023UnisimNeural,drechsler2024VehicleDynamics,guo2025ScenedmConsistent,ganesan2025BridgingSimtoreal,huang2025CadreControllable,steinhauser2025DataQuality,neis2025DatadrivenLane,yan2025DistributionallyConsistent,gao2026DrivenceRealistic,simakov2025EffectsSimulator,zhang2025EndtoendLearning,naing2025FinegrainedTrajectory,li2025GenerativeReconstruction,wang2025GenerativeTesting,zhang2025HighriskTrajectories,wang2025RadeLearning,jung2025RealvirtualSynchronized,elmquist2025MethodologyQuantify,wu2025ToolBenchmarking,jing2025EfficientHighrisk,kolb2025AutomaticallyImproving,lee2025ModelcentricQuantitative,oh2024AdvilsImplementation,lambertenghi2024AssessingQuality,cao2024ReinforcementLearning,duminil2024ComprehensiveExploration,magosi2023NovelApproach,wang2025DigitalTwinbased,martinserrano2025BehaviouralGap,liang2025DttfsimDigital,drechsler2022DynamicVehicleintheloop,magosi2022EvaluationMethodology,gao2025MsietransformerNovel,li2024MultilevelMetrics,daza2023SimtorealTransfer,song2026VisionlanguageModelbased,dauner2024Navsim,wang2022Cadsim,yang2020RecoveringSimulating,reisgys2021ArgumentationADAS,deter2021SimulatingAutonomous} \\
\hangindent=1em\hangafter=1 Feature or image-quality comparison & \ResultStudies{24}{132} & \cite{oladimeji2024EnhancedApproaches,reway2022SimulationbasedTest,cao2024AresTextdriven,duminil2024AssessingFidelity,hsiang2022DevelopmentSimulationbased,thieling2019NeuralNetworks,soni2021DeepLearning,chen2025SnerfAutonomous,steinhauser2025DataQuality,baresi2025EfficientDomain,zhou2026HugsimRealtime,kuroda2025AnalysisSnow,eberhardt2025BeyondSmudge,hess2025SplatadRealtime,yu2026TrajectorybasedVideo,schulz2025VirtualKitti,li2025XldCrosslane,gadipudi2022SyntheticReal,argui2025ExploringAugmentation,tian2026HgsimHighfidelity,wang2022Cadsim,leung2025RoadCode,reisgys2021ArgumentationADAS,guo2021SeamlessModel} \\
\hangindent=1em\hangafter=1 Human or qualitative assessment & \ResultStudies{21}{132} & \cite{langer2021CalibrationAssessment,lichtle2024FromSim,kim2022QueryingLabelled,zofka2018TrafficParticipants,yeo2019MaximMixedreality,yuan2024FromNaturalistic,zhou2024EvaluatingAutonomous,ulhas2024GanbasedDomain,sun2024DrivescenegenGenerating,wu2024RealityBites,durukal2022DigitalTwin,che2021TestMethod,fu2024ImmersiveTraffic,baresi2025EfficientDomain,zhang2025EndtoendLearning,buzdugan2025HumancenteredSimulations,abbasi2025Acr2xodrFrom,yu2026TrajectorybasedVideo,cao2024ReinforcementLearning,khaire2025AugmentedRealityenhanced,argui2025ExploringAugmentation} \\
\hangindent=1em\hangafter=1 No substantive evaluation & \ResultStudies{16}{132} & \cite{liu2021ApplicationDigital,lu2022DesignVehicle,kim2024DesigningNaturalistic,samak2024TowardsValidation,szalai2020MixedReality,szalay2019ProofConcept,xinxin2020CsgCritical,zofka2023PavingWays,rezaldi20213dOrthomosaic,long2024GeneratingRealistic,liu2024IntelligentCosimulation,samak2024MetaverseSafer,capdevila2024SmartsafeMobility,hu2025SimulationbasedTesting,wang2023WhenSimulator,heikkinen2019HardwareintheloopPlatform} \\
\bottomrule
\end{tabular}
\end{table}

\subsection{Evaluation}
\label{sec:results-validation}

This group covers evaluation procedures, the execution level of the evaluation reference, and reported metrics. A study can use several evaluation procedures or metric categories in RQ\textsubscript{3.1} and RQ\textsubscript{3.3}. RQ\textsubscript{3.2} reports one execution level per component for the evaluation reference.

\subsubsection{RQ\textsubscript{3.1}: Evaluation Procedure}

Nearly all studies report a validation or measurement procedure: 222 of 235 (94.5\%), while 13 (5.5\%) report none. The evaluation techniques and the corresponding primary studies are shown in~\autoref{tab:rq3-1}. Direct or matched simulation-to-reality comparison (132 studies) and task/SUT-performance comparison (130) are the two most common techniques. They are followed by quantitative or distributional comparison (65), feature or image-quality comparison (24), and human or qualitative assessment (21).
Sixteen studies have no substantive evaluation technique. This includes the 13 studies reporting no evaluation procedure and three that report an activity, such as a functional demonstration, without a comparison covered by the coded evaluation techniques.

Most studies combine evaluation techniques. Ninety-eight studies (41.7\%) use exactly one, 92 (39.1\%) combine two, and 29 (12.3\%) combine three or more. The most common exact combination is task/SUT-performance comparison with direct or matched simulation-to-reality comparison (49 studies, 20.9\%). This is followed by task/SUT-performance comparison alone (38, 16.2\%) and direct or matched simulation-to-reality comparison alone (35, 14.9\%). 

The next most common exact combinations are quantitative or distributional comparison + direct or matched sim-to-real comparison (17); quantitative or distributional comparison + task/SUT performance comparison + direct or matched sim-to-real comparison (16); no substantive evaluation (16); quantitative or distributional comparison (11); quantitative or distributional comparison + task/SUT performance comparison (11); feature or image-quality comparison (8); human or qualitative assessment (6); feature or image-quality comparison + task/SUT performance comparison (4); and human or qualitative assessment + direct or matched sim-to-real comparison (4).

Evaluation techniques differ across study objectives. Mitigate only studies most often use task/SUT-performance comparison (53 of 103, 51.5\%), followed by direct or matched simulation-to-reality comparison (45, 43.7\%). Direct or matched comparison is the most common technique among Measure and mitigate studies (54 of 76, 71.1\%). Among Measure only studies, task/SUT-performance comparison and direct or matched comparison occur at similar rates, with 35 of 56 studies (62.5\%) and 33 (58.9\%), respectively. Quantitative or distributional comparison occurs in 20.4\%, 32.9\%, and 33.9\% of the three groups, respectively.

No substantive evaluation occurs in 15 Mitigate only studies, no Measure and mitigate studies, and one Measure only study. For mitigation studies, this means that the effectiveness of the proposed mitigation is not explicitly assessed. For the Measure only study, no substantive procedure for measuring the gap is reported.

\subsubsection{RQ\textsubscript{3.2}: Execution Level of the Evaluation Reference}

\begin{figure*}[tbp]
\centering
\includegraphics[width=0.85\linewidth]{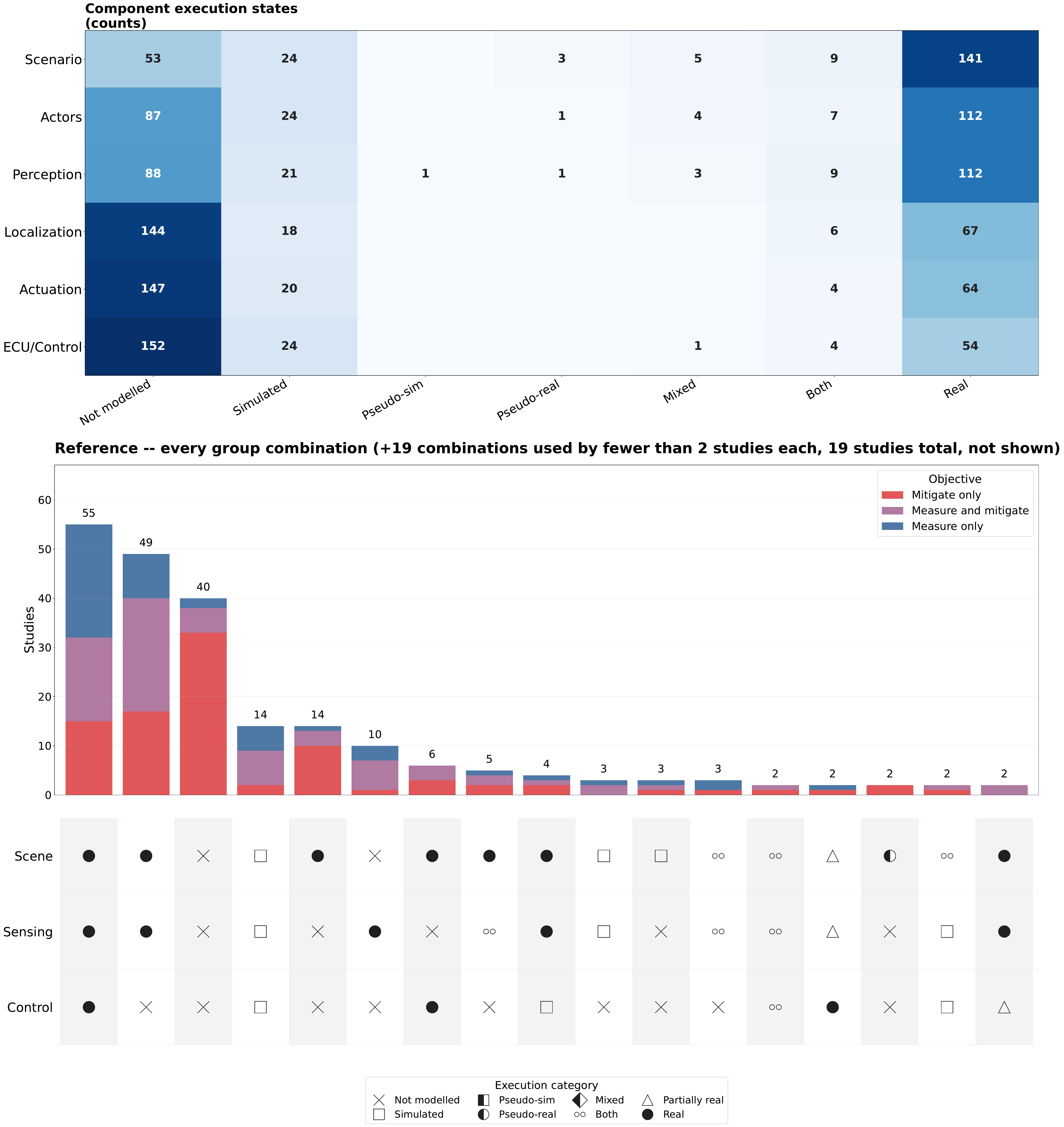}
\caption{RQ\textsubscript{3.2} execution level of the evaluation: six-component execution-state counts (top), and the observed functional-group state combinations ranked by frequency (bottom). In the lower panel, colours encode study objective, and icons encode execution category.}
\label{fig:rq3-2}
\end{figure*}

\autoref{fig:rq3-2} shows the execution level of the evaluation reference across the six components and the three functional groups. At the functional-group level, the most common reference profile has real content across Scene, Sensing, and Control (55 of 235 studies, 23.4\%). It is followed by real Scene and Sensing with Control not modelled (49, 20.9\%), and by references in which all three groups are absent (40, 17.0\%). Only 14 studies (6.0\%) have a fully simulated reference. In total, 36 distinct profiles occur, 19 of them in a single study, as the lower panel of the figure shows.

At the component level, references are mostly real or absent. Of the 1,410 component classifications, 550 (39.0\%) are Real, and 671 (47.6\%) are Not modelled, while only 131 (9.3\%) are Simulated. The intermediate levels are rare: Both (39, 2.8\%), Mixed (13, 0.9\%), Pseudo-real (5, 0.4\%), and Pseudo-sim (1, 0.1\%). By contrast, only 176 of the 1,410 classifications for the proposed approach (12.5\%) are Real.

The distribution differs between components. The component counts coded Real are Scenario (141), Actors (112), Perception (112), Localization (67), Actuation (64), ECU/Control (54). Reference content is absent for Scenario (53), Actors (87), Perception (88), Localization (144), Actuation (147), and ECU/Control (152), so Localization, Actuation, and ECU/Control are Not modelled in more than half of the studies. The counts coded Simulated are Scenario (24), Actors (24), Perception (21), Localization (18), Actuation (20), and ECU/Control (24). Both is the most frequent intermediate level, with Scenario (9), Actors (7), Perception (9), Localization (6), Actuation (4), ECU/Control (4).

To see how these references are obtained, we count the study-level reference sources among the studies whose proposed approach includes each component. A source may support only part of the setup, so these counts are not attributed to individual components, and sources may overlap. A real dataset is the most frequent source for Scenario (98 of 224, 43.8\%), Actors (89 of 195, 45.6\%), and Perception (73 of 184, 39.7\%). A real vehicle is the most frequent source for Localization (56 of 130, 43.1\%), Actuation (58 of 136, 42.6\%), and ECU/Control (61 of 143, 42.7\%). Real hardware is rarely the source, in 7 to 18 studies per component. A simulator is the reference source in 17.7\% to 22.8\% of these studies, and no executed reference is reported in 16.1\% to 22.3\%.

Of the 55 studies with real content across all three reference groups, 50 include physical vehicle or hardware experiments and 5 use recorded real data without a physical vehicle or hardware experiment. A real-content profile is therefore not a count of live tests. Recorded images or point clouds can support a perception comparison; a closed-loop driving claim additionally requires evidence about the interaction of commands, motion, and subsequent observations.

\subsubsection{RQ\textsubscript{3.3}: Evaluation Metrics}\label{sec:metrics}

Seven categories capture the reported metrics, while an eighth records studies reporting no metric. The categories, their most frequent specific metrics, and the corresponding primary studies are shown in~\autoref{tab:rq3-3}; all specific metrics are listed in the Appendix. Sensor/perception comparison (107 studies) and ADS behaviour/trajectory comparison (96) are the two most common categories. They are followed by Vehicle dynamics/actuation comparison (37), Safety/event outcome (34), No metric reported (23), Runtime/computational (19), Qualitative assessment (16), and Human/participant behavioural comparison (1).

\head{Sensor/perception comparison (107 studies)} Detection accuracy/rate is the most common metric with 30 studies, followed by average precision/mAP/AP with 27, IoU/segmentation overlap with 20, precision/recall/F1 with 19, and FID/FVD/KID with 18. Image-similarity metrics include SSIM (15), RMSE/MAE-based error (13), LPIPS (12), and PSNR (11). Manivasagam et al.~\cite{manivasagam2023TowardsZero} report detection average precision to quantify how closely a realistic LiDAR simulator reproduces real sensing conditions.

\head{ADS behaviour/trajectory comparison (96 studies)} Speed/velocity comparison is the most common metric with 36 studies, followed by trajectory error using ADE/FDE/RMSE/ATE (15), trajectory/path distance (13), and RMSE/MAE-based error (11). Lane keeping/car following behaviour appears in 9 studies and distribution-distance tests in 6. Inter-vehicle distance and direct real-versus-simulated driving-behaviour comparison each appear in 5 studies. Bui et al.~\cite{bui2024ValidationAdaptive} validate a model-predictive adaptive cruise controller by comparing commanded and real vehicle speed in a real-time system.

\head{Vehicle dynamics/actuation comparison (37 studies)} Acceleration is the most common measure with 17 studies, followed by steering (8), jerk (6), RMSE/MAE-based error (6), distribution-distance tests (5), and yaw (4). Stadler et al.~\cite{stadler2022CredibilityAssessment} include acceleration among the metrics used to assess a scenario-based virtual test relative to reality.

\head{Safety/event outcome (34 studies)} Collision/crash rate is the most common measure with 26 studies, followed by time to collision or post-encroachment time (PET) with 12. Pass/fail, or disengagement rate, appears in 8 studies. Safety distance/near miss appears in 3 studies, while a composite scenario criticality score and traffic-safety distributional statistics appear in one study each. Bai et al.~\cite{bai2024ArbasedMeta} measure collision rate in their AR-based vehicle-road cooperation testing system.

\head{Runtime/computational (19 studies)} Memory/compute utilization appears in 7 studies, latency/runtime in 6, frame rate/FPS in 3, and inference time and network/communication load in 2 each. Several other measures occur once.

\head{Qualitative assessment (16 studies)} Visual or qualitative inspection is the most common approach with 11 studies, followed by Likert-scale ratings (3), human/user studies (2), and subjective realism scoring (1).

\head{Human/participant behavioural comparison (1 study)} Mart\'in-Serrano et al.~\cite{martinserrano2025BehaviouralGap} compare human driver behaviour in a real vehicle with behaviour in a virtual-reality driving scenario.

\head{No metric reported (23 studies)} These studies report no evaluation metric. This includes two studies that report only a general ``comparability of results'' between two setups without naming a specific quantity.

\begin{table}[tbp]
\caption{RQ\textsubscript{3.3} evaluation metric categories and their eight most frequent specific metrics (ties alphabetical), with primary studies.}
\label{tab:rq3-3}
\centering
\footnotesize
\renewcommand{\arraystretch}{0.9}
\setlength{\ResultBarMax}{1.2cm}
\begin{tabular}{@{}p{0.35\textwidth} p{0.13\textwidth} p{0.47\textwidth}@{}}
\toprule
\textbf{Metric category / metric} & \textbf{Studies} & \textbf{Primary studies} \\
\midrule
\hangindent=1em\hangafter=1 \textbf{Sensor/\hspace{0pt}perception comparison} & \ResultStudies{107}{107} &  \\
\hangindent=2em\hangafter=1\hspace{1em}Detection accuracy/\hspace{0pt}rate & \ResultStudies{30}{107} & \cite{rosenberger2019BenchmarkingFunctional,oladimeji2024EnhancedApproaches,reway2020TestMethod,manivasagam2023TowardsZero,bashetty2020DeepcrashtestTurning,chen2022AnalysisRealtime,hsiang2022DevelopmentSimulationbased,schyr2022VehicleintheloopTestinga,vonbernuth2018RenderingPhysically,stadler2022CredibilityAssessment,barker2018LowcostHardwareintheloop,yang2023UnisimNeural,li2025DatadrivenHeadlight,simakov2025EffectsSimulator,peng2025MethodApplications,saad2025DigitalTwin,wang2026UtdtBaseline,otto2025ValidationAibased,haider2025ModelingSimulation,schabus2025ObjectbasedLidar,manivasagam2020LidarsimRealistic,vargasrivero2022EffectSpray,duminil2024ComprehensiveExploration,balm2025BenchmarkingDomain,argui2025ExploringAugmentation,tian2026HgsimHighfidelity,segonne2023QualificationPavin,wang2022Cadsim,yang2020RecoveringSimulating,deter2021SimulatingAutonomous} \\
\hangindent=2em\hangafter=1\hspace{1em}Average precision (mAP/\hspace{0pt}AP) & \ResultStudies{27}{107} & \cite{manivasagam2023TowardsZero,bateman2024ExploringReal,chen2022AnalysisRealtime,hsiang2022DevelopmentSimulationbased,vonbernuth2018RenderingPhysically,talwar2020EvaluatingValidity,xu2021ReliabilityGan,zhang2023BuildingCritical,amini2024BridgingGap,yang2023UnisimNeural,guo2025ScenedmConsistent,fu2025ControllablePedestrian,zeng2025ControllableRainy,steinhauser2025DataQuality,li2025DatadrivenHeadlight,zhou2026DiffusionModels,nosheen2025EnhancingSynthetic,peng2025MethodApplications,zuo2025RaladBridging,gao2025DatacentricImage,haider2025ModelingSimulation,petrovic2025SuitabilitySmallsized,manivasagam2020LidarsimRealistic,huch2023QuantifyingLidar,tian2026HgsimHighfidelity,yang2020RecoveringSimulating,leung2025RoadCode} \\
\hangindent=2em\hangafter=1\hspace{1em}IoU/\hspace{0pt}segmentation overlap & \ResultStudies{20}{107} & \cite{argui2023BuildingVisionbased,tim2020SimulationReal,talwar2020EvaluatingValidity,zhang2023BuildingCritical,ngo2021MultilayeredApproach,yang2023UnisimNeural,steinhauser2025DataQuality,pasios2025EnhancingPhotorealism,zhou2026HugsimRealtime,gaddam2025LidarSensor,zuo2025RaladBridging,elmquist2025MethodologyQuantify,lambertenghi2025MultimodalityEvaluation,haghighi2026UnifiedGenerative,manivasagam2020LidarsimRealistic,shen2023SimonwheelsPhysical,tian2026HgsimHighfidelity,wang2022Cadsim,yang2020RecoveringSimulating,leung2025RoadCode} \\
\hangindent=2em\hangafter=1\hspace{1em}Precision/\hspace{0pt}recall/\hspace{0pt}F1 & \ResultStudies{19}{107} & \cite{saleh2019DomainAdaptation,reway2022SimulationbasedTest,reway2020TestMethod,manivasagam2023TowardsZero,bateman2024ExploringReal,brogle2019HardwareintheloopAutonomous,chen2022AnalysisRealtime,drechsler2022MireMixed,hsiang2022DevelopmentSimulationbased,tim2020SimulationReal,barker2018LowcostHardwareintheloop,steinhauser2025DataQuality,hong2025EffectiveLearning,nosheen2025EnhancingSynthetic,li2025GenerativeReconstruction,gao2025DatacentricImage,huch2023QuantifyingLidar,yang2020RecoveringSimulating,leung2025RoadCode} \\
\hangindent=2em\hangafter=1\hspace{1em}Distributional image/\hspace{0pt}video realism (FID/\hspace{0pt}FVD/\hspace{0pt}KID) & \ResultStudies{18}{107} & \cite{pahk2023EffectsSim2real,pahk2024LaneSegmentation,cao2024AresTextdriven,soni2021DeepLearning,sun2024DrivescenegenGenerating,yang2023UnisimNeural,chen2025SnerfAutonomous,zeng2025ControllableRainy,steinhauser2025DataQuality,zhou2026DiffusionModels,wang2025DrivegenCoevaluating,nosheen2025EnhancingSynthetic,wang2025GenerativeTesting,zhou2026HugsimRealtime,haghighi2026UnifiedGenerative,yu2026TrajectorybasedVideo,tian2026HgsimHighfidelity,wang2022Cadsim} \\
\hangindent=2em\hangafter=1\hspace{1em}Structural similarity (SSIM) & \ResultStudies{15}{107} & \cite{reway2022SimulationbasedTest,hsiang2022DevelopmentSimulationbased,hasirlioglu2018ModelbasedApproach,yang2023UnisimNeural,chen2025SnerfAutonomous,steinhauser2025DataQuality,li2025DgnrDensityguided,nosheen2025EnhancingSynthetic,zhou2026HugsimRealtime,zhao2025TachographbasedApproach,shen2023SimonwheelsPhysical,argui2025ExploringAugmentation,tian2026HgsimHighfidelity,wang2022Cadsim,leung2025RoadCode} \\
\hangindent=2em\hangafter=1\hspace{1em}RMSE/\hspace{0pt}MAE-family error & \ResultStudies{13}{107} & \cite{reway2022SimulationbasedTest,stadler2022CredibilityAssessment,amini2024BridgingGap,yang2023UnisimNeural,steinhauser2025DataQuality,li2025GenerativeReconstruction,gaddam2025LidarSensor,peng2025MethodApplications,zhao2025TachographbasedApproach,haghighi2026UnifiedGenerative,argui2025ExploringAugmentation,wang2022Cadsim,leung2025RoadCode} \\
\hangindent=2em\hangafter=1\hspace{1em}Learned perceptual similarity (LPIPS) & \ResultStudies{12}{107} & \cite{pahk2024LaneSegmentation,yang2023UnisimNeural,chen2025SnerfAutonomous,steinhauser2025DataQuality,li2025DgnrDensityguided,zhou2026DiffusionModels,nosheen2025EnhancingSynthetic,zhou2026HugsimRealtime,yu2026TrajectorybasedVideo,shen2023SimonwheelsPhysical,wang2022Cadsim,leung2025RoadCode} \\ [0.5em]
\hangindent=1em\hangafter=1 \textbf{ADS behaviour/\hspace{0pt}trajectory comparison} & \ResultStudies{96}{107} &  \\
\hangindent=2em\hangafter=1\hspace{1em}Speed/\hspace{0pt}velocity comparison & \ResultStudies{36}{107} & \cite{langer2021CalibrationAssessment,jayapalgowdu2022EvaluationScenariobased,lichtle2024FromSim,peintner2021MixedReality,wachtel2021ValidationRadar,bui2024ValidationAdaptive,dona2022VirtualTesting,drechsler2022HowSimulation,gao2020HardwareintheloopSimulation,gupta2023XintheloopXil,gutierrezmoreno2024DecisionMaking,tettamanti2018VehicleintheloopTest,zhou2024EvaluatingAutonomous,stadler2022CredibilityAssessment,barker2018LowcostHardwareintheloop,haider2024MethodologyVerifying,solmaz2020VehicleintheloopMethodology,fei2024CorrelationSoftwareintheloop,suo2023MixsimHierarchical,guo2025ScenedmConsistent,ganesan2025BridgingSimtoreal,ameli2025DesignPreparation,yan2025DistributionallyConsistent,verma2025EgoVehicle,liu2026EnhancingAutonomous,zhang2025HighriskTrajectories,buzdugan2025HumancenteredSimulations,allamaa2025LearningbasedNmpc,darocha2025MicroautowareAutoware,wang2025RadeLearning,drechsler2025AcceleratingApproval,jing2025EfficientHighrisk,hess2025SplatadRealtime,oh2024AdvilsImplementation,drechsler2022DynamicVehicleintheloop,coppola2025VirtualTesting} \\
\hangindent=2em\hangafter=1\hspace{1em}Trajectory error (ADE/\hspace{0pt}FDE/\hspace{0pt}RMSE/\hspace{0pt}ATE) & \ResultStudies{15}{107} & \cite{pahk2023EffectsSim2real,peintner2021MixedReality,manivasagam2023TowardsZero,bergamini2021SimnetLearning,jiang2021HighaccuracyFramework,suo2023MixsimHierarchical,reckenzaun2024TransnationalTesting,yang2023UnisimNeural,ganesan2025BridgingSimtoreal,wang2025DrivegenCoevaluating,zhang2025EndtoendLearning,wu2025FusingTemporal,jung2025RealvirtualSynchronized,drechsler2025AcceleratingApproval,gao2025MsietransformerNovel} \\
\hangindent=2em\hangafter=1\hspace{1em}Trajectory/\hspace{0pt}path point-wise distance (sim vs. real) & \ResultStudies{13}{107} & \cite{krecht2023ValidationProcess,amini2020LearningRobust,bergamini2021SimnetLearning,bui2024ValidationAdaptive,fremont2020FormalScenariobased,gao2020HardwareintheloopSimulation,gupta2023XintheloopXil,tenbrock2021ConscendDataset,queiroz2024DrivervehicleModel,lambertenghi2025MultimodalityEvaluation,li2025AutonomousDriving,oh2024AdvilsImplementation,cao2024ReinforcementLearning} \\
\hangindent=2em\hangafter=1\hspace{1em}RMSE/\hspace{0pt}MAE-family error & \ResultStudies{11}{107} & \cite{dona2022VirtualTesting,yang2021OnlineAdaptation,stadler2022CredibilityAssessment,fei2024CorrelationSoftwareintheloop,yang2023UnisimNeural,ganesan2025BridgingSimtoreal,wu2025FusingTemporal,allamaa2025LearningbasedNmpc,drechsler2025AcceleratingApproval,jing2025EfficientHighrisk,wang2022Cadsim} \\
\hangindent=2em\hangafter=1\hspace{1em}Lane-keeping/\hspace{0pt}car-following behaviour & \ResultStudies{9}{107} & \cite{langer2021CalibrationAssessment,lichtle2024FromSim,schyr2022VehicleintheloopTestinga,stadler2022CredibilityAssessment,solmaz2020VehicleintheloopMethodology,hong2025EffectiveLearning,li2025AutonomousDriving,li2025XldCrosslane,wang2023WhenSimulator} \\
\hangindent=2em\hangafter=1\hspace{1em}Distribution-distance test & \ResultStudies{6}{107} & \cite{langer2021CalibrationAssessment,stadler2022CredibilityAssessment,lan2024AssessingReliability,suo2023MixsimHierarchical,liu2026EnhancingAutonomous,cao2024ReinforcementLearning} \\
\hangindent=2em\hangafter=1\hspace{1em}Inter-vehicle/\hspace{0pt}relative-spacing distance & \ResultStudies{5}{107} & \cite{peintner2021MixedReality,fei2024CorrelationSoftwareintheloop,suo2023MixsimHierarchical,guo2025ScenedmConsistent,wang2025RadeLearning} \\
\hangindent=2em\hangafter=1\hspace{1em}Real-vs-simulated driving-behaviour comparison & \ResultStudies{5}{107} & \cite{shuguang2022VehicleintheloopIntelligent,yang2021OnlineAdaptation,reckenzaun2024TransnationalTesting,samak2026Sim2realDiffusion,amini2022Vista20} \\ [0.5em]
\hangindent=1em\hangafter=1 \textbf{Vehicle dynamics/\hspace{0pt}actuation comparison} & \ResultStudies{37}{107} &  \\
\hangindent=2em\hangafter=1\hspace{1em}Acceleration & \ResultStudies{17}{107} & \cite{langer2021CalibrationAssessment,lichtle2024FromSim,peintner2021MixedReality,bui2024ValidationAdaptive,dona2022VirtualTesting,stadler2022CredibilityAssessment,durukal2022DigitalTwin,solmaz2020VehicleintheloopMethodology,fei2024CorrelationSoftwareintheloop,suo2023MixsimHierarchical,guo2025ScenedmConsistent,verma2025EgoVehicle,liu2026EnhancingAutonomous,zhang2025HighriskTrajectories,buzdugan2025HumancenteredSimulations,wei2025InteractiveCritical,cao2024ReinforcementLearning} \\
\hangindent=2em\hangafter=1\hspace{1em}Steering & \ResultStudies{8}{107} & \cite{peintner2021MixedReality,gutierrezmoreno2024DecisionMaking,solmaz2020VehicleintheloopMethodology,ganesan2025BridgingSimtoreal,darocha2025MicroautowareAutoware,kim2025AutonomousVehicle,oh2024AdvilsImplementation,stocco2023MindGap} \\
\hangindent=2em\hangafter=1\hspace{1em}Jerk & \ResultStudies{6}{107} & \cite{gao2026DrivenceRealistic,zhang2025EndtoendLearning,liu2026EnhancingAutonomous,zhang2025HighriskTrajectories,wei2025InteractiveCritical,cao2024ReinforcementLearning} \\
\hangindent=2em\hangafter=1\hspace{1em}RMSE/\hspace{0pt}MAE-family error & \ResultStudies{6}{107} & \cite{dona2022VirtualTesting,stadler2022CredibilityAssessment,durukal2022DigitalTwin,fei2024CorrelationSoftwareintheloop,ganesan2025BridgingSimtoreal,drechsler2025AcceleratingApproval} \\
\hangindent=2em\hangafter=1\hspace{1em}Distribution-distance test & \ResultStudies{5}{107} & \cite{langer2021CalibrationAssessment,stadler2022CredibilityAssessment,suo2023MixsimHierarchical,liu2026EnhancingAutonomous,cao2024ReinforcementLearning} \\
\hangindent=2em\hangafter=1\hspace{1em}Braking/\hspace{0pt}deceleration & \ResultStudies{4}{107} & \cite{tettamanti2018VehicleintheloopTest,ameli2025DesignPreparation,shen2023SimonwheelsPhysical,reisgys2021ArgumentationADAS} \\
\hangindent=2em\hangafter=1\hspace{1em}Yaw & \ResultStudies{4}{107} & \cite{solmaz2020VehicleintheloopMethodology,gao2026DrivenceRealistic,buzdugan2025HumancenteredSimulations,oh2024AdvilsImplementation} \\
\hangindent=2em\hangafter=1\hspace{1em}Correlation coefficient & \ResultStudies{3}{107} & \cite{stadler2022CredibilityAssessment,durukal2022DigitalTwin,fei2024CorrelationSoftwareintheloop} \\ [0.5em]
\hangindent=1em\hangafter=1 \textbf{Safety/\hspace{0pt}event outcome} & \ResultStudies{34}{107} &  \\
\hangindent=2em\hangafter=1\hspace{1em}Collision/\hspace{0pt}crash rate & \ResultStudies{26}{107} & \cite{bai2024ArbasedMeta,bashetty2020DeepcrashtestTurning,bergamini2021SimnetLearning,chen2020MixedTest,fremont2020FormalScenariobased,schyr2022VehicleintheloopTestinga,xu2023BitsBilevel,zhou2024EvaluatingAutonomous,stadler2022CredibilityAssessment,suo2023MixsimHierarchical,guo2025ScenedmConsistent,hong2025EffectiveLearning,zhang2025EndtoendLearning,liu2026EnhancingAutonomous,zhang2025HighriskTrajectories,zhou2026HugsimRealtime,wei2025InteractiveCritical,wang2025RadeLearning,jing2025EfficientHighrisk,sun2025TestingAutonomous,wei2025GeneratingRisky,dauner2024Navsim,reisgys2021ArgumentationADAS,guo2021SeamlessModel,sigl2021ExperiencingDatadriven,wang2023WhenSimulator} \\
\hangindent=2em\hangafter=1\hspace{1em}Time-to-collision/\hspace{0pt}PET & \ResultStudies{12}{107} & \cite{stadler2022CredibilityAssessment,guo2025ScenedmConsistent,liu2026EnhancingAutonomous,zhang2025HighriskTrajectories,zhou2026HugsimRealtime,wei2025InteractiveCritical,wang2025RadeLearning,jing2025EfficientHighrisk,dauner2024Navsim,reisgys2021ArgumentationADAS,guo2021SeamlessModel,sigl2021ExperiencingDatadriven} \\
\hangindent=2em\hangafter=1\hspace{1em}Pass/\hspace{0pt}fail or disengagement rate & \ResultStudies{8}{107} & \cite{zhang2025HighriskTrajectories,zhou2026HugsimRealtime,stangherlin2025XintheloopApproach,stocco2023MindGap,arango2026ClosingSim2real,steimle2025HumanvehicleintheloopTest,dauner2024Navsim,sigl2021ExperiencingDatadriven} \\
\hangindent=2em\hangafter=1\hspace{1em}Safety distance/\hspace{0pt}near-miss & \ResultStudies{3}{107} & \cite{durukal2022DigitalTwin,suo2023MixsimHierarchical,song2026VisionlanguageModelbased} \\
\hangindent=2em\hangafter=1\hspace{1em}Scenario criticality score (composite risk indicators) & \ResultStudies{1}{107} & \cite{hu2025SimulationbasedTesting} \\
\hangindent=2em\hangafter=1\hspace{1em}Traffic-safety distributional statistics & \ResultStudies{1}{107} & \cite{yan2025DistributionallyConsistent} \\ [0.5em]
\hangindent=1em\hangafter=1 \textbf{No metric reported} & \ResultStudies{23}{107} & \cite{liu2021ApplicationDigital,lu2022DesignVehicle,kim2024DesigningNaturalistic,nalic2019DevelopmentCosimulation,samak2024TowardsValidation,zofka2018SleepwalkerFramework,szalai2020MixedReality,szalay2019ProofConcept,xinxin2020CsgCritical,ulhas2024GanbasedDomain,chen2019NovelIntegrated,long2024GeneratingRealistic,fu2024ImmersiveTraffic,hwang2024ImplementationSimulationbased,liu2024IntelligentCosimulation,samak2024MetaverseSafer,capdevila2024SmartsafeMobility,lee2024StudyFull,zhang2025PixellevelImage,khaire2025AugmentedRealityenhanced,toth2023DevelopmentFunctional,park2020VehicleintheloopGlobal,heikkinen2019HardwareintheloopPlatform} \\ [0.5em]
\hangindent=1em\hangafter=1 \textbf{Runtime/\hspace{0pt}computational} & \ResultStudies{19}{107} &  \\
\hangindent=2em\hangafter=1\hspace{1em}Memory/\hspace{0pt}compute utilization & \ResultStudies{7}{107} & \cite{hoppe2022AgentbasedAutonomous,li2025DgnrDensityguided,simakov2025EffectsSimulator,baresi2025EfficientDomain,nosheen2025EnhancingSynthetic,gaddam2025LidarSensor,yu2026TrajectorybasedVideo} \\
\hangindent=2em\hangafter=1\hspace{1em}Latency/\hspace{0pt}runtime & \ResultStudies{6}{107} & \cite{chen2022AnalysisRealtime,neis2025DatadrivenLane,gao2026DrivenceRealistic,hong2025EffectiveLearning,darocha2025MicroautowareAutoware,shen2023SimonwheelsPhysical} \\
\hangindent=2em\hangafter=1\hspace{1em}Frame rate (FPS) & \ResultStudies{3}{107} & \cite{li2025DgnrDensityguided,gaddam2025LidarSensor,petrovic2025SuitabilitySmallsized} \\
\hangindent=2em\hangafter=1\hspace{1em}Inference time & \ResultStudies{2}{107} & \cite{brogle2019HardwareintheloopAutonomous,nosheen2025EnhancingSynthetic} \\
\hangindent=2em\hangafter=1\hspace{1em}Network/\hspace{0pt}communication load & \ResultStudies{2}{107} & \cite{zofka2023PavingWays,shoukat2024AutonomousDriving} \\
\hangindent=2em\hangafter=1\hspace{1em}Accuracy-vs-compute tradeoff (Pareto efficiency) & \ResultStudies{1}{107} & \cite{simakov2025EffectsSimulator} \\
\hangindent=2em\hangafter=1\hspace{1em}Model training time & \ResultStudies{1}{107} & \cite{zuo2025RaladBridging} \\
\hangindent=2em\hangafter=1\hspace{1em}ODD-coverage \& ADS-failure testing overhead & \ResultStudies{1}{107} & \cite{baresi2025EfficientDomain} \\ [0.5em]
\hangindent=1em\hangafter=1 \textbf{Qualitative assessment} & \ResultStudies{16}{107} &  \\
\hangindent=2em\hangafter=1\hspace{1em}Visual/\hspace{0pt}qualitative inspection & \ResultStudies{11}{107} & \cite{zofka2018TrafficParticipants,krecht2023ValidationProcess,wagner2019TowardsCrossverification,yuan2024FromNaturalistic,rezaldi20213dOrthomosaic,hasirlioglu2018ModelbasedApproach,che2021TestMethod,wachtel2024SimulationValidation,stocco2023MindGap,huch2023QuantifyingLidar,deter2021SimulatingAutonomous} \\
\hangindent=2em\hangafter=1\hspace{1em}Likert-scale rating & \ResultStudies{3}{107} & \cite{zeng2025ControllableRainy,zhang2025EndtoendLearning,abbasi2025Acr2xodrFrom} \\
\hangindent=2em\hangafter=1\hspace{1em}Human/\hspace{0pt}user study & \ResultStudies{2}{107} & \cite{kim2022QueryingLabelled,petrovic2025SuitabilitySmallsized} \\
\hangindent=2em\hangafter=1\hspace{1em}Whiteness/\hspace{0pt}subjective realism score & \ResultStudies{1}{107} & \cite{petrovic2025SuitabilitySmallsized} \\ [0.5em]
\hangindent=1em\hangafter=1 \textbf{Human/\hspace{0pt}participant behavioural comparison} & \ResultStudies{1}{107} & \cite{martinserrano2025BehaviouralGap} \\
\bottomrule
\end{tabular}
\end{table}

\subsection{Experimental Components}
\label{sec:results-experimental-components}

\subsubsection{RQ\textsubscript{4.1}: Simulation Platforms}

\autoref{tab:rq4-1} lists the simulation platforms and datasets named by more than four studies. The complete list can be found in our Appendix.
CARLA is the most common named simulator with 45 studies. Nineteen studies report no named simulation platform or reference dataset. Unity and IPG CarMaker follow with 16 studies each, while every other named platform appears in 13 studies or fewer. 
Among the reference datasets, KITTI is the most common with 18 studies. nuScenes follows with 10 studies, Cityscapes and the Waymo Open Dataset with 9 each, and Virtual KITTI with 8. 

\begin{table}[t]
\caption{RQ\textsubscript{4.1} simulation platforms and reference datasets reported by at least five studies.}
\label{tab:rq4-1}
\centering

\footnotesize
\setlength{\ResultBarMax}{1.6cm}
\begin{tabular}{@{}>{\raggedright\arraybackslash}p{\dimexpr0.9\columnwidth-\dimexpr\ResultBarMax+2.3em\relax-2\tabcolsep\relax} p{\dimexpr\ResultBarMax+2.3em\relax}@{}}
\toprule
\textbf{Simulation platform / dataset} & \textbf{Studies} \\
\midrule
\hangindent=1em\hangafter=1 \textbf{Simulation platforms} & \ResultStudies{182}{182} \\
\hangindent=2em\hangafter=1\hspace{1em}CARLA & \ResultStudies{45}{182} \\
\hangindent=2em\hangafter=1\hspace{1em}IPG CarMaker & \ResultStudies{16}{182} \\
\hangindent=2em\hangafter=1\hspace{1em}Unity & \ResultStudies{16}{182} \\
\hangindent=2em\hangafter=1\hspace{1em}MATLAB/\hspace{0pt}Simulink & \ResultStudies{13}{182} \\
\hangindent=2em\hangafter=1\hspace{1em}SUMO & \ResultStudies{12}{182} \\
\hangindent=2em\hangafter=1\hspace{1em}Custom/\hspace{0pt}research simulator & \ResultStudies{11}{182} \\
\hangindent=2em\hangafter=1\hspace{1em}VIRES VTD & \ResultStudies{9}{182} \\
\hangindent=2em\hangafter=1\hspace{1em}SVL/\hspace{0pt}LGSVL & \ResultStudies{7}{182} \\
\hangindent=2em\hangafter=1\hspace{1em}Gazebo & \ResultStudies{7}{182} \\
\hangindent=2em\hangafter=1\hspace{1em}LidarSim & \ResultStudies{6}{182} \\
\hangindent=2em\hangafter=1\hspace{1em}Autoware simulator & \ResultStudies{6}{182} \\
\midrule
\hangindent=1em\hangafter=1 \textbf{Dataset/\hspace{0pt}reference data} & \ResultStudies{66}{182} \\
\hangindent=2em\hangafter=1\hspace{1em}KITTI & \ResultStudies{18}{182} \\
\hangindent=2em\hangafter=1\hspace{1em}nuScenes & \ResultStudies{10}{182} \\
\hangindent=2em\hangafter=1\hspace{1em}Cityscapes & \ResultStudies{9}{182} \\
\hangindent=2em\hangafter=1\hspace{1em}Waymo Open Dataset & \ResultStudies{9}{182} \\
\hangindent=2em\hangafter=1\hspace{1em}Virtual KITTI & \ResultStudies{8}{182} \\
\hangindent=2em\hangafter=1\hspace{1em}Custom/\hspace{0pt}study-collected dataset & \ResultStudies{7}{182} \\
\hangindent=2em\hangafter=1\hspace{1em}Traffic-trajectory dataset (NGSIM/\hspace{0pt}HighD/\hspace{0pt}rounD/\hspace{0pt}INTERACTION) & \ResultStudies{7}{182} \\
\midrule
\hangindent=1em\hangafter=1 \textbf{No simulation platform or reference dataset reported} & \ResultStudies{19}{182} \\
\bottomrule
\end{tabular}
\end{table}

\subsubsection{RQ\textsubscript{4.2}: Hardware Platforms}

RQ\textsubscript{4.2} records the physical equipment used by each study, from full-size or small-scale vehicles to individual sensors, controllers, and compute hardware. A study can include several pieces of equipment. The hardware categories and the corresponding primary studies are shown in~\autoref{tab:rq4-2}. The No hardware/unspecified category represents the absence of reported hardware.

No hardware or unspecified hardware is the most common category with 91 studies. These studies use simulation or existing data without reporting physical equipment. Among studies reporting hardware, full-size vehicles are the most common platform (76), followed by cameras (48), LiDAR (44), localization/positioning hardware such as GNSS/IMU units (44), and radar (19). Compute hardware, such as embedded GPUs or in-vehicle processing units, is reported by 25 studies and ECU/embedded controllers by 20.

The remaining categories are sensor-stimulation equipment (11); roadside/communication infrastructure, VR/MR or motion-capture equipment, and small-scale vehicles/robots (9 each); human driving controls (8); testbench/HIL equipment (6); drone/aerial platforms (4); and atmospheric test chambers (3). Atmospheric test chambers generate artificial rain, fog, or spray water to test real sensors or vehicles under controlled adverse-weather conditions. They are used by Hasirlioglu and Riener~\cite{hasirlioglu2018ModelbasedApproach}, Vargas Rivero et al.~\cite{vargasrivero2022EffectSpray}, and the PAVIN-platform study~\cite{segonne2023QualificationPavin}. Lu et al.~\cite{lu2022DesignVehicle} build a vehicle-in-the-loop test system around a real full-size autopilot-equipped vehicle. Pahk et al.~\cite{pahk2023EffectsSim2real} evaluate a lane-keeping-assist system using real camera hardware after translating simulated images toward the real domain.

\begin{table}[tbp]
\caption{RQ\textsubscript{4.2} hardware platforms and supporting primary studies.}
\label{tab:rq4-2}
\centering
\footnotesize
\begin{tabular}{@{}p{0.28\textwidth} p{0.20\textwidth} p{0.47\textwidth}@{}}
\toprule
\textbf{Hardware platform} & \textbf{Studies} & \textbf{Primary studies} \\
\midrule
\hangindent=1em\hangafter=1 No hardware or unspecified & \ResultStudies{91}{91} & \cite{liu2021ApplicationDigital,langer2021CalibrationAssessment,kim2024DesigningNaturalistic,nalic2019DevelopmentCosimulation,saleh2019DomainAdaptation,pahk2024LaneSegmentation,kim2022QueryingLabelled,argui2023BuildingVisionbased,bashetty2020DeepcrashtestTurning,bateman2024ExploringReal,bergamini2021SimnetLearning,cao2024AresTextdriven,duminil2024AssessingFidelity,gambi2019GeneratingEffective,gu2023CopulabasedInteractive,briand-offline-emse,hoppe2022AgentbasedAutonomous,hsiang2022DevelopmentSimulationbased,yang2021OnlineAdaptation,vonbernuth2018RenderingPhysically,szalai2020MixedReality,wagner2019TowardsCrossverification,thieling2019NeuralNetworks,tim2020SimulationReal,xinxin2020CsgCritical,talwar2020EvaluatingValidity,tenbrock2021ConscendDataset,xu2021ReliabilityGan,soni2021DeepLearning,zhang2022SelfdrivingSimulation,stadler2022TowardsMultivariate,shi2022IntegratedTraffic,zhu2022MillimeterwaveRadar,zhang2022TrajgenGenerating,xu2023BitsBilevel,zhang2023BuildingCritical,zhou2024EvaluatingAutonomous,sun2024DrivescenegenGenerating,wu2024RealityBites,stadler2022CredibilityAssessment,queiroz2024DrivervehicleModel,lan2024AssessingReliability,amini2024BridgingGap,long2024GeneratingRealistic,suo2023MixsimHierarchical,chen2025SnerfAutonomous,guo2025ScenedmConsistent,huang2025CadreControllable,fu2025ControllablePedestrian,steinhauser2025DataQuality,zhou2026DiffusionModels,yan2025DistributionallyConsistent,wang2025DrivegenCoevaluating,simakov2025EffectsSimulator,pasios2025EnhancingPhotorealism,naing2025FinegrainedTrajectory,wu2025FusingTemporal,wang2025RadeLearning,gao2025DatacentricImage,elmquist2025MethodologyQuantify,zhang2025PixellevelImage,wu2025ToolBenchmarking,haghighi2026UnifiedGenerative,abbasi2025Acr2xodrFrom,jing2025EfficientHighrisk,kolb2025AutomaticallyImproving,eberhardt2025BeyondSmudge,harhammer2025SimtorealDomain,samak2026Sim2realDiffusion,hess2025SplatadRealtime,sun2025TestingAutonomous,bi2025TowardFullstack,yu2026TrajectorybasedVideo,huang2026VersatileBehavior,schulz2025VirtualKitti,lee2025ModelcentricQuantitative,schabus2025ObjectbasedLidar,lambertenghi2024AssessingQuality,cao2024ReinforcementLearning,gadipudi2022SyntheticReal,hu2025SimulationbasedTesting,wang2022AutomaticTraffic,liang2025DttfsimDigital,argui2025ExploringAugmentation,wei2025GeneratingRisky,gao2025MsietransformerNovel,dauner2024Navsim,wang2022Cadsim,leung2025RoadCode,deter2021SimulatingAutonomous,wang2023WhenSimulator} \\
\hangindent=1em\hangafter=1 Full-size vehicle & \ResultStudies{76}{91} & \cite{ngo2021DeepEvaluation,jayapalgowdu2022EvaluationScenariobased,lichtle2024FromSim,peintner2021MixedReality,reway2020TestMethod,samak2024TowardsValidation,manivasagam2023TowardsZero,zofka2018TrafficParticipants,wachtel2021ValidationRadar,parra2020ValidationRealtime,krecht2023ValidationProcess,shuguang2022VehicleintheloopIntelligent,bai2024ArbasedMeta,brogle2019HardwareintheloopAutonomous,chen2020MixedTest,dona2022VirtualTesting,drechsler2022MireMixed,drechsler2022HowSimulation,fremont2020FormalScenariobased,gao2020HardwareintheloopSimulation,gutierrezmoreno2024DecisionMaking,schyr2022VehicleintheloopTestinga,tettamanti2018VehicleintheloopTest,szalay2019ProofConcept,yeo2019MaximMixedreality,zofka2023PavingWays,durukal2022DigitalTwin,huang2024HierarchicalLidar,jiang2021HighaccuracyFramework,haider2024MethodologyVerifying,hasirlioglu2018ModelbasedApproach,ngo2021MultilayeredApproach,chen2019NovelIntegrated,oh2024ReliabilityEvaluation,che2021TestMethod,solmaz2020VehicleintheloopMethodology,fei2024CorrelationSoftwareintheloop,hwang2024ImplementationSimulationbased,liu2024IntelligentCosimulation,samak2024MetaverseSafer,capdevila2024SmartsafeMobility,lee2024StudyFull,reckenzaun2024TransnationalTesting,yang2023UnisimNeural,drechsler2024VehicleDynamics,ganesan2025BridgingSimtoreal,ameli2025DesignPreparation,allamaa2025LearningbasedNmpc,peng2025MethodApplications,zuo2025RaladBridging,jung2025RealvirtualSynchronized,zhao2025TachographbasedApproach,drechsler2025AcceleratingApproval,oh2024AdvilsImplementation,manivasagam2020LidarsimRealistic,shen2023SimonwheelsPhysical,vargasrivero2022EffectSpray,amini2022Vista20,duminil2024ComprehensiveExploration,magosi2023NovelApproach,khaire2025AugmentedRealityenhanced,shoukat2024AutonomousDriving,cardozo2025CarlaHardwareintheloop,arango2026ClosingSim2real,drechsler2022DynamicVehicleintheloop,magosi2022EvaluationMethodology,tian2026HgsimHighfidelity,steimle2025HumanvehicleintheloopTest,li2024MultilevelMetrics,coppola2025VirtualTesting,daza2023SimtorealTransfer,jekl2025SpeedPlanner,park2020VehicleintheloopGlobal,yang2020RecoveringSimulating,reisgys2021ArgumentationADAS,sigl2021ExperiencingDatadriven} \\
\hangindent=1em\hangafter=1 Camera & \ResultStudies{48}{91} & \cite{pahk2023EffectsSim2real,oladimeji2024EnhancedApproaches,lichtle2024FromSim,reway2022SimulationbasedTest,reway2020TestMethod,shuguang2022VehicleintheloopIntelligent,amini2020LearningRobust,bai2024ArbasedMeta,brogle2019HardwareintheloopAutonomous,chen2020MixedTest,drechsler2022MireMixed,drechsler2022HowSimulation,fremont2020FormalScenariobased,gupta2023XintheloopXil,schyr2022VehicleintheloopTestinga,tettamanti2018VehicleintheloopTest,barker2018LowcostHardwareintheloop,haider2024MethodologyVerifying,hasirlioglu2018ModelbasedApproach,chen2019NovelIntegrated,hwang2024ImplementationSimulationbased,liu2024IntelligentCosimulation,samak2024MetaverseSafer,ulreich2024NovelTest,capdevila2024SmartsafeMobility,lee2024StudyFull,reckenzaun2024TransnationalTesting,yang2023UnisimNeural,ganesan2025BridgingSimtoreal,li2025DatadrivenHeadlight,ameli2025DesignPreparation,hong2025EffectiveLearning,jung2025RealvirtualSynchronized,lambertenghi2025MultimodalityEvaluation,zhao2025TachographbasedApproach,li2025AutonomousDriving,li2025TowardIntelligent,wang2026UtdtBaseline,ulreich2025ValidatingCamera,otto2025ValidationAibased,li2025XldCrosslane,petrovic2025SuitabilitySmallsized,shen2023SimonwheelsPhysical,amini2022Vista20,wang2025DigitalTwinbased,cardozo2025CarlaHardwareintheloop,song2026VisionlanguageModelbased,yang2020RecoveringSimulating} \\
\hangindent=1em\hangafter=1 LiDAR & \ResultStudies{44}{91} & \cite{rosenberger2018AnalysisReal,rosenberger2019BenchmarkingFunctional,kettelgerdes2023InfieldMeasurement,manivasagam2023TowardsZero,zofka2018TrafficParticipants,wachtel2021ValidationRadar,bai2024ArbasedMeta,brogle2019HardwareintheloopAutonomous,chen2022AnalysisRealtime,chen2020MixedTest,fremont2020FormalScenariobased,gupta2023XintheloopXil,tettamanti2018VehicleintheloopTest,zofka2018SleepwalkerFramework,szalay2019ProofConcept,wallace2022ValidatingSimulation,yuan2024FromNaturalistic,huang2024HierarchicalLidar,haider2024MethodologyVerifying,hasirlioglu2018ModelbasedApproach,chen2019NovelIntegrated,hwang2024ImplementationSimulationbased,samak2024MetaverseSafer,capdevila2024SmartsafeMobility,lee2024StudyFull,reckenzaun2024TransnationalTesting,yang2023UnisimNeural,ganesan2025BridgingSimtoreal,peng2025MethodApplications,saad2025DigitalTwin,lambertenghi2025MultimodalityEvaluation,anand2026TowardClosing,wang2026UtdtBaseline,petrovic2025SuitabilitySmallsized,oh2024AdvilsImplementation,shen2023SimonwheelsPhysical,huch2023QuantifyingLidar,vargasrivero2022EffectSpray,amini2022Vista20,wang2025DigitalTwinbased,balm2025BenchmarkingDomain,hofrichter2025IntroducingDevelopment,li2024MultilevelMetrics,yang2020RecoveringSimulating} \\
\hangindent=1em\hangafter=1 Localization/\hspace{0pt}positioning hardware & \ResultStudies{44}{91} & \cite{ngo2021DeepEvaluation,lu2022DesignVehicle,pahk2023EffectsSim2real,peintner2021MixedReality,reway2020TestMethod,parra2020ValidationRealtime,krecht2023ValidationProcess,shuguang2022VehicleintheloopIntelligent,amini2020LearningRobust,bai2024ArbasedMeta,drechsler2022MireMixed,drechsler2022HowSimulation,fremont2020FormalScenariobased,gupta2023XintheloopXil,gutierrezmoreno2024DecisionMaking,zofka2023PavingWays,haider2024MethodologyVerifying,hasirlioglu2018ModelbasedApproach,ngo2021MultilayeredApproach,chen2019NovelIntegrated,che2021TestMethod,solmaz2020VehicleintheloopMethodology,fu2024ImmersiveTraffic,hwang2024ImplementationSimulationbased,liu2024IntelligentCosimulation,samak2024MetaverseSafer,lee2024StudyFull,reckenzaun2024TransnationalTesting,drechsler2024VehicleDynamics,ganesan2025BridgingSimtoreal,verma2025EgoVehicle,jung2025RealvirtualSynchronized,zhao2025TachographbasedApproach,petrovic2025SuitabilitySmallsized,oh2024AdvilsImplementation,shen2023SimonwheelsPhysical,magosi2023NovelApproach,wang2025DigitalTwinbased,shoukat2024AutonomousDriving,toth2023DevelopmentFunctional,steimle2025HumanvehicleintheloopTest,li2024MultilevelMetrics,guo2021SeamlessModel,sigl2021ExperiencingDatadriven} \\
\hangindent=1em\hangafter=1 Compute hardware & \ResultStudies{25}{91} & \cite{barker2018LowcostHardwareintheloop,zeng2025ControllableRainy,neis2025DatadrivenLane,ameli2025DesignPreparation,li2025DgnrDensityguided,gao2026DrivenceRealistic,hong2025EffectiveLearning,baresi2025EfficientDomain,zhang2025EndtoendLearning,liu2026EnhancingAutonomous,nosheen2025EnhancingSynthetic,li2025GenerativeReconstruction,wang2025GenerativeTesting,zhang2025HighriskTrajectories,zhou2026HugsimRealtime,buzdugan2025HumancenteredSimulations,wei2025InteractiveCritical,gaddam2025LidarSensor,darocha2025MicroautowareAutoware,zuo2025RaladBridging,stangherlin2025XintheloopApproach,petrovic2025SuitabilitySmallsized,tian2026HgsimHighfidelity,steimle2025HumanvehicleintheloopTest,heikkinen2019HardwareintheloopPlatform} \\
\hangindent=1em\hangafter=1 ECU/\hspace{0pt}embedded controller & \ResultStudies{20}{91} & \cite{reway2022SimulationbasedTest,reway2020TestMethod,parra2020ValidationRealtime,shuguang2022VehicleintheloopIntelligent,bui2024ValidationAdaptive,dona2022VirtualTesting,zofka2023PavingWays,chen2019NovelIntegrated,oh2024ReliabilityEvaluation,solmaz2020VehicleintheloopMethodology,capdevila2024SmartsafeMobility,reckenzaun2024TransnationalTesting,drechsler2024VehicleDynamics,allamaa2025LearningbasedNmpc,peng2025MethodApplications,darocha2025MicroautowareAutoware,stangherlin2025XintheloopApproach,oh2024AdvilsImplementation,shen2023SimonwheelsPhysical,cardozo2025CarlaHardwareintheloop} \\
\hangindent=1em\hangafter=1 Radar & \ResultStudies{19}{91} & \cite{ngo2021DeepEvaluation,jayapalgowdu2022EvaluationScenariobased,bai2024ArbasedMeta,chen2020MixedTest,gupta2023XintheloopXil,schyr2022VehicleintheloopTestinga,haider2024MethodologyVerifying,hasirlioglu2018ModelbasedApproach,ngo2021MultilayeredApproach,chen2019NovelIntegrated,hwang2024ImplementationSimulationbased,samak2024MetaverseSafer,wachtel2024SimulationValidation,lee2024StudyFull,ganesan2025BridgingSimtoreal,haider2025ModelingSimulation,shen2023SimonwheelsPhysical,magosi2023NovelApproach,magosi2022EvaluationMethodology} \\
\hangindent=1em\hangafter=1 Sensor-stimulation equipment & \ResultStudies{11}{91} & \cite{jayapalgowdu2022EvaluationScenariobased,reway2022SimulationbasedTest,fremont2020FormalScenariobased,gupta2023XintheloopXil,barker2018LowcostHardwareintheloop,ulreich2024NovelTest,capdevila2024SmartsafeMobility,lee2024StudyFull,kuroda2025AnalysisSnow,shoukat2024AutonomousDriving,steimle2025HumanvehicleintheloopTest} \\
\hangindent=1em\hangafter=1 Roadside/\hspace{0pt}communication infrastructure & \ResultStudies{9}{91} & \cite{peintner2021MixedReality,shuguang2022VehicleintheloopIntelligent,szalay2019ProofConcept,che2021TestMethod,liu2024IntelligentCosimulation,verma2025EgoVehicle,li2025TowardIntelligent,wang2026UtdtBaseline,shoukat2024AutonomousDriving} \\
\hangindent=1em\hangafter=1 Small-scale vehicle/\hspace{0pt}robot & \ResultStudies{9}{91} & \cite{stocco2023ModelVs,yuan2024FromNaturalistic,ulhas2024GanbasedDomain,barker2018LowcostHardwareintheloop,hong2025EffectiveLearning,lambertenghi2025MultimodalityEvaluation,petrovic2025SuitabilitySmallsized,waheed2025QuantifyingSim2real,stocco2023MindGap} \\
\hangindent=1em\hangafter=1 VR/\hspace{0pt}MR or motion-capture equipment & \ResultStudies{9}{91} & \cite{zofka2018TrafficParticipants,drechsler2022MireMixed,gupta2023XintheloopXil,yeo2019MaximMixedreality,fu2024ImmersiveTraffic,samak2024MetaverseSafer,lambertenghi2025MultimodalityEvaluation,martinserrano2025BehaviouralGap,steimle2025HumanvehicleintheloopTest} \\
\hangindent=1em\hangafter=1 Human driving controls & \ResultStudies{8}{91} & \cite{fu2024ImmersiveTraffic,samak2024MetaverseSafer,lee2024StudyFull,reckenzaun2024TransnationalTesting,buzdugan2025HumancenteredSimulations,wang2025DigitalTwinbased,steimle2025HumanvehicleintheloopTest,guo2021SeamlessModel} \\
\hangindent=1em\hangafter=1 Testbench/\hspace{0pt}HIL equipment & \ResultStudies{6}{91} & \cite{dona2022VirtualTesting,lee2024StudyFull,drechsler2024VehicleDynamics,drechsler2025AcceleratingApproval,kim2025AutonomousVehicle,elgharbawy2019OntologybasedAdaptive} \\
\hangindent=1em\hangafter=1 Drone/\hspace{0pt}aerial platform & \ResultStudies{4}{91} & \cite{rezaldi20213dOrthomosaic,larter2022HierarchicalPedestrian,reckenzaun2024TransnationalTesting,li2025AutonomousDriving} \\
\hangindent=1em\hangafter=1 Atmospheric test chamber & \ResultStudies{3}{91} & \cite{hasirlioglu2018ModelbasedApproach,vargasrivero2022EffectSpray,segonne2023QualificationPavin} \\
\bottomrule
\end{tabular}
\end{table}


Hardware use differs across research targets. Studies with no reported hardware most frequently address camera perception and traffic/NPC realism, with 30 studies each. Full-size vehicles occur most often in ego vehicle dynamics/actuation (28) and SUT behaviour realism (19). These two targets also account for most uses of ECU/embedded controllers, with 10 and 9 studies, and nearly all uses of testbench/HIL equipment, with 5 and 1 of the 6 studies. LiDAR hardware occurs mainly in LiDAR perception (16) and SUT behaviour realism (11), followed by multi-sensor perception and ego vehicle dynamics/actuation (9 each). Camera hardware occurs across camera perception (13), multi-sensor perception (8), traffic/NPC realism (9), ego vehicle dynamics/actuation (12), and SUT behaviour realism (11). Localization/positioning hardware occurs most often in ego vehicle dynamics/actuation (13) and SUT behaviour realism (12). The complete set of of data is in our Appendix.

\begin{table}[tbp]
\caption{RQ\textsubscript{4.3} systems under test and specific technologies.}
\label{tab:rq4-3}
\centering
\footnotesize
\setlength{\ResultBarMax}{1.6cm}
\begin{tabular}{@{}>{\raggedright\arraybackslash}p{\dimexpr0.9\columnwidth-\dimexpr\ResultBarMax+2.3em\relax-2\tabcolsep\relax} p{\dimexpr\ResultBarMax+2.3em\relax}@{}}
\toprule
\textbf{System under test} & \textbf{Studies} \\
\midrule
\hangindent=1em\hangafter=1 \textbf{No SUT} & \ResultStudies{89}{89} \\ [0.5em]
\hangindent=1em\hangafter=1 \textbf{Camera perception} & \ResultStudies{40}{89} \\
\hangindent=2em\hangafter=1\hspace{1em}YOLO-family detector & \ResultStudies{15}{89} \\
\hangindent=2em\hangafter=1\hspace{1em}Object detector & \ResultStudies{12}{89} \\
\hangindent=2em\hangafter=1\hspace{1em}Segmentation model & \ResultStudies{7}{89} \\
\hangindent=2em\hangafter=1\hspace{1em}BEV-based detector/\hspace{0pt}feature extractor & \ResultStudies{4}{89} \\
\hangindent=2em\hangafter=1\hspace{1em}Lane detection & \ResultStudies{3}{89} \\
\hangindent=2em\hangafter=1\hspace{1em}VRU (pedestrian/\hspace{0pt}cyclist) detection \& driver-warning system & \ResultStudies{3}{89} \\
\hangindent=2em\hangafter=1\hspace{1em}Faster R-CNN detector & \ResultStudies{2}{89} \\
\hangindent=2em\hangafter=1\hspace{1em}3D human pose estimation & \ResultStudies{1}{89} \\
\hangindent=2em\hangafter=1\hspace{1em}ADAS smart front camera (sensor and embedded software) & \ResultStudies{1}{89} \\
\hangindent=2em\hangafter=1\hspace{1em}Depth estimation & \ResultStudies{1}{89} \\
\hangindent=2em\hangafter=1\hspace{1em}Object detection + tracking & \ResultStudies{1}{89} \\ [0.5em]
\hangindent=1em\hangafter=1 \textbf{Full ADS} & \ResultStudies{37}{89} \\
\hangindent=2em\hangafter=1\hspace{1em}Modular & \ResultStudies{29}{89} \\
\hangindent=2em\hangafter=1\hspace{1em}End-to-end & \ResultStudies{8}{89} \\
\hangindent=2em\hangafter=1\hspace{1em}Black-box AV (architecture undisclosed) & \ResultStudies{1}{89} \\
\hangindent=2em\hangafter=1\hspace{1em}Simplified driving function with PID vehicle controller (path following, TTC-based braking) & \ResultStudies{1}{89} \\ [0.5em]
\hangindent=1em\hangafter=1 \textbf{Ego-vehicle control/\hspace{0pt}actuation} & \ResultStudies{36}{89} \\
\hangindent=2em\hangafter=1\hspace{1em}Lane keeping/\hspace{0pt}changing control & \ResultStudies{11}{89} \\
\hangindent=2em\hangafter=1\hspace{1em}Adaptive cruise control & \ResultStudies{10}{89} \\
\hangindent=2em\hangafter=1\hspace{1em}Automatic emergency braking & \ResultStudies{6}{89} \\
\hangindent=2em\hangafter=1\hspace{1em}Steering control & \ResultStudies{4}{89} \\
\hangindent=2em\hangafter=1\hspace{1em}Vehicle control stack/\hspace{0pt}ECU & \ResultStudies{3}{89} \\
\hangindent=2em\hangafter=1\hspace{1em}Model predictive control & \ResultStudies{2}{89} \\
\hangindent=2em\hangafter=1\hspace{1em}Ground-robot dynamics control system & \ResultStudies{1}{89} \\
\hangindent=2em\hangafter=1\hspace{1em}Integrated trajectory-planner + vehicle controller & \ResultStudies{1}{89} \\
\hangindent=2em\hangafter=1\hspace{1em}PID trajectory-tracking controller & \ResultStudies{1}{89} \\
\hangindent=2em\hangafter=1\hspace{1em}Rule-based reactive control policy & \ResultStudies{1}{89} \\
\hangindent=2em\hangafter=1\hspace{1em}Waypoint-following control & \ResultStudies{1}{89} \\ [0.5em]
\hangindent=1em\hangafter=1 \textbf{Planning/\hspace{0pt}trajectory prediction} & \ResultStudies{18}{89} \\
\hangindent=2em\hangafter=1\hspace{1em}Path/\hspace{0pt}motion planning & \ResultStudies{14}{89} \\
\hangindent=2em\hangafter=1\hspace{1em}Decision-making/\hspace{0pt}planning policy & \ResultStudies{3}{89} \\
\hangindent=2em\hangafter=1\hspace{1em}Trajectory/\hspace{0pt}motion prediction & \ResultStudies{2}{89} \\
\hangindent=2em\hangafter=1\hspace{1em}Automated valet-parking planner & \ResultStudies{1}{89} \\ [0.5em]
\hangindent=1em\hangafter=1 \textbf{LiDAR perception} & \ResultStudies{17}{89} \\
\hangindent=2em\hangafter=1\hspace{1em}3D object detection & \ResultStudies{7}{89} \\
\hangindent=2em\hangafter=1\hspace{1em}PV-RCNN & \ResultStudies{3}{89} \\
\hangindent=2em\hangafter=1\hspace{1em}BEV feature extractor (LiDAR/\hspace{0pt}camera fusion) & \ResultStudies{2}{89} \\
\hangindent=2em\hangafter=1\hspace{1em}LiDAR object detection + tracking & \ResultStudies{2}{89} \\
\hangindent=2em\hangafter=1\hspace{1em}PointPillars & \ResultStudies{2}{89} \\
\hangindent=2em\hangafter=1\hspace{1em}Sensor simulation/\hspace{0pt}generative model & \ResultStudies{2}{89} \\
\hangindent=2em\hangafter=1\hspace{1em}LiDAR intensity simulation model & \ResultStudies{1}{89} \\
\hangindent=2em\hangafter=1\hspace{1em}Point-cloud segmentation & \ResultStudies{1}{89} \\
\hangindent=2em\hangafter=1\hspace{1em}PointNet(++) & \ResultStudies{1}{89} \\
\hangindent=2em\hangafter=1\hspace{1em}PolarNet & \ResultStudies{1}{89} \\
\hangindent=2em\hangafter=1\hspace{1em}RangeNet++ & \ResultStudies{1}{89} \\
\hangindent=2em\hangafter=1\hspace{1em}VoxelNet & \ResultStudies{1}{89} \\ [0.5em]
\hangindent=1em\hangafter=1 \textbf{Radar perception} & \ResultStudies{7}{89} \\ [0.5em]
\hangindent=1em\hangafter=1 \textbf{Localization} & \ResultStudies{5}{89} \\
\hangindent=2em\hangafter=1\hspace{1em}Modular-ADS localization component & \ResultStudies{2}{89} \\
\hangindent=2em\hangafter=1\hspace{1em}GNSS/\hspace{0pt}GPS-based localization & \ResultStudies{1}{89} \\
\hangindent=2em\hangafter=1\hspace{1em}NDT-based mapping/\hspace{0pt}localization & \ResultStudies{1}{89} \\
\hangindent=2em\hangafter=1\hspace{1em}SLAM-based localization & \ResultStudies{1}{89} \\ [0.5em]
\hangindent=1em\hangafter=1 \textbf{LLM/\hspace{0pt}foundation-model reasoning agent} & \ResultStudies{3}{89} \\ [0.5em]
\hangindent=1em\hangafter=1 \textbf{Multi-sensor fusion perception/\hspace{0pt}tracking} & \ResultStudies{2}{89} \\
\bottomrule
\end{tabular}
\end{table}

\subsubsection{RQ\textsubscript{4.3}: Systems Under Test}

The systems under test are grouped into ten categories. The categories and specific technologies are shown in~\autoref{tab:rq4-3}. 

\head{No SUT (89 studies)} These studies evaluate simulation or testbed fidelity without a vehicle or ADS-side driving function. This category also includes studies in which the testing infrastructure, environment-side behaviour model, or human-participant comparison is the object of study rather than a downstream vehicle-side function.

\head{Camera perception (40 studies)} These studies evaluate camera-based perception systems. The table distinguishes YOLO-family detectors (15), generic object detectors (12), segmentation models (7), BEV-based detectors or feature extractors (4), lane detection and VRU/pedestrian detection (3 each), and several techniques occurring in one study.

\head{Full ADS (37 studies)} These studies evaluate complete driving systems from sensing to actuation. The table distinguishes modular architectures (29), end-to-end architectures (8), and black-box or undisclosed architectures (1). Samak et al.~\cite{samak2024TowardsValidation} validate a modular autonomous driving stack across simulation and real scales using an integrated digital-twin framework.

\head{Ego vehicle control/actuation (36 studies)} These studies evaluate low-level control systems for steering, braking, or acceleration. The table distinguishes lane keeping/changing control (11), adaptive cruise control (10), automatic emergency braking (6), steering control (4), vehicle control stack/ECU (3), model predictive control (2), and several techniques occurring in one study.

\head{Planning/trajectory prediction (18 studies)} These studies evaluate decision-making, motion-planning, or trajectory-prediction systems. The table distinguishes path/motion planning (14), decision-making/planning policy (3), trajectory/motion prediction (2), and automated valet parking planning (1).

\head{LiDAR perception (17 studies)} These studies evaluate LiDAR-based perception systems. The table includes 3D object detection (7), PV-RCNN (3), LiDAR object detection and tracking (2), PointPillars (2), sensor simulation or generative models evaluated as systems under test in their own right (2), and several architectures occurring in one study.

\head{Radar perception (7 studies)} These studies evaluate radar-based perception systems.

\head{Localization (5 studies)} These studies evaluate localization and pose-estimation systems. The table distinguishes modular-ADS localization components (2), GNSS/GPS-based localization (1), NDT-based mapping/localization (1), and SLAM-based localization (1).

\head{LLM/foundation model reasoning agent (3 studies)} These studies evaluate driving systems based on LLM or foundation-model reasoning.

\head{Multi-sensor perception and tracking (2 studies)} These studies evaluate perception systems combining multiple sensing modalities.

\begin{table}[tbp]
\caption{RQ\textsubscript{4.4} test-case origin and supporting primary studies.}
\label{tab:rq4-4}
\centering

\footnotesize
\begin{tabular}{@{}p{0.20\textwidth} p{0.20\textwidth} p{0.55\textwidth}@{}}
\toprule
\textbf{Test-case origin} & \textbf{Studies} & \textbf{Primary studies} \\
\midrule
\hangindent=1em\hangafter=1 Simulation-generated & \ResultStudies{157}{157} & \cite{rosenberger2019BenchmarkingFunctional,lu2022DesignVehicle,nalic2019DevelopmentCosimulation,saleh2019DomainAdaptation,pahk2023EffectsSim2real,oladimeji2024EnhancedApproaches,jayapalgowdu2022EvaluationScenariobased,pahk2024LaneSegmentation,peintner2021MixedReality,reway2022SimulationbasedTest,reway2020TestMethod,samak2024TowardsValidation,manivasagam2023TowardsZero,zofka2018TrafficParticipants,parra2020ValidationRealtime,amini2020LearningRobust,bashetty2020DeepcrashtestTurning,bergamini2021SimnetLearning,cao2024AresTextdriven,chen2020MixedTest,drechsler2022HowSimulation,duminil2024AssessingFidelity,fremont2020FormalScenariobased,gu2023CopulabasedInteractive,gupta2023XintheloopXil,gutierrezmoreno2024DecisionMaking,hoppe2022AgentbasedAutonomous,hsiang2022DevelopmentSimulationbased,schyr2022VehicleintheloopTestinga,tettamanti2018VehicleintheloopTest,yang2021OnlineAdaptation,vonbernuth2018RenderingPhysically,zofka2018SleepwalkerFramework,szalai2020MixedReality,wagner2019TowardsCrossverification,thieling2019NeuralNetworks,szalay2019ProofConcept,yeo2019MaximMixedreality,tim2020SimulationReal,xinxin2020CsgCritical,talwar2020EvaluatingValidity,xu2021ReliabilityGan,soni2021DeepLearning,zhang2022SelfdrivingSimulation,wallace2022ValidatingSimulation,shi2022IntegratedTraffic,zhu2022MillimeterwaveRadar,zhang2022TrajgenGenerating,zofka2023PavingWays,zhang2023BuildingCritical,yuan2024FromNaturalistic,zhou2024EvaluatingAutonomous,ulhas2024GanbasedDomain,sun2024DrivescenegenGenerating,huang2024HierarchicalLidar,larter2022HierarchicalPedestrian,barker2018LowcostHardwareintheloop,haider2024MethodologyVerifying,hasirlioglu2018ModelbasedApproach,ngo2021MultilayeredApproach,chen2019NovelIntegrated,oh2024ReliabilityEvaluation,solmaz2020VehicleintheloopMethodology,lan2024AssessingReliability,amini2024BridgingGap,fei2024CorrelationSoftwareintheloop,long2024GeneratingRealistic,fu2024ImmersiveTraffic,hwang2024ImplementationSimulationbased,liu2024IntelligentCosimulation,samak2024MetaverseSafer,suo2023MixsimHierarchical,ulreich2024NovelTest,wachtel2024SimulationValidation,capdevila2024SmartsafeMobility,lee2024StudyFull,reckenzaun2024TransnationalTesting,yang2023UnisimNeural,chen2025SnerfAutonomous,guo2025ScenedmConsistent,fu2025ControllablePedestrian,li2025DatadrivenHeadlight,neis2025DatadrivenLane,zhou2026DiffusionModels,yan2025DistributionallyConsistent,wang2025DrivegenCoevaluating,hong2025EffectiveLearning,simakov2025EffectsSimulator,zhang2025EndtoendLearning,liu2026EnhancingAutonomous,pasios2025EnhancingPhotorealism,nosheen2025EnhancingSynthetic,li2025GenerativeReconstruction,wang2025GenerativeTesting,zhang2025HighriskTrajectories,zhou2026HugsimRealtime,buzdugan2025HumancenteredSimulations,wei2025InteractiveCritical,gaddam2025LidarSensor,peng2025MethodApplications,wang2025RadeLearning,zuo2025RaladBridging,jung2025RealvirtualSynchronized,gao2025DatacentricImage,elmquist2025MethodologyQuantify,lambertenghi2025MultimodalityEvaluation,zhang2025PixellevelImage,wu2025ToolBenchmarking,haghighi2026UnifiedGenerative,abbasi2025Acr2xodrFrom,jing2025EfficientHighrisk,kolb2025AutomaticallyImproving,kim2025AutonomousVehicle,eberhardt2025BeyondSmudge,harhammer2025SimtorealDomain,samak2026Sim2realDiffusion,hess2025SplatadRealtime,sun2025TestingAutonomous,stangherlin2025XintheloopApproach,bi2025TowardFullstack,anand2026TowardClosing,yu2026TrajectorybasedVideo,huang2026VersatileBehavior,haider2025ModelingSimulation,schabus2025ObjectbasedLidar,waheed2025QuantifyingSim2real,shen2023SimonwheelsPhysical,huch2023QuantifyingLidar,cao2024ReinforcementLearning,gadipudi2022SyntheticReal,vargasrivero2022EffectSpray,amini2022Vista20,duminil2024ComprehensiveExploration,wang2025DigitalTwinbased,khaire2025AugmentedRealityenhanced,shoukat2024AutonomousDriving,balm2025BenchmarkingDomain,cardozo2025CarlaHardwareintheloop,arango2026ClosingSim2real,liang2025DttfsimDigital,drechsler2022DynamicVehicleintheloop,magosi2022EvaluationMethodology,tian2026HgsimHighfidelity,hofrichter2025IntroducingDevelopment,coppola2025VirtualTesting,elgharbawy2019OntologybasedAdaptive,segonne2023QualificationPavin,daza2023SimtorealTransfer,jekl2025SpeedPlanner,song2026VisionlanguageModelbased,wang2022Cadsim,yang2020RecoveringSimulating,leung2025RoadCode,guo2021SeamlessModel,deter2021SimulatingAutonomous,wang2023WhenSimulator,heikkinen2019HardwareintheloopPlatform} \\
\hangindent=1em\hangafter=1 Real-world executed & \ResultStudies{87}{157} & \cite{rosenberger2018AnalysisReal,ngo2021DeepEvaluation,kim2024DesigningNaturalistic,lichtle2024FromSim,kettelgerdes2023InfieldMeasurement,reway2020TestMethod,wachtel2021ValidationRadar,parra2020ValidationRealtime,krecht2023ValidationProcess,shuguang2022VehicleintheloopIntelligent,amini2020LearningRobust,bateman2024ExploringReal,brogle2019HardwareintheloopAutonomous,chen2022AnalysisRealtime,chen2020MixedTest,dona2022VirtualTesting,drechsler2022MireMixed,drechsler2022HowSimulation,fremont2020FormalScenariobased,gao2020HardwareintheloopSimulation,gupta2023XintheloopXil,gutierrezmoreno2024DecisionMaking,stocco2023ModelVs,schyr2022VehicleintheloopTestinga,tettamanti2018VehicleintheloopTest,zofka2018SleepwalkerFramework,szalai2020MixedReality,wagner2019TowardsCrossverification,yeo2019MaximMixedreality,xu2021ReliabilityGan,zhang2022TrajgenGenerating,zofka2023PavingWays,yuan2024FromNaturalistic,stadler2022CredibilityAssessment,durukal2022DigitalTwin,queiroz2024DrivervehicleModel,jiang2021HighaccuracyFramework,barker2018LowcostHardwareintheloop,haider2024MethodologyVerifying,ngo2021MultilayeredApproach,oh2024ReliabilityEvaluation,che2021TestMethod,solmaz2020VehicleintheloopMethodology,fei2024CorrelationSoftwareintheloop,hwang2024ImplementationSimulationbased,liu2024IntelligentCosimulation,samak2024MetaverseSafer,suo2023MixsimHierarchical,ulreich2024NovelTest,capdevila2024SmartsafeMobility,reckenzaun2024TransnationalTesting,drechsler2024VehicleDynamics,ganesan2025BridgingSimtoreal,huang2025CadreControllable,hong2025EffectiveLearning,verma2025EgoVehicle,zuo2025RaladBridging,jung2025RealvirtualSynchronized,lambertenghi2025MultimodalityEvaluation,zhao2025TachographbasedApproach,abbasi2025Acr2xodrFrom,bi2025TowardFullstack,ulreich2025ValidatingCamera,otto2025ValidationAibased,lee2025ModelcentricQuantitative,haider2025ModelingSimulation,petrovic2025SuitabilitySmallsized,waheed2025QuantifyingSim2real,oh2024AdvilsImplementation,stocco2023MindGap,manivasagam2020LidarsimRealistic,shen2023SimonwheelsPhysical,vargasrivero2022EffectSpray,magosi2023NovelApproach,hu2025SimulationbasedTesting,shoukat2024AutonomousDriving,arango2026ClosingSim2real,liang2025DttfsimDigital,toth2023DevelopmentFunctional,magosi2022EvaluationMethodology,wei2025GeneratingRisky,tian2026HgsimHighfidelity,hofrichter2025IntroducingDevelopment,li2024MultilevelMetrics,park2020VehicleintheloopGlobal,reisgys2021ArgumentationADAS,sigl2021ExperiencingDatadriven} \\
\hangindent=1em\hangafter=1 Real-world reconstruction & \ResultStudies{73}{157} & \cite{liu2021ApplicationDigital,langer2021CalibrationAssessment,ngo2021DeepEvaluation,lu2022DesignVehicle,pahk2023EffectsSim2real,oladimeji2024EnhancedApproaches,jayapalgowdu2022EvaluationScenariobased,peintner2021MixedReality,reway2020TestMethod,samak2024TowardsValidation,manivasagam2023TowardsZero,zofka2018TrafficParticipants,wachtel2021ValidationRadar,bashetty2020DeepcrashtestTurning,cao2024AresTextdriven,dona2022VirtualTesting,fremont2020FormalScenariobased,gambi2019GeneratingEffective,briand-offline-emse,stocco2023ModelVs,schyr2022VehicleintheloopTestinga,xinxin2020CsgCritical,tenbrock2021ConscendDataset,zhang2022SelfdrivingSimulation,stadler2022TowardsMultivariate,zhang2023BuildingCritical,yuan2024FromNaturalistic,zhou2024EvaluatingAutonomous,rezaldi20213dOrthomosaic,queiroz2024DrivervehicleModel,larter2022HierarchicalPedestrian,haider2024MethodologyVerifying,chen2019NovelIntegrated,che2021TestMethod,solmaz2020VehicleintheloopMethodology,hwang2024ImplementationSimulationbased,yang2023UnisimNeural,chen2025SnerfAutonomous,steinhauser2025DataQuality,liu2026EnhancingAutonomous,li2025GenerativeReconstruction,zhang2025HighriskTrajectories,zhou2026HugsimRealtime,wei2025InteractiveCritical,allamaa2025LearningbasedNmpc,gaddam2025LidarSensor,peng2025MethodApplications,zuo2025RaladBridging,saad2025DigitalTwin,drechsler2025AcceleratingApproval,li2025AutonomousDriving,stangherlin2025XintheloopApproach,li2025TowardIntelligent,wang2026UtdtBaseline,ulreich2025ValidatingCamera,otto2025ValidationAibased,schulz2025VirtualKitti,lee2025ModelcentricQuantitative,oh2024AdvilsImplementation,stocco2023MindGap,lambertenghi2024AssessingQuality,vargasrivero2022EffectSpray,amini2022Vista20,wang2025DigitalTwinbased,wang2022AutomaticTraffic,martinserrano2025BehaviouralGap,magosi2022EvaluationMethodology,argui2025ExploringAugmentation,wang2022Cadsim,yang2020RecoveringSimulating,leung2025RoadCode,sigl2021ExperiencingDatadriven,deter2021SimulatingAutonomous} \\
\hangindent=1em\hangafter=1 Recorded/\hspace{0pt}replayed & \ResultStudies{63}{157} & \cite{kim2024DesigningNaturalistic,saleh2019DomainAdaptation,lichtle2024FromSim,kim2022QueryingLabelled,krecht2023ValidationProcess,amini2020LearningRobust,bergamini2021SimnetLearning,dona2022VirtualTesting,briand-offline-emse,hsiang2022DevelopmentSimulationbased,vonbernuth2018RenderingPhysically,tim2020SimulationReal,talwar2020EvaluatingValidity,soni2021DeepLearning,stadler2022TowardsMultivariate,zhang2022TrajgenGenerating,xu2023BitsBilevel,yuan2024FromNaturalistic,zhou2024EvaluatingAutonomous,ulhas2024GanbasedDomain,sun2024DrivescenegenGenerating,wu2024RealityBites,queiroz2024DrivervehicleModel,larter2022HierarchicalPedestrian,jiang2021HighaccuracyFramework,haider2024MethodologyVerifying,ngo2021MultilayeredApproach,chen2019NovelIntegrated,oh2024ReliabilityEvaluation,solmaz2020VehicleintheloopMethodology,lan2024AssessingReliability,fu2024ImmersiveTraffic,suo2023MixsimHierarchical,wachtel2024SimulationValidation,yang2023UnisimNeural,drechsler2024VehicleDynamics,guo2025ScenedmConsistent,ganesan2025BridgingSimtoreal,steinhauser2025DataQuality,ameli2025DesignPreparation,li2025DgnrDensityguided,verma2025EgoVehicle,naing2025FinegrainedTrajectory,wu2025FusingTemporal,zhang2025HighriskTrajectories,zuo2025RaladBridging,saad2025DigitalTwin,zhang2025PixellevelImage,kolb2025AutomaticallyImproving,eberhardt2025BeyondSmudge,hess2025SplatadRealtime,huang2026VersatileBehavior,li2025XldCrosslane,gadipudi2022SyntheticReal,amini2022Vista20,wang2025DigitalTwinbased,balm2025BenchmarkingDomain,tian2026HgsimHighfidelity,gao2025MsietransformerNovel,elgharbawy2019OntologybasedAdaptive,dauner2024Navsim,yang2020RecoveringSimulating,guo2021SeamlessModel} \\
\hangindent=1em\hangafter=1 Manual/\hspace{0pt}predefined & \ResultStudies{37}{157} & \cite{rosenberger2018AnalysisReal,rosenberger2019BenchmarkingFunctional,bateman2024ExploringReal,brogle2019HardwareintheloopAutonomous,bui2024ValidationAdaptive,chen2022AnalysisRealtime,dona2022VirtualTesting,drechsler2022MireMixed,drechsler2022HowSimulation,gao2020HardwareintheloopSimulation,gupta2023XintheloopXil,gutierrezmoreno2024DecisionMaking,hoppe2022AgentbasedAutonomous,schyr2022VehicleintheloopTestinga,tettamanti2018VehicleintheloopTest,stadler2022CredibilityAssessment,jiang2021HighaccuracyFramework,barker2018LowcostHardwareintheloop,haider2024MethodologyVerifying,hwang2024ImplementationSimulationbased,liu2024IntelligentCosimulation,samak2024MetaverseSafer,ulreich2024NovelTest,lee2024StudyFull,reckenzaun2024TransnationalTesting,yang2023UnisimNeural,drechsler2024VehicleDynamics,hong2025EffectiveLearning,buzdugan2025HumancenteredSimulations,darocha2025MicroautowareAutoware,lambertenghi2025MultimodalityEvaluation,drechsler2025AcceleratingApproval,kuroda2025AnalysisSnow,otto2025ValidationAibased,lee2025ModelcentricQuantitative,oh2024AdvilsImplementation,deter2021SimulatingAutonomous} \\
\hangindent=1em\hangafter=1 Hybrid/\hspace{0pt}mixed & \ResultStudies{16}{157} & \cite{argui2023BuildingVisionbased,bai2024ArbasedMeta,drechsler2022MireMixed,drechsler2022HowSimulation,gao2020HardwareintheloopSimulation,szalay2019ProofConcept,shi2022IntegratedTraffic,haider2024MethodologyVerifying,fu2024ImmersiveTraffic,ulreich2024NovelTest,zeng2025ControllableRainy,gao2026DrivenceRealistic,baresi2025EfficientDomain,arango2026ClosingSim2real,steimle2025HumanvehicleintheloopTest,coppola2025VirtualTesting} \\
\bottomrule
\end{tabular}
\end{table}

\subsubsection{RQ\textsubscript{4.4}: Test Case Generation}

Test cases are obtained or generated in several ways. The test-case origins and the corresponding primary studies are shown in~\autoref{tab:rq4-4}.

Simulation-generated test cases are the most common, appearing in 157 studies. They are followed by real-world execution (87 studies), real-world reconstruction (73), and recorded or replayed test cases (63). Manual or predefined test cases appear in 37 studies and hybrid or mixed generation in 16.

Nali\'c et al.~\cite{nalic2019DevelopmentCosimulation} generate test scenarios within a co-simulation framework. Kettelgerdes et al.~\cite{kettelgerdes2023InfieldMeasurement} take in-field measurements on a real test track to study precipitation effects on solid-state LiDAR sensors. Liu et al.~\cite{liu2021ApplicationDigital} construct a digital-twin reconstruction of real-world driving conditions. Kim et al.~\cite{kim2024DesigningNaturalistic} extract and replay naturalistic driving situations recorded from real traffic. Brogle et al.~\cite{brogle2019HardwareintheloopAutonomous} run predefined test cases in a hardware-in-the-loop simulation without real-time constraints. Argui et al.~\cite{argui2023BuildingVisionbased} combine real and simulated navigation elements in a vision-based mixed-reality framework.

\begin{table}[t]
\footnotesize
\centering
\caption{RQ\textsubscript{4.5} testing artefacts and citations.} \label{tab:rq4-5-cites}
\begin{tabular}{@{}p{0.12\textwidth} p{0.15\textwidth} p{0.68\textwidth}@{}}
\toprule
\textbf{Category} & \textbf{Subcategory} & \textbf{Citations} \\
\midrule
Testing artefacts & None/unspecified (188) & \cite{rosenberger2018AnalysisReal,liu2021ApplicationDigital,rosenberger2019BenchmarkingFunctional,langer2021CalibrationAssessment,ngo2021DeepEvaluation,lu2022DesignVehicle,kim2024DesigningNaturalistic,nalic2019DevelopmentCosimulation,saleh2019DomainAdaptation,pahk2023EffectsSim2real,oladimeji2024EnhancedApproaches,jayapalgowdu2022EvaluationScenariobased,lichtle2024FromSim,kettelgerdes2023InfieldMeasurement,pahk2024LaneSegmentation,peintner2021MixedReality,kim2022QueryingLabelled,reway2022SimulationbasedTest,reway2020TestMethod,manivasagam2023TowardsZero,zofka2018TrafficParticipants,wachtel2021ValidationRadar,parra2020ValidationRealtime,krecht2023ValidationProcess,shuguang2022VehicleintheloopIntelligent,bai2024ArbasedMeta,bashetty2020DeepcrashtestTurning,bateman2024ExploringReal,brogle2019HardwareintheloopAutonomous,bui2024ValidationAdaptive,cao2024AresTextdriven,chen2022AnalysisRealtime,chen2020MixedTest,dona2022VirtualTesting,drechsler2022MireMixed,drechsler2022HowSimulation,duminil2024AssessingFidelity,fremont2020FormalScenariobased,gao2020HardwareintheloopSimulation,gu2023CopulabasedInteractive,gupta2023XintheloopXil,gutierrezmoreno2024DecisionMaking,hoppe2022AgentbasedAutonomous,hsiang2022DevelopmentSimulationbased,schyr2022VehicleintheloopTestinga,tettamanti2018VehicleintheloopTest,yang2021OnlineAdaptation,vonbernuth2018RenderingPhysically,zofka2018SleepwalkerFramework,szalai2020MixedReality,wagner2019TowardsCrossverification,thieling2019NeuralNetworks,szalay2019ProofConcept,yeo2019MaximMixedreality,tim2020SimulationReal,xinxin2020CsgCritical,talwar2020EvaluatingValidity,xu2021ReliabilityGan,soni2021DeepLearning,zhang2022SelfdrivingSimulation,wallace2022ValidatingSimulation,stadler2022TowardsMultivariate,shi2022IntegratedTraffic,zhu2022MillimeterwaveRadar,zhang2022TrajgenGenerating,zofka2023PavingWays,yuan2024FromNaturalistic,zhou2024EvaluatingAutonomous,ulhas2024GanbasedDomain,sun2024DrivescenegenGenerating,rezaldi20213dOrthomosaic,stadler2022CredibilityAssessment,durukal2022DigitalTwin,queiroz2024DrivervehicleModel,huang2024HierarchicalLidar,larter2022HierarchicalPedestrian,jiang2021HighaccuracyFramework,barker2018LowcostHardwareintheloop,haider2024MethodologyVerifying,hasirlioglu2018ModelbasedApproach,chen2019NovelIntegrated,oh2024ReliabilityEvaluation,che2021TestMethod,solmaz2020VehicleintheloopMethodology,lan2024AssessingReliability,long2024GeneratingRealistic,fu2024ImmersiveTraffic,hwang2024ImplementationSimulationbased,liu2024IntelligentCosimulation,ulreich2024NovelTest,wachtel2024SimulationValidation,capdevila2024SmartsafeMobility,lee2024StudyFull,reckenzaun2024TransnationalTesting,yang2023UnisimNeural,drechsler2024VehicleDynamics,chen2025SnerfAutonomous,guo2025ScenedmConsistent,huang2025CadreControllable,fu2025ControllablePedestrian,zeng2025ControllableRainy,steinhauser2025DataQuality,ameli2025DesignPreparation,li2025DgnrDensityguided,zhou2026DiffusionModels,yan2025DistributionallyConsistent,wang2025DrivegenCoevaluating,hong2025EffectiveLearning,simakov2025EffectsSimulator,verma2025EgoVehicle,zhang2025EndtoendLearning,liu2026EnhancingAutonomous,wu2025FusingTemporal,li2025GenerativeReconstruction,wang2025GenerativeTesting,buzdugan2025HumancenteredSimulations,gaddam2025LidarSensor,peng2025MethodApplications,wang2025RadeLearning,jung2025RealvirtualSynchronized,gao2025DatacentricImage,saad2025DigitalTwin,elmquist2025MethodologyQuantify,zhang2025PixellevelImage,zhao2025TachographbasedApproach,drechsler2025AcceleratingApproval,abbasi2025Acr2xodrFrom,jing2025EfficientHighrisk,kolb2025AutomaticallyImproving,li2025AutonomousDriving,kim2025AutonomousVehicle,eberhardt2025BeyondSmudge,harhammer2025SimtorealDomain,samak2026Sim2realDiffusion,sun2025TestingAutonomous,stangherlin2025XintheloopApproach,anand2026TowardClosing,li2025TowardIntelligent,yu2026TrajectorybasedVideo,ulreich2025ValidatingCamera,otto2025ValidationAibased,schulz2025VirtualKitti,lee2025ModelcentricQuantitative,haider2025ModelingSimulation,schabus2025ObjectbasedLidar,petrovic2025SuitabilitySmallsized,waheed2025QuantifyingSim2real,oh2024AdvilsImplementation,manivasagam2020LidarsimRealistic,shen2023SimonwheelsPhysical,gadipudi2022SyntheticReal,vargasrivero2022EffectSpray,duminil2024ComprehensiveExploration,magosi2023NovelApproach,wang2025DigitalTwinbased,hu2025SimulationbasedTesting,khaire2025AugmentedRealityenhanced,wang2022AutomaticTraffic,shoukat2024AutonomousDriving,martinserrano2025BehaviouralGap,balm2025BenchmarkingDomain,cardozo2025CarlaHardwareintheloop,liang2025DttfsimDigital,toth2023DevelopmentFunctional,drechsler2022DynamicVehicleintheloop,magosi2022EvaluationMethodology,wei2025GeneratingRisky,tian2026HgsimHighfidelity,steimle2025HumanvehicleintheloopTest,hofrichter2025IntroducingDevelopment,gao2025MsietransformerNovel,li2024MultilevelMetrics,coppola2025VirtualTesting,elgharbawy2019OntologybasedAdaptive,segonne2023QualificationPavin,daza2023SimtorealTransfer,jekl2025SpeedPlanner,park2020VehicleintheloopGlobal,song2026VisionlanguageModelbased,wang2022Cadsim,yang2020RecoveringSimulating,leung2025RoadCode,reisgys2021ArgumentationADAS,guo2021SeamlessModel,sigl2021ExperiencingDatadriven,deter2021SimulatingAutonomous,wang2023WhenSimulator,heikkinen2019HardwareintheloopPlatform} \\
 & Code (26) & \cite{amini2020LearningRobust,argui2023BuildingVisionbased,bergamini2021SimnetLearning,gambi2019GeneratingEffective,stocco2023ModelVs,zhang2023BuildingCritical,wu2024RealityBites,samak2024MetaverseSafer,ganesan2025BridgingSimtoreal,li2025DatadrivenHeadlight,pasios2025EnhancingPhotorealism,nosheen2025EnhancingSynthetic,naing2025FinegrainedTrajectory,zhang2025HighriskTrajectories,zhou2026HugsimRealtime,wei2025InteractiveCritical,darocha2025MicroautowareAutoware,zuo2025RaladBridging,wu2025ToolBenchmarking,haghighi2026UnifiedGenerative,hess2025SplatadRealtime,wang2026UtdtBaseline,huang2026VersatileBehavior,stocco2023MindGap,lambertenghi2024AssessingQuality,dauner2024Navsim} \\
 & Other artefact (14) & \cite{samak2024TowardsValidation,xu2023BitsBilevel,ngo2021MultilayeredApproach,amini2024BridgingGap,fei2024CorrelationSoftwareintheloop,suo2023MixsimHierarchical,neis2025DatadrivenLane,gao2026DrivenceRealistic,allamaa2025LearningbasedNmpc,lambertenghi2025MultimodalityEvaluation,kuroda2025AnalysisSnow,bi2025TowardFullstack,amini2022Vista20,arango2026ClosingSim2real} \\
 & Dataset (5) & \cite{briand-offline-emse,tenbrock2021ConscendDataset,huch2023QuantifyingLidar,cao2024ReinforcementLearning,argui2025ExploringAugmentation} \\
 & Code and dataset (2) & \cite{baresi2025EfficientDomain,li2025XldCrosslane} \\ \bottomrule
\end{tabular}
\end{table}

\subsubsection{RQ\textsubscript{4.5}: Testing Artefacts}

Public testing artefacts reported by the studies are listed with citations in~\autoref{tab:rq4-5-cites}.
Most studies report no public artefact (188 studies, 80.0\%). Source code is reported by 26 studies (11.1\%), another type of artefact by 14 (6.0\%), a public dataset by 5 (2.1\%), and both code and a dataset by 2 (0.9\%). The 14 studies with another type of artefact are heterogeneous: they report open-source tools, platforms, or replication packages recorded without a repository link (6), a tool released together with the data it generates (1), demonstration videos or project web pages (3), and scenario or measurement data that is described as available without a confirmed release (4). One study classified under Source code~\cite{bergamini2021SimnetLearning} reports a repository that no longer resolves. We retain the Source code classification because the paper reports the repository, although the artefact is no longer retrievable.

\section{Discussion}
\label{discussion}

The central implication of this mapping is that the importance of the reality gap depends on how simulation results are used. Failure discovery, comparison of ADS versions, pass/fail assessment, and estimation of operational risk require different forms of evidence and may be affected by different discrepancies.
A discrepancy may affect ADS behaviour and change the testing conclusion. Future studies should therefore specify the decision being supported, the ADS and scenario conditions considered, the reference used for comparison, and the criteria used to evaluate correspondence. This would help identify which discrepancies matter, what level of fidelity is sufficient, and which additional measurements can reduce uncertainty most effectively. The following research directions address these questions based on the findings of our mapping.

\subsection{Gap Mitigation and Evaluation Adequacy}
\label{discussion-credibility}

Of the 235 studies, 179 address mitigation, alone or alongside measurement. However, improving over an unmitigated baseline does not mean that the remaining discrepancy is acceptable. Reducing one discrepancy also leaves others unresolved: a more accurate sensor model can coexist with unrealistic actor responses that affect ADS behaviour. The value of a mitigation for testing therefore depends on which aspect of the reality gap it addresses and how it interacts with the remaining discrepancies.

Likewise, different evaluation setups provide different types of evidence. Recorded observations allow sensor-level comparisons, but a fixed recording cannot capture how ADS actions affect subsequent observations. Physical execution provides evidence only for the aspects and conditions that were actually compared. Using physical equipment or describing a system as a digital twin does not, by itself, establish that the evaluation demonstrates the claimed correspondence.

Studies of simulation fidelity investigate which aspects of a model require greater detail~\cite{11623850}. For example, Sagmeister et al.~\cite{agendaFidelity} compare vehicle models against real-world data and examine how model simplifications affect closed-loop motion-control evaluation. Such studies help identify fidelity requirements for specific applications, but do not establish that greater fidelity always leads to more reliable testing conclusions.
Multi-simulator approaches investigate whether combining models improves testing. Digital siblings~\cite{biagiola2023better} combine predictions and evaluate them against a digital twin of a scaled vehicle, while MultiSim~\cite{agendaMultiSim} searches for failures across simulators. However, the reported benefits do not establish that agreement between simulators implies agreement with physical outcomes. Further research should investigate when using multiple simulators improves the prediction of real-world test outcomes and how dependencies between simulators affect the reliability of these predictions.

Researchers should evaluate improvements in simulation fidelity and combinations of simulators against an independent reference and a clearly defined testing objective. These evaluations should distinguish between finding failures, obtaining consistent results across simulators, and agreement with physical outcomes. This research is relevant to ISO21448~\cite{ISO21448} (SOTIF), which addresses safety assessment, and the UNECE NATM guidelines~\cite{agendaUneceNATM}, which address the evaluation of virtual toolchains. Both provide a context for assessing simulation evidence in relation to the claims it supports. Future work should translate these assessment objectives into measurable criteria for specific claims, defining which discrepancies are acceptable and how these limits can be established experimentally. Tool developers can support this process by documenting model assumptions, the conditions under which models are applicable, and dependencies shared between simulators.

A concrete research objective is to determine how much model fidelity is needed for simulation and physical testing to support the same engineering decision, such as selecting between ADS versions or applying a pass/fail criterion, using independent physical measurements within a defined class of scenarios.

\subsection{From Discrepancy Measurements to Testing Consequences}
\label{testing-consequences}

Our metric analysis identifies 61 studies using at least two of the four metric families, nine using at least three, and three using all four. However, using multiple metric families does not necessarily mean that a study investigates whether discrepancies in sensor observations affect ADS behaviour and, in turn, change test verdicts. Our extraction does not allow us to determine how many studies establish these relationships.

These relationships form a causal chain~\cite{10.1145/3803437.3805585}: a discrepancy may affect ADS behaviour, which may in turn change the test verdict. Each causal link requires empirical evidence. An image-quality improvement can leave behaviour unchanged, while a small timing discrepancy can change a verdict near a braking threshold. Evaluation should therefore establish whether, and under which conditions, a discrepancy propagates to the testing conclusion~\cite{10638595}. This resembles fault propagation analysis, which investigates whether a fault leads to a detectable failure~\cite{44380}. Controlled interventions and sensitivity analysis could help identify consequential discrepancy types and magnitudes.

Future studies should investigate this causal chain in a single experimental setting, with a clearly defined scenario class, ADS under test, and acceptance criterion. Where feasible, simulated and physical executions should use the same ADS, corresponding scenarios, and consistent evaluation criteria. Comparisons should account for differences caused by closed-loop divergence, variability between executions, time alignment, and oracle implementation. Reporting how frequently discrepancies change test verdicts would help quantify their impact under the evaluated conditions, provided that the procedure used to select discrepancies is also reported.

Researchers should distinguish between cases where an ADS passes in simulation but fails in reality and cases where it fails in simulation but passes in reality. The first case indicates potential false confidence, while the second may reflect simulator artefacts, execution variability, or differences between the simulated and physical experiments. These errors have different consequences: the former can leave physical failures undetected, while the latter can lead to unnecessary rejection, investigation, or retesting. Evaluations should report the frequency of both types of disagreement and specify which error type a method aims to minimise. When false confidence has more serious consequences, stricter acceptance criteria may be appropriate, but the resulting increase in false alarms should also be reported.

This motivates the creation of new benchmarks of paired simulated and physical executions that report the full pass/fail contingency table, define conditional disagreement rates, and distinguish routine scenarios from cases near an acceptance boundary~\cite{2026-Naziri-ICSE}.

\subsection{Uncertainty About the Remaining Gap}
\label{discussion-uncertainty}

The reality gap does not need to be completely eliminated for simulation results to be useful, but its effect on testing conclusions must be considered. Uncertainty estimates should distinguish between variability within the simulation and uncertainty about how well the simulation corresponds to reality. Repeating simulations can reduce sampling uncertainty but does not eliminate systematic modelling errors.
Werling et al.\cite{agendaWerlingSafety} combine uncertainty quantification, stochastic simulation, and sensitivity analysis to assess system-level risk. Sim2Val\cite{agendaSim2Val} uses paired measurements and control variates to improve estimates of real-world metrics. These approaches provide ways to use imperfect simulation models under explicit assumptions, but do not provide a general confidence score for simulation results.

Researchers should first clarify what an uncertainty estimate represents. Confidence in an estimate of physical performance, the probability that simulation and physical testing produce the same verdict, and the maturity of an evaluation process are different concepts. Predictions of agreement between simulated and physical verdicts should be evaluated using independent paired executions, including cases near acceptance thresholds. Uncertainty intervals should be evaluated against physical measurements to determine whether they achieve the expected coverage and how informative they are. Their applicability under changed conditions should be assessed separately.
Conformal prediction is one possible approach for constructing prediction sets or intervals using paired simulated and physical executions~\cite{angelopoulos2023conformal}. Its standard finite-sample coverage guarantees require exchangeability and generally apply across test cases rather than to each scenario individually. Future research should investigate appropriate calibration procedures and how dependencies between test cases and changes in data distributions affect these guarantees.

Assessors should also consider how uncertainty about the remaining reality gap affects the evidence supporting a claim. Building on the physical qualification of simulation evidence in the UNECE NATM guidelines and Euro NCAP protocols~\cite{agendaUneceNATM,agendaEuroNCAP2026}, researchers and assessment bodies should develop reporting conventions that connect uncertainty estimates to acceptance criteria, applicability conditions, and unresolved verdicts. This would make probabilistic evidence easier to interpret and compare without treating a confidence value as a substitute for its supporting assumptions.

Tool developers should report uncertainty alongside simulation results and the conditions under which they are applicable. When plausible discrepancies could change a pass/fail decision, the test verdict may remain unresolved. ADS developers can use these cases to prioritise additional measurements, model improvements, or physical testing.

Future work should investigate how physical experiments can be selected to reduce uncertainty about a stated claim under a fixed budget, comparing adaptive selection with random or coverage-based selection using an independent evaluation set.

\subsection{ADS Architecture and Gap Propagation}
\label{modular-e2e-testing}

Our architectural comparison includes 29 studies of modular ADS and eight of end-to-end ADS, with two studies included in both groups. Direct or matched comparisons are more common in studies of modular ADS, while comparisons based on task or SUT performance are more common in studies of end-to-end ADS. However, the small number of studies and the overlap between groups do not allow general conclusions about evaluation quality.

Explicit interfaces provide a methodological distinction. In modular ADS, detections, trajectories, and commands can be compared or substituted at defined boundaries, making some discrepancies easier to isolate. These interfaces also allow evaluations that bypass relevant parts of the reality gap: a planner evaluated with idealised detections leaves the effect of realistic perception errors unresolved. However, component-level correspondence does not necessarily establish system-level correspondence because timing, correlated errors, and feedback affect component interactions.

Learned components introduce an additional challenge. Even when synthetic and real inputs look similar, they can produce different internal representations and downstream responses~\cite{11624013}. Comparing inputs alone may therefore fail to identify differences that affect ADS behaviour. Comparing internal representations can complement behavioural measurements, but these differences should also be evaluated for their effect on ADS decisions and test verdicts.

End-to-end ADS do not necessarily expose conventional functional boundaries, making it harder to identify which functions are responsible for observed discrepancies. Large backbones and vision-language planners further complicate this analysis~\cite{2026-Gao-OJ-ITS}, as architectural labels alone do not indicate which internal representations or interfaces can be evaluated. Controlled perturbations can help investigate how changes in simulated inputs affect ADS behaviour without requiring access to internal states. However, these effects must also be compared with those observed under corresponding physical conditions~\cite{lambertenghi2025MultimodalityEvaluation,2025-Lambertenghi-ICST}.

Two research directions emerge: determining when correspondence between individual components also leads to correspondence at the system level, and developing methods to identify discrepancies that affect testing outcomes without relying on modular architectures. Applying comparable perturbations to different ADS architectures would help investigate how discrepancies propagate through these systems. Modular simulation environments such as AlpaSim~\cite{agendaAlpaSim} can support these experiments even when evaluating end-to-end ADS. Tool developers should therefore provide synchronised observations, reproducible configurations, and controlled model substitution to help ADS developers identify discrepancies and determine their effects on testing conclusions.

Future work should have controlled experiments that vary discrepancies individually and jointly to establish when their effects amplify or cancel, and whether component-level correspondence remains sufficient to predict system behaviour and test verdicts.

\subsection{Evidence for Credibility and Real-World Claims}
\label{discussion-assessment}

Regulators and assessors need to determine whether a testing campaign provides sufficient evidence to support a specific claim about real-world performance. Reducing a reality gap metric is only useful for this purpose if its relationship to the claim is established. Agreement between simulation and reality is only one part of this assessment, as requirements, scenario selection, test oracles, and conditions not covered by testing also affect the conclusions that can be drawn.

The UNECE NATM guidelines recommend virtual--physical comparisons and justification of extrapolation beyond validation scenarios~\cite{agendaUneceNATM}. Euro NCAP's virtual-testing protocol combines model qualification with physical verification and permits idealised perception within its specified procedure~\cite{agendaEuroNCAP2026}. Such qualification supports a bounded assessment; it does not establish perception robustness or unrestricted ADS performance. NVIDIA's Halos SAFE documentation and Bosch's credibility framework similarly relate simulation assessment to the claim and decision it supports~\cite{agendaNvidiaArchitecture,agendaNvidiaCredibility,agendaBoschCredibility}.

These approaches raise three research questions: which physical comparisons are sufficient to evaluate the discrepancies relevant to a specific claim, how can evidence be extended beyond the conditions used for comparison, and how should the remaining uncertainty influence acceptance decisions? Researchers should develop experimental designs and methods for generalising results to address these questions. Assessment bodies can define what evidence and justification are required, while industry provides the comparisons, applicability conditions, and traceability needed to evaluate whether these requirements are met.

Campaign-level conclusions additionally require a sampling argument. Scenarios selected to expose failures do not directly estimate their operational frequency. Waymo's collision-avoidance methodology connects scenarios and acceptance criteria to a defined objective~\cite{agendaWaymoCAT}, while the Safety First for Automated Driving white paper identifies representativeness and simulation uncertainty as challenges for statistical safety arguments~\cite{agendaSaFAD}. Results cannot be treated as independent evidence when tests share modelling errors, and repeated success does not resolve conditions omitted from the campaign.

A clearly defined operational design domain helps establish the scope of a testing campaign and interpret its coverage. ODD specifications and structured scenario descriptions~\cite{reich2023concept} should therefore guide scenario generation, selection, and reporting. However, coverage alone is not sufficient to establish confidence in a performance estimate. Operational exposure, sampling probabilities, dependencies between tests, and the remaining reality gap must also be considered. Researchers should investigate how these factors affect the estimated performance, while assessors should distinguish between evidence that a testing campaign can identify failures and evidence that supports estimates of operational risk.

A future research objective is to develop and evaluate claim-specific qualification protocols that link comparison references, acceptance criteria, sampling assumptions, and applicability conditions to engineering decisions.

\subsection{Maintaining Evidence as the Reality (Gap) Changes}
\label{discussion-continuous-credibility}

The reality gap depends on the ADS, simulation models, and operating conditions being compared. Changes such as retraining the ADS, replacing a sensor model, or modifying the vehicle can affect which discrepancies are relevant. Previous evidence of correspondence between simulation and reality may therefore remain valid for some claims but require reassessment for others. Real-world observations may also reveal conditions that were not considered in the original evaluation.
Learned simulators introduce an additional concern: matching recorded observations does not establish correspondence when the ADS departs from the recorded trajectory. Controllable multi-sensor generation, as described in Waymo's world-model announcement~\cite{agendaWaymoWorldModel}, motivates evaluating novel viewpoints, actor reactions, temporal consistency, and cross-sensor agreement. Evaluation data should remain distinct from data used to train or calibrate the policy and simulator.

Simulator updates can improve correspondence in some conditions while changing it in others. The applicability of earlier evidence must therefore be reconsidered for claims affected by the update~\cite{11029921}. Continual learning also raises questions about retaining previously established behaviour~\cite{2021-Stocco-JSEP,2020-Stocco-GAUSS}. Researchers should develop change-impact methods that identify which evidence remains applicable and which claims require new paired measurements, rather than assume either complete reuse or complete revalidation.

Changes in operating conditions are harder to identify than versioned software updates. Distribution monitoring and misbehaviour detection~\cite{stocco2020misbehaviour,2022-Stocco-ASE,2024-Grewal-ICST} offer starting points for identifying conditions that warrant reassessment. However, a distribution shift does not necessarily change a verdict, and a consequential discrepancy may escape an aggregate distance measure. Monitoring methods should therefore be evaluated for their ability to detect changes relevant to supported claims, including missed changes, false alarms, and detection delay.

For tool developers, versioned configurations, data provenance, and records linking claims to supporting comparisons would enable targeted reassessment. ASAM's work on exchanging simulation-quality information provides a relevant standardisation direction~\cite{agendaAsamQSQ}; the relationship between exchanged metrics and testing validity still requires evaluation. For ADS developers, field observations and unresolved disagreements should guide additional measurements and physical comparisons. Shared paired executions and documented changes would help assess whether these procedures preserve credible evidence as the reality gap evolves.

Future work should compare targeted and full revalidation after controlled changes to the ADS, simulator, or sensors, measuring both the physical-testing effort saved and the consequential verdict disagreements missed on an independent evaluation set.
\section{Related Work}
\label{related-work}

Existing secondary studies examine ADS testing, simulation, and transfer between simulated and real environments from complementary perspectives. \autoref{tab:related-work-positioning} compares their scope with this mapping, focusing on whether they systematically characterise reality gap targets, mitigation, evaluation, execution configurations, and component-level reference evidence.

\head{ADS and scenario-based testing}
Tang et al.~\cite{ADS_testing_survey_2022} organise ADS testing into module-level and system-level techniques, while Zhao et al.~\cite{ZHAO2024122836} place testing within a broader overview of autonomous driving architectures and functions. Zhong et al.~\cite{zhong2021survey} focus on scenario-based testing in simulation, formalising the approach and classifying the corresponding literature. These surveys establish the broader testing landscape, but do not systematically relate reality gap claims to the simulated and real components used in their evaluation.

\head{Testing ML-based systems}
Riccio et al.~\cite{RiccioEMSE20} map test adequacy criteria, input generation, and oracles across ML-based systems, including autonomous driving. Zhang et al.~\cite{zhangTSE22} organise the field around testing properties, components, and workflows. These studies provide relevant perspectives on evaluating learned components, but their scope extends across application domains and does not centre on correspondence between simulated and physical ADS execution.

\head{Simulation platforms}
Kaur et al.~\cite{DBLP:journals/corr/abs-2101-05337} compare driving simulators against defined requirements. Rosique et al.~\cite{s19030648} review perception systems and simulation capabilities for autonomous vehicle research. Their unit of analysis is primarily the platform or sensing technology. Our mapping instead examines how individual studies configure simulated and real components and which references and measurements they report.

\head{Sim-to-real transfer and digital twins}
Hu et al.~\cite{hu2023sim2real} provide the closest perspective to our study, reviewing technologies for bridging the reality gap in autonomous driving, including transfer methods, digital twins, and parallel intelligence. Their synthesis is organised primarily around technologies for transferring knowledge and constructing virtual representations. We complement this perspective by examining the reality gap as a testing problem: which discrepancies are targeted, how studies measure or mitigate them, and how the proposed approach differs from its evaluation reference. Gu et al.~\cite{gu2025xrtesting} map 34 studies on software testing for extended reality (XR) applications, covering testing techniques, objectives, evaluation methodologies, and supporting tools and datasets. Their study concerns the correctness and quality of applications operating in virtual, augmented, or mixed environments. Our mapping addresses a different relationship: whether simulated representations and executions correspond to their real-world counterparts in ADS testing. Mixed reality can form part of an ADS test configuration without making the XR application itself the system under test.

The contribution of this mapping is therefore the joint analysis of research claims, execution configurations, and reported evaluation evidence. Applying the same component-level classification to an approach and its reference makes their relationship explicit, while distinguishing the presence of real-world content from a comparison that addresses the targeted discrepancy. This characterisation supports interpretation of the reported evidence; it does not rank study quality or establish whether discrepancies propagate to changed ADS behaviour or test verdicts.

\begin{table*}[tbp]
\caption{Scope of related secondary studies and this mapping. ``Yes'' indicates systematic classification, ``Partial'' discussion without classification, and ``--'' a dimension outside the scope or not systematically addressed.}
\label{tab:related-work-positioning}
\centering
\scriptsize
\renewcommand{\arraystretch}{1.08}
\setlength{\tabcolsep}{2.0pt}
\begin{tabularx}{\textwidth}{p{4.75cm}XXXXXX}
\toprule
\textbf{Secondary study scope} &
\textbf{reality gap focus} &
\textbf{Gap targets} &
\textbf{Mitigation} &
\textbf{Evaluation} &
\textbf{Execution} &
\textbf{Component evidence} \\
\midrule
ADS testing surveys~\cite{ADS_testing_survey_2022,ZHAO2024122836} &
-- & Partial & Partial & Partial & -- & -- \\
Scenario-based testing surveys~\cite{zhong2021survey} &
-- & Partial & Partial & -- & -- & -- \\
Testing ML-based systems mappings~\cite{RiccioEMSE20,zhangTSE22} &
-- & Partial & -- & Partial & -- & -- \\
Simulation platform surveys~\cite{DBLP:journals/corr/abs-2101-05337,s19030648} &
-- & -- & Partial & -- & -- & -- \\
Sim-to-real and digital-twin surveys~\cite{hu2023sim2real} &
Partial & Partial & Yes & Partial & -- & -- \\
\textbf{This mapping} &
Yes & Yes & Yes & Yes & Yes & Yes \\
\bottomrule
\end{tabularx}
\end{table*}

\section{Conclusion}
\label{sec:conclusion}

This systematic mapping study examined how ADS research measures and mitigates the simulation-to-reality gap in testing and validation. We analysed 235 primary studies selected from 4,195 publications, characterising their objectives, targets, mitigation techniques, experimental setups, and evaluation evidence. Our component-level classification distinguishes the proposed approach from its evaluation reference across scenario, actors, perception, localization, actuation, and ECU/control.

The literature concentrates on mitigation, addressed by 179 studies alone or alongside measurement, with scenario and actor realism and perception as the most common targets. Experimental setups combine simulated models, recorded observations, and physical equipment in ways that a single simulated-versus-real label cannot adequately describe. These combinations must instead be interpreted against the discrepancy targeted and the claim evaluated.
Reducing a discrepancy does not by itself establish credible test results or ADS trustworthiness. Recorded observations can support sensor or perception comparisons, whereas closed-loop claims require evidence addressing feedback between system actions and subsequent observations. Similarly, physical equipment and multiple metric families do not establish that discrepancies are traced through ADS behaviour to test verdicts. Our mapping characterises the reported evidence; it does not determine how frequently these causal relationships are established or rank individual studies by quality.

Future research should connect gap mitigation to its testing consequences, quantify uncertainty about the remaining gap, and establish when evidence remains applicable across architectures, configurations, and operating conditions. Where feasible, matched simulated and physical executions should assess behavioural correspondence and verdict agreement. The classification and study mappings provide a basis for organising these comparisons. The central challenge is to establish which discrepancies can change a testing conclusion and what evidence justifies relying on that conclusion in the real world.

\section*{Acknowledgements}
\addcontentsline{toc}{section}{Acknowledgements}
This research was funded by the Bavarian Ministry of Economic Affairs, Regional Development and Energy. We thank Donghwan Shin, Neil Walkinshaw, Christopher M. Poskitt, and Amir Soltani for insightful discussions that helped shape parts of the discussion in this paper.

\bibliographystyle{ACM-Reference-Format}
\bibliography{paper,primary_studies}


\end{document}